\documentclass[english, 12pt]{article}

\usepackage[bookmarks=true, colorlinks=true, linkcolor=blue, citecolor=blue, breaklinks=true]{hyperref}
\usepackage{natbib}
\usepackage[T1]{fontenc}
\usepackage[utf8]{inputenc}
\usepackage{babel}
\usepackage{float}
\usepackage{amsthm}
\usepackage{amsmath}
\usepackage{amssymb}
\usepackage{mathtools}

\usepackage{graphicx, graphics}
\usepackage{setspace}
\usepackage{esint}
\usepackage{algorithm}
\usepackage[noend]{algpseudocode}
\usepackage{comment}
\usepackage{siunitx}
\usepackage[flushleft]{threeparttable}
\usepackage{multirow}
\usepackage{subcaption}
\usepackage[title,titletoc]{appendix}
\usepackage{fullpage}
\theoremstyle{plain}

\usepackage{placeins}

\newcommand{\resultfig}[3]{
	\begin{figure}[H]
		\centering
		\includegraphics[width=0.75\textwidth]{#1}
		\caption{#2}
		\label{#3}
	\end{figure}
}

\begin{document}
	
	\title{Deep Learning for Dynamic Programming with Recursive Utility Using First-order Conditions\thanks{The research is supported by the National Natural Science Foundation of China (Grant No. 72573011).}}
	
	\author{Xianhua Peng\thanks{HSBC Business School, Peking University, University Town, Nanshan District, Shenzhen, 518055, China. Email: xianhuapeng@pku.edu.cn.}
	\and  Wu Guo\thanks{HSBC Business School, Peking University, University Town, Nanshan District, Shenzhen, 518055, China. Email: wuguo@stu.pku.edu.cn.}
    \and  Songyan Wang\thanks{HSBC Business School, Peking University, University Town, Nanshan District, Shenzhen, 518055, China. Email: songyan812@stu.pku.edu.cn.}
    \and  Jianfei Zhu\thanks{HSBC Business School, Peking University, University Town, Nanshan District, Shenzhen, 518055, China. Email: jfzhu@stu.pku.edu.cn.}
	}
	
	\date{This version: \today}
	
	\maketitle

		\begin{abstract}
        This paper proposes the certainty-equivalent first-order learning (CEFOL) algorithm, a deep learning algorithm for solving discrete-time dynamic programming problems with recursive utility. Dynamic programming with recursive utility is challenging because nonlinear certainty equivalent appears in the Bellman equation and the first-order optimality conditions but is difficult to evaluate. By introducing a separate neural network to represent the certainty equivalent, CEFOL enables the exploitation of the Bellman and model-specific first-order optimality conditions.            
        In addition to certainty equivalent, 
        CEFOL also uses neural networks to learn the        
        value functions, policy functions, and Lagrange multipliers by using model-specific first-order conditions to construct residuals for minimization. 
        By using first-order and KKT residuals to learn the policy, CEFOL directly accommodates general equality and inequality constraints on the controls, including occasionally binding constraints, without requiring penalty functions or problem-specific reformulations. We apply the algorithm to risk-sensitive and Epstein--Zin consumption-saving problems, a small-noise robust-control problem, and a DSGE model with recursive preferences and stochastic volatility. Across these applications, out-of-sample Bellman diagnostics and model-specific optimality residuals, including Euler or first-order residuals where applicable, are generally of order $10^{-4}$ to $10^{-3}$ over the relevant state regions, with larger values mainly near binding constraints, and the learned value and policy functions closely match VFI benchmarks when available. The CEFOL algorithm also works for dynamic programming problems with expected utility, as expected utility is a special case of recursive utility.

		
	\end{abstract}
    \section{Introduction}
    \label{sec:introduction}
    
    Discrete-time dynamic programming problems are central to economics and finance, with applications in stochastic growth, real business cycle theory, portfolio choice, optimal execution, asset pricing, climate policy, and industrial organization. Dynamic programming methods provide a unified way to study intertemporal decisions under uncertainty; see, for example, \citet{stokey1989recursive}, \citet{Hansen-Sargent-2013}, \citet{Ljungqvist-Sargent-2018_updated}, and \citet{miao2020economic}. In numerical applications, however, these problems are difficult when the state space is high-dimensional, the policy rule is nonlinear, the model contains multiple sources of risk, and the relevant ergodic region is hard to represent with a fixed grid.
    
    Traditional grid-based numerical methods suffer from the curse of dimensionality: the number of grid points grows exponentially with the number of state variables. This limitation affects not only the value-function approximation, but also the conditional-expectation evaluation, policy improvement, interpolation, and the representation of the state region visited by the model. As a result, conventional grid-based methods are often limited to low- or moderate-dimensional settings. Recent deep learning methods use neural networks to approximate value functions, policy functions, conditional expectations, and model-specific quantities in high-dimensional dynamic models. These methods show that deep learning can be effective in expected-utility and time-additive settings.
    
    However, many economic and financial applications require preference specifications beyond standard expected utility. Recursive preferences, such as Epstein--Zin, risk-sensitive, and robust-control preferences, separate risk aversion from intertemporal substitution and allow agents to respond to long-run risk, model uncertainty, and stochastic volatility. Expected utility is nested in this recursive formulation as a special case: when the certainty-equivalent transformation is linear and the time aggregator is additively separable, the continuation term reduces to a standard conditional expectation. Recursive utility has therefore been widely used in asset pricing, long-horizon portfolio choice, robust control, climate policy, and other dynamic decision problems under uncertainty.
    
    Despite their theoretical and practical appeal, discrete-time dynamic programming problems with recursive utility are difficult to solve numerically. First, recursive utility generally has no explicit expression in terms of future states and controls. Unlike expected utility, where the value function can sometimes be represented as a discounted stream of future utilities once a policy is fixed, recursive utility is defined implicitly through the Bellman equation. Second, the Bellman equation contains a certainty-equivalent term, which is a nonlinear transformation of a conditional expectation of a nonlinear function of next period value. This certainty-equivalent value is a nonlinear function of both the current state and the current control. Third, because nonlinear transformations and conditional expectations do not commute, a simple sample average of next period values does not recover the certainty-equivalent value. Fourth, policy and multiplier updates must evaluate certainty-equivalent values at candidate controls, not only along the policy path. In a direct differentiated formulation, the first-order conditions also involve derivatives of the certainty-equivalent value and the future value gradient, which are computationally demanding to approximate. Fifth, value approximation, certainty-equivalent approximation, and policy updates are tightly coupled: errors in one component can feed into the others and change the distribution of simulated states. These challenges motivate treating the state-control certainty-equivalent value as a distinct learned component.

    These computational difficulties are compounded by a general limitation of gradient-based deep learning methods. Neural-network training typically involves nonconvex optimization, so the computed solution is not guaranteed to be globally optimal. This makes independent accuracy evaluation especially important. For recursive utility, the learned solution must jointly satisfy the Bellman equation, the nonlinear certainty-equivalent relation, and the model-specific optimality conditions. This motivates our use of VFI comparisons, out-of-sample Bellman errors, and first-order residuals in the numerical evaluation.
    
    This paper proposes the certainty-equivalent first-order learning (CEFOL) algorithm, a deep learning algorithm for solving discrete-time dynamic programming problems with recursive utility and general constraints on the controls. CEFOL approximates policy functions, value functions, multipliers, and state-control certainty-equivalent values with neural networks. The policy and multiplier networks are trained using model-specific first-order residuals, while the value and certainty-equivalent networks maintain Bellman consistency and certainty-equivalent consistency. We use the term first-order residual broadly: Euler residuals are intertemporal special cases, and KKT terms are included when constraints are explicitly imposed, allowing the method to handle a general feasible set defined by equality constraints and inequality constraints. The method is therefore policy-focused and disciplines the learned decision rule through economically meaningful marginal conditions while using a learned certainty-equivalent value to handle the nonlinear certainty-equivalent term.
    
    The key idea is to decouple the nonlinear certainty-equivalent component from the policy update. CEFOL introduces a neural network for the state-control certainty-equivalent value. This network is not an additional value function; instead, it serves as an auxiliary state-control component for constructing both the Bellman target and the model-specific optimality residuals evaluated at candidate controls. By learning this component directly, CEFOL amortizes the nonlinear certainty-equivalent calculation and avoids repeatedly solving an inner nonlinear expectation problem during each policy update.
    
    We evaluate CEFOL on several discrete-time dynamic programming problems with recursive utility: risk-sensitive and Epstein--Zin consumption-saving models, a small-noise robust-control model, and an Epstein--Zin DSGE model with stochastic volatility. Across these applications, CEFOL produces stable value, policy, multiplier, and certainty-equivalent approximations. The learned solutions generate small out-of-sample Bellman errors and first-order residuals, and closely match VFI benchmarks when such benchmarks are available. When a model-specific first-order condition takes the form of a standard intertemporal Euler equation, we report the corresponding diagnostic as an Euler residual.
    
    This paper makes six main contributions. First, the paper develops a deep learning framework for discrete-time dynamic programming with recursive utility. The framework is written for a general recursive utility specification with a nonlinear time aggregator and a nonlinear certainty-equivalent transformation, and can be specialized to Epstein--Zin preferences, risk-sensitive preferences, robust-control preferences, and related recursive formulations.
    
    Second, CEFOL combines certainty-equivalent approximation with model-specific optimality conditions. Existing deep learning methods can use first-order residuals or residuals from model-specific systems of equations in standard expected-utility models. Recursive utility is more difficult because the optimality system itself contains nonlinear state-control certainty-equivalent values. CEFOL addresses this difficulty by learning this component directly and feeding it into the model-specific first-order residuals.
    
    Third, CEFOL incorporates simulation-based conditional-moment estimation into the policy and multiplier updates under recursive utility. The learned state-control certainty-equivalent value enters the model-specific first-order residuals, and conditionally independent residual evaluations are used to estimate the squared conditional mean of these residuals. This adapts residual-based simulation methods to recursive utility settings in which the optimality conditions contain nonlinear certainty-equivalent terms.
    
    Fourth, the algorithm provides a modular architecture for constrained dynamic decision problems. The value network learns Bellman consistency, the certainty-equivalent network learns the nonlinear certainty-equivalent value, the multiplier network learns constraint multipliers, and the policy network learns the control rule through first-order residuals, with KKT terms used when constraints are explicitly imposed. This separation allows the method to handle vector controls, equality constraints, inequality constraints, and occasionally binding constraints.
    
    Fifth, the paper evaluates the learned solutions using out-of-sample residual diagnostics computed from the original Bellman equation and model-specific optimality conditions. We report Bellman errors together with first-order and constraint residuals on independently simulated test states. Euler residuals are reported when the first-order condition takes the standard intertemporal Euler-equation form. These diagnostics provide separate evidence on value-function consistency, certainty-equivalent accuracy, and policy optimality.
    
    Sixth, the framework covers expected-utility dynamic programming as a special case. When the certainty-equivalent transformation is linear and the time aggregator is additively separable, the recursive specification reduces to the expected-utility formulation. In this case, the certainty-equivalent network represents the conditional expectation of next period value. This shows that CEFOL provides a unified architecture for both recursive-utility and expected-utility dynamic programming problems.

    CEFOL differs from existing simulation-based deep learning methods in how it handles the certainty-equivalent value. The framework of \citet{maliar2021deep} provides an important conditional-moment approach to dynamic economic models. Applying this idea directly to recursive utility, however, would still require evaluating nonlinear certainty-equivalent terms inside the first-order residuals or residuals from model-specific systems of equations. CEFOL instead learns the state-control certainty-equivalent value separately and uses conditionally independent residual evaluations only when updating the policy and multiplier networks.
    
    To the best of our knowledge, \citet{friedl2023deep} is the closest discrete-time deep learning study for recursive utility. They solve an integrated assessment model with Epstein--Zin utility, Bayesian learning, and climate tipping risk. Their approach builds on \citet{azinovic2022deep}, approximates policy functions, Lagrange multipliers, and the value function with neural networks, and solves the model by minimizing a joint residual system formed from model-specific equilibrium conditions.

    CEFOL differs from this approach in several important respects. First, CEFOL introduces a separate certainty-equivalent network that directly represents the conditional state-control certainty-equivalent value generated by the nonlinear transformation of next-period value. In contrast, \citet{friedl2023deep} evaluate the certainty-equivalent term inside a joint model-specific residual system rather than learning a separate certainty-equivalent approximator. This distinction also separates CEFOL from the companion CEL algorithm \citep{peng2026deeplearningdynamicprogramming} discussed below: CEL uses the learned certainty-equivalent value in the Bellman objective for policy improvement, whereas CEFOL feeds the learned certainty-equivalent value into model-specific first-order and KKT residuals for policy and multiplier learning. Second, CEFOL separates certainty-equivalent approximation, Bellman consistency, and policy and multiplier updates into distinct training components. In contrast, \citet{friedl2023deep} approximate several model-specific quantities within a deep equilibrium network and update them through a single joint residual loss. Third, CEFOL is designed as a general deep learning framework for recursive utility models with vector controls, equality constraints, inequality constraints, and occasionally binding constraints, whereas \citet{friedl2023deep} implement a model-specific system of equations for an integrated assessment application. Fourth, CEFOL updates the value function through a separate objective, while \citet{friedl2023deep} include the value function as one output of their deep equilibrium network, together with choice variables and Lagrange multipliers, and discipline it within the joint residual loss. Fifth, CEFOL evaluates the learned solution using out-of-sample Bellman errors together with first-order and constraint residuals, providing separate diagnostics for value-function consistency and policy optimality. Sixth, CEFOL incorporates target networks and delayed policy updates, which improve the stability of training by stabilizing target values and avoiding overly frequent policy updates. In contrast, \citet{friedl2023deep} optimize the model-specific objectives jointly through a single residual loss without these separate stabilization devices.

    The CEFOL algorithm complements the certainty-equivalent learning  (CEL)  algorithm proposed in \cite{peng2026deeplearningdynamicprogramming}. Both algorithms share the idea of learning the state-control certainty-equivalent value by neural networks, but they have essential difference. CEL learns the optimal policy through directly maximizing the right-hand side of Bellman equation without using first-order or KKT conditions; in contrast,  CEFOL uses first-order and KKT conditions to construct residuals and updates the policy and multiplier networks by minimizing these residuals. 
    
    The rest of the paper is organized as follows. Section~\ref{subsec:literature_review} reviews related literature. Section~\ref{sec:model_setting} presents the general recursive utility model and the first-order/KKT representation. Section~\ref{sec:algorithm} develops the CEFOL algorithm, including the neural-network architectures, certainty-equivalent and Bellman losses, first-order/KKT residuals, stabilization techniques, and training procedure. Section~\ref{sec:numerical_results} reports numerical results for the recursive consumption-saving models, the small-noise robust-control model, and the DSGE model with recursive preferences and stochastic volatility.
    
    \subsection{Literature Review}
    \label{subsec:literature_review}
    
    This subsection reviews three related strands of literature: recursive utility theory and applications, traditional numerical methods for dynamic programming, and machine-learning methods for dynamic economic models.
    
    Recursive utility builds on the seminal work of \citet{krepsPorteus1978}, \citet{epstein1989substitution}, and \citet{Weil1990}. By separating risk aversion from the elasticity of intertemporal substitution, recursive preferences provide a flexible tool for studying intertemporal decisions under risk and uncertainty. They have been widely used in asset pricing, long-run risk, portfolio choice, robust control, fiscal policy, and other dynamic decision problems \citep[see, e.g.,][]{WEIL1989,bansal2004risks,DuffieEpstein,schroder1999optimal,hansen1995discounted,hansen2008robustness,Hansen-Sargent-2013,kaplan2014model}. Recursive utility has also been applied to strategic asset allocation with time-varying investment opportunities \citep[see, e.g.,][]{campbell1999consumption,campbell2003multivariate}, integrated assessment models with climate risk \citep[see, e.g.,][]{Cai-Lontzek-2019,zhao2023SCC}, and consumption-investment problems in incomplete markets \citep[see, e.g.,][]{kraft2013consumption,kraft2017optimal,xing2017consumption,matoussi2018convex,feng2021optimal,feng2024consumption}. Broader discussions of recursive preferences and their implications can be found in \citet{Backus2016}.
    
    A related theoretical literature studies the dynamic programming foundations of recursive utility. \citet{hansen2012recursive} analyze recursive utility in Markov environments with stochastic growth, and \citet{dumas1998efficient} study efficient allocations with recursive preferences. More recent work establishes existence, uniqueness, and recursive representations in models with unbounded states, stochastic volatility, and other general Markov environments \citep[see, e.g.,][]{christensen2022existence,pohl2024existence,stachurski2024asset,stachurski2021dynamic,ma2021dynamic,ma2022unbounded,Jaskiewicz2024recursive}. \citet{Bloise2024Koopmans} show that value multiplicity reflects non-uniqueness in the recursive utilities induced by the Koopmans aggregator, rather than ambiguity inherent in the Bellman operator. These papers provide the theoretical basis for using Bellman equations. Our focus is different: we develop a deep learning method for recursive utility models that learns policies, multipliers, and state-control certainty-equivalent values jointly.
    
    Classical numerical methods for dynamic programming include value function iteration, policy iteration, projection methods, perturbation methods, and Euler-equation methods \citep[see, e.g.,][]{Judd-1998,Miranda-Fackler-2002,carroll2006,Carroll2011,Carroll2022_updated,miao2020economic}. Projection methods approximate value or policy functions with basis expansions \citep[see, e.g.,][]{judd1992projection}, while endogenous-grid methods provide efficient procedures for standard consumption-saving models \citep[see, e.g.,][]{carroll2006}. Perturbation methods are widely used in DSGE models with recursive preferences \citep[see, e.g.,][]{tallarini2000risk,andreasen2012effects,caldara2012computing}. These methods are powerful in many low- and moderate-dimensional settings, but recursive utility creates additional difficulties because nonlinear certainty-equivalent terms enter the Bellman equation and, after differentiation, the optimality conditions. Accuracy evaluation in this literature often relies on residual diagnostics; \citet{santos2000accuracy} provides theoretical foundations for using Euler equation residuals to assess numerical solutions.
    
    A growing machine-learning literature uses flexible function approximation to solve dynamic economic models. \citet{HanE2016} use deep learning to approximate optimal controls in high-dimensional finite-horizon stochastic control problems. For representative-agent and macro-finance models, neural networks and Gaussian processes have been used to approximate value functions, policy functions, conditional expectations, and expectation terms in Euler equations \citep[see, e.g.,][]{Renner-Scheidegger-2018,scheidegger-Bilionis-2019,Lepetyuk2020,valaitis2024machine}. \citet{maliar2021deep} propose a general deep learning framework based on lifetime utility maximization, Euler-equation error minimization, and Bellman-residual minimization, and introduce the all-in-one expectation operator for high-dimensional integration. CEFOL is related to this simulation-based perspective, but the proposed method targets a different difficulty: with recursive utility, the residuals themselves involve nonlinear state-control certainty-equivalent values, so this quantity must be learned and inserted into the model-specific first-order system.
    
    Related work applies deep learning to solve finite-horizon, overlapping-generations, and heterogeneous-agent models. These studies use neural networks to approximate value functions, policy functions, and other model-specific quantities, and train them by minimizing residuals from model-specific systems of equations \citep[see, e.g.,][]{duarte2021simple,azinovic2022deep,Azinovic2023,HallHoffarth2023}. Approximate dynamic programming and reinforcement learning provide another broad set of tools \citep[see, e.g.,][]{Sutton-Barto-1998,Powell-2011,Bertsekas-2012}, with applications in economics and finance \citep[see, e.g.,][]{feng2023optimal,aboussalah2022value,jiang2025high,dixon2023time}. Most of these methods are designed for expected-utility or time-additive objectives, where the expectation terms entering the Bellman equation or first-order conditions are simpler than the nonlinear certainty-equivalent terms generated by recursive utility.
    
    A related continuous-time machine-learning literature studies HJB equations, PDEs, BSDEs, and stochastic control problems. The closest continuous-time study for recursive utility is \citet{duarte2024machine}, who solve finance models with Epstein--Zin preferences by minimizing Hamilton--Jacobi--Bellman residuals. Related continuous-time methods, mostly under expected utility or time-additive objectives, use deep learning to solve high-dimensional PDEs, BSDEs, HJB equations, and stochastic control problems \citep[see, e.g.,][]{EHan2017_published,BeckEJ2019_published,Hure-Pham-Bachouch-Lang-2021,Bachouch_2021,Reppen2023}. These continuous-time approaches are based on continuous-time differential or stochastic-differential representations. CEFOL instead focuses on discrete-time dynamic programming problems with recursive utility, where the state-control certainty-equivalent value enters both the Bellman target and the model-specific first-order residuals.
    
    Overall, this paper contributes to the literature by extending deep learning methods for discrete-time dynamic programming to recursive utility models with nonlinear certainty-equivalent values. The proposed method is designed for vector controls, high-dimensional state variables, stochastic dynamics, and occasionally binding constraints. The proposed method provides a simulation-based computational tool for dynamic decision problems in which both Bellman consistency and policy accuracy are central.

	\section{Model Setting}
	\label{sec:model_setting}

    This section introduces the general recursive utility environment used to define the CEFOL residuals. Section~\ref{subsec:problem_formulation} states the discrete-time dynamic programming problem, the certainty-equivalent value, and the feasible set. Section~\ref{subsec:cefol_general_foc_kkt} gives the general first-order/KKT representation used to construct the policy and multiplier updates in CEFOL. Section~\ref{subsec:cefol_computational_challenges} explains why recursive utility is computationally different from expected utility.

	\subsection{Problem Formulation and Recursive Preferences}
	\label{subsec:problem_formulation}
	
	We consider a discrete-time infinite-horizon dynamic programming problem. Time is indexed by $t=0,1,2,\ldots$. At each date, the agent observes a state vector
	\begin{equation}
		s_t\in\mathcal{S}\subseteq\mathbb{R}^{n_s}
	\end{equation}
	and chooses a control vector
	\begin{equation}
		c_t\in\mathcal{A}(s_t)\subseteq\mathbb{R}^{n_c},
	\end{equation}
	where $\mathcal{A}(s_t)$ denotes the feasible set of controls. The decision rule is Markovian and is written as
	\begin{equation}
		c_t=c(s_t).
		\label{eq:cefol_markov_policy}
	\end{equation}
	The state evolves according to
	\begin{equation}
		s_{t+1}=\psi(s_t,c_t,z_{t+1}),
		\label{eq:cefol_general_transition}
	\end{equation}
	where $z_{t+1}$ is an exogenous innovation. Conditional on the current state-control pair $(s_t,c_t)$, the next state is independent of the past history. Additional history dependence can be incorporated by augmenting the state vector.
	
	We adopt a general recursive utility specification and use $V(s_t)$ to denote the value at state $s_t$. The primitive nonlinear transformation is denoted by $f(x)$, where $x$ is a realized next period value. For a given current pair $(s_t,c_t)$, the certainty-equivalent value is defined as
	\begin{equation}
		\mathcal{C}(s_t,c_t)
		=
		f^{-1}
		\left(
		\mathbb{E}
		\left[
		f\left(V(s_{t+1})\right)
		\mid s_t,c_t
		\right]
		\right),
		\qquad
		s_{t+1}=\psi(s_t,c_t,z_{t+1}).
		\label{eq:cefol_state_control_ce}
	\end{equation}
	Thus, $\mathcal{C}(s_t,c_t)$ is a state-control certainty-equivalent value that can be evaluated at candidate controls during policy and multiplier updates in CEFOL.
	
    Recursive utility is represented by
    \begin{equation}
    V(s_t)
    =
    w\left(s_t,c_t,\mathcal{C}(s_t,c_t)\right),
    \label{eq:cefol_recursive_utility_general}
    \end{equation}
    where ($w$) is the time aggregator and \(\mathcal{C}(s_t,c_t)\) is the state-control certainty-equivalent value. This generic recursive representation nests a broad class of preference specifications and stochastic recursive dynamic programs; see \citet{krepsPorteus1978} and \citet{Backus2016}. The associated Bellman equation is
	\begin{equation}
		V(s)
		=
		\max_{c\in\mathcal{A}(s)}
		w\left(
		s,
		c,
		f^{-1}
		\left(
		\mathbb{E}
		\left[
		f(V(s'))
		\mid s,c
		\right]
		\right)
		\right).
		\label{eq:cefol_general_bellman}
	\end{equation}
	The first-order and KKT representation used for policy learning is derived in Section~\ref{subsec:cefol_general_foc_kkt}.
	
	The formulation covers a range of recursive preferences. In the risk-sensitive specification used below, which follows the risk-sensitive and robust-control formulations of \citet{hansen1995discounted,hansen2008robustness,Hansen-Sargent-2013}, the nonlinear transformation can be written as
	\begin{equation}
		f(x)=\exp(-\sigma\beta x),
		\label{eq:cefol_risk_sensitive_f}
	\end{equation}
	where $\beta\in(0,1)$ is the subjective discount factor and $\sigma>0$ controls risk sensitivity, so that the certainty-equivalent value is
	\begin{equation}
		\mathcal{C}(s_t,c_t)
		=
		-\frac{1}{\sigma\beta}
		\log
		\mathbb{E}
		\left[
		\exp\left(-\sigma\beta V(s_{t+1})\right)
		\mid s_t,c_t
		\right].
		\label{eq:cefol_risk_sensitive_example}
	\end{equation}
	In Epstein--Zin preferences \citep{epstein1989substitution,Weil1990}, for positive values, one may use
	\begin{equation}
		f(x)=x^{1-\gamma},
		\label{eq:cefol_ez_f}
	\end{equation}
	where $\gamma>0$ is the coefficient of relative risk aversion, which gives
	\begin{equation}
		\mathcal{C}(s_t,c_t)
		=
		\left(
		\mathbb{E}
		\left[
		V(s_{t+1})^{1-\gamma}
		\mid s_t,c_t
		\right]
		\right)^{\frac{1}{1-\gamma}}.
		\label{eq:cefol_ez_example}
	\end{equation}
	Both specifications fit the same state-control certainty-equivalent notation and differ only in the transformation $f$.
	
	\paragraph{General control constraints}
	\label{par:cefol_general_constraints}
	In the most general case, the control vector may be subject to multiple constraints. These constraints may be simple bounds, such as non-negativity constraints, or more complex restrictions involving several controls jointly, such as budget constraints, portfolio-share constraints, leverage constraints, collateral constraints, or market-clearing restrictions. We write the feasible set as
	\begin{equation}
		\mathcal{A}(s_t)
		=
		\left\{
		c_t:
		q_r(s_t,c_t)=0,
		r=1,\ldots,M_q,
		\quad
		g_m(s_t,c_t)\geq 0,
		m=1,\ldots,M_g
		\right\}.
		\label{eq:cefol_general_feasible_set}
	\end{equation}
	Here $q_r(s_t,c_t)=0$ are equality constraints and $g_m(s_t,c_t)\geq0$ are inequality constraints. This formulation includes simple borrowing constraints as special cases, but it also allows for more general restrictions involving several controls simultaneously. Some constraints can be imposed directly through the output transformation of the policy network. For example, a consumption ratio can be mapped into $(0,1)$ by a sigmoid function, and portfolio weights can be mapped onto a simplex by a softmax transformation. This hard-constraint approach is numerically stable for simple constraints. However, when constraints are economically meaningful or involve several controls jointly, it is often preferable to keep them explicitly and impose them through KKT residuals. Keeping these constraints explicit yields multiplier functions and diagnostics for binding constraints.

    \subsection{General First-Order and KKT Conditions}
	\label{subsec:cefol_general_foc_kkt}
    
    The optimal policy can often be characterized by first-order conditions and Kuhn--Tucker conditions. In the general recursive utility problem, these conditions provide the basis for the policy and multiplier losses in CEFOL. Recursive utility affects these conditions through the state-control certainty-equivalent value, so the residuals may involve current controls, future states, future controls, future values, certainty-equivalent values, and relevant derivatives of the transition function, value function, certainty-equivalent transformation, and time aggregator.

    This subsection has two parts. First, we differentiate the Bellman equation to define the model-specific integrand $F_{t+1,k}$ and present the full KKT system, consisting of the stationarity condition, complementarity conditions, and equality constraints. Second, we show that under a structural condition on the transition function, the unobservable $\partial V(s_{t+1})/\partial s_{t+1}$ can be eliminated from $F_{t+1,k}$.
    
	Write the control vector as
	\[
	c_t=(c_{t,1},c_{t,2},\ldots,c_{t,n_c})^{\top}.
	\]
    Assume that the relevant functions are differentiable and that the derivative can be passed through the conditional expectation. Differentiating the right-hand side of \eqref{eq:cefol_general_bellman} with respect to $c_{t,k}$ gives
	\begin{equation}
		\frac{\partial w\left(s_t,c_t,\mathcal{C}(s_t,c_t)\right)} {\partial c_{t, k} } = \mathbb{E}_t[F_{t+1,k}],
        \qquad
		k=1,\ldots,n_c ,
	\end{equation}
    where
	\begin{equation}
		\begin{aligned}
			F_{t+1,k}
			={}&
			\left.
			\frac{\partial w(s_t,c_t,C)}{\partial c_{t,k}}
			\right|_{C=\mathcal C(s_t,c_t)}
			\\
			&+
			\left.
			\frac{\partial w(s_t,c_t,C)}{\partial C}
			\right|_{C=\mathcal C(s_t,c_t)}
			\frac{1}{f'\left(\mathcal C(s_t,c_t)\right)}
			f'\left(V(s_{t+1})\right)
			\left(
			\frac{\partial V(s_{t+1})}{\partial s_{t+1}}
			\right)^{\top}
			\frac{\partial \psi(s_t,c_t,z_{t+1})}{\partial c_{t,k}}.
		\end{aligned}
		\label{eq:cefol_general_foc_integrand}
	\end{equation}

    Let $\lambda_m(s_t)\geq0$ denote the multiplier associated with the inequality constraint $g_m(s_t,c_t)\geq0$, and let $\nu_r(s_t)$ denote the multiplier associated with the equality constraint $q_r(s_t,c_t)=0$. The KKT conditions can be written as
	\begin{equation}
		\mathbb{E}_t[F_{t+1,k}]
		+
		\sum_{m=1}^{M_g}
		\lambda_m(s_t)
		\frac{\partial g_m(s_t,c_t)}{\partial c_{t,k}}
		+
		\sum_{r=1}^{M_q}
		\nu_r(s_t)
		\frac{\partial q_r(s_t,c_t)}{\partial c_{t,k}}
		=0,
		\qquad
		k=1,\ldots,n_c ,
		\label{eq:cefol_general_stationarity_condition}
	\end{equation}
	\begin{equation}
		g_m(s_t,c_t)\geq0,
		\qquad
		\lambda_m(s_t)\geq0,
		\qquad
		g_m(s_t,c_t)\lambda_m(s_t)=0,
		\qquad
		m=1,\ldots,M_g,
		\label{eq:cefol_general_complementarity}
	\end{equation}
	\begin{equation}
		q_r(s_t,c_t)=0,
		\qquad
		r=1,\ldots,M_q,
		\label{eq:cefol_general_equality_constraints}
	\end{equation}
	
	Euler equations are included as a special case when a first-order condition equates current marginal utility to a discounted conditional expectation of future marginal values. The formulation applies to recursive utility problems with vector controls, nonlinear certainty-equivalent values, and general constraint structures.

    \paragraph{Simplifying $F_{t+1,k}$ under a structural condition.}

Formally, the structural condition is stated as follows. Suppose there exists a scalar function $y_{t+1}=y(s_{t+1},c_{t+1})$ such that the transition function can be written as
\begin{equation}
    \psi_i(s_{t+1},c_{t+1},z_{t+2})
    =
    \widetilde\psi_i\!\left(y_{t+1},\,z_{t+2}\right),
    \qquad i=1,\ldots,n_s,
    \label{eq:cefol_reduced_transition}
\end{equation}
for some functions $\widetilde\psi_i$. Under \eqref{eq:cefol_reduced_transition}, the ordinary chain rule gives
\begin{equation}
    \frac{\partial\psi_i}{\partial s_{t+1,j}}
    =
    \frac{\partial\widetilde\psi_i}{\partial y_{t+1}}\,
    \frac{\partial y_{t+1}}{\partial s_{t+1,j}},
    \qquad
    \frac{\partial\psi_i}{\partial c_{t+1,k}}
    =
    \frac{\partial\widetilde\psi_i}{\partial y_{t+1}}\,
    \frac{\partial y_{t+1}}{\partial c_{t+1,k}} .
    \label{eq:cefol_intermediate_variable}
\end{equation}
The simplification of $F_{t+1,k}$ then proceeds as follows. To eliminate $\partial V(s_{t+1})/\partial s_{t+1}$, we need the derivative of the certainty-equivalent value (which enters both the stationarity condition and the envelope), the envelope theorem at $t+1$, and a relation linking $\partial\mathcal{C}/\partial s_{t+1,j}$ to $\partial\mathcal{C}/\partial c_{t+1,k}$. Let $x_t$ denote a scalar current variable, either a control component or a state component, that affects the next-period state through the transition rule. Differentiating \eqref{eq:cefol_state_control_ce} gives
	\begin{equation}
		\frac{\partial \mathcal C(s_t,c_t)}{\partial x_t}
		=
		\mathbb E_t
		\left[
		\chi_{t+1}^{f}
		\left(
		\frac{\partial V(s_{t+1})}{\partial s_{t+1}}
		\right)^{\top}
		\frac{\partial s_{t+1}}{\partial x_t}
		\right],
		\label{eq:cefol_ce_derivative_general}
	\end{equation}
	where
	\begin{equation}
		\chi_{t+1}^{f}
		=
		\frac{
			f'\left(V(s_{t+1})\right)
		}{
			f'\left(\mathcal C(s_t,c_t)\right)
		}.
		\label{eq:cefol_general_distortion}
	\end{equation}
	In \eqref{eq:cefol_ce_derivative_general}, gradients with respect to vector states are treated as column vectors, so the last two terms form an inner product. Equation \eqref{eq:cefol_ce_derivative_general} is the link between the nonlinear certainty-equivalent term and marginal-value conditions. Setting $x_t=c_{t,k}$ gives the derivative of the certainty-equivalent value that enters the stationarity condition; setting $x_t=s_{t,j}$ gives the derivative that enters the envelope condition.

Let $c_{t+1}=c(s_{t+1})$. Applying the envelope theorem at the next-period state gives, for $j=1,\ldots,n_s$,
	\begin{equation}
	\begin{aligned}
		\frac{\partial V(s_{t+1})}{\partial s_{t+1,j}}
		={}&
		\left.
		\frac{\partial w(s_{t+1},c_{t+1},C)}{\partial s_{t+1,j}}
		\right|_{C=\mathcal C(s_{t+1},c_{t+1})}
		+
		\left.
		\frac{\partial w(s_{t+1},c_{t+1},C)}{\partial C}
		\right|_{C=\mathcal C(s_{t+1},c_{t+1})}
		\frac{\partial \mathcal C(s_{t+1},c_{t+1})}{\partial s_{t+1,j}}
		\\
		&+
		\sum_{m=1}^{M_g}
		\lambda_m(s_{t+1})
		\frac{\partial g_m(s_{t+1},c_{t+1})}{\partial s_{t+1,j}}
		+
		\sum_{r=1}^{M_q}
		\nu_r(s_{t+1})
		\frac{\partial q_r(s_{t+1},c_{t+1})}{\partial s_{t+1,j}} .
	\end{aligned}
	\label{eq:cefol_next_period_envelope}
	\end{equation}
	Derivatives with respect to $s_{t+1,j}$ hold the next-period control fixed at the next-period optimizer.

The envelope \eqref{eq:cefol_next_period_envelope} contains the term $(\partial w/\partial C)(\partial\mathcal{C}/\partial s_{t+1,j})$, which still involves $\partial V(s_{t+2})/\partial s_{t+2}$ through $\partial\mathcal{C}/\partial s_{t+1,j}$. To eliminate it, evaluate \eqref{eq:cefol_ce_derivative_general} at $t+1$ for both a state component and a control component:
    \begin{align}
        \frac{\partial\mathcal C(s_{t+1},c_{t+1})}{\partial s_{t+1,j}}
        &=
        \frac{1}{f'(\mathcal C(s_{t+1},c_{t+1}))}
        \mathbb E_{t+1}
        \!\left[
        f'(V(s_{t+2}))
        \sum_{i=1}^{n_s}
        \frac{\partial V(s_{t+2})}{\partial s_{t+2,i}}
        \frac{\partial\psi_i}{\partial s_{t+1,j}}
        \right],
        \label{eq:cefol_general_ce_state} \\[6pt]
        \frac{\partial\mathcal C(s_{t+1},c_{t+1})}{\partial c_{t+1,k}}
        &=
        \frac{1}{f'(\mathcal C(s_{t+1},c_{t+1}))}
        \mathbb E_{t+1}
        \!\left[
        f'(V(s_{t+2}))
        \sum_{i=1}^{n_s}
        \frac{\partial V(s_{t+2})}{\partial s_{t+2,i}}
        \frac{\partial\psi_i}{\partial c_{t+1,k}}
        \right].
    \end{align}
    Both have the form $(1/f'(\mathcal C))\mathbb{E}_{t+1}[f'(V)\sum_i(\partial V/\partial s_{t+2,i})(\partial\psi_i/\partial\cdot)]$, where $\partial\psi_i/\partial\cdot$ is $\partial\psi_i/\partial s_{t+1,j}$ in the first and $\partial\psi_i/\partial c_{t+1,k}$ in the second. Under \eqref{eq:cefol_intermediate_variable}, the deterministic factors $\partial y_{t+1}/\partial s_{t+1,j}$ and $\partial y_{t+1}/\partial c_{t+1,k}$ factor out of their respective expectations, and the remaining expectations are identical. Taking the ratio therefore gives
\begin{equation}
    \frac{\partial\mathcal C}{\partial s_{t+1,j}}
    =
    \frac{\partial\mathcal C}{\partial c_{t+1,k}}
    \frac{\partial y_{t+1}/\partial s_{t+1,j}}{\partial y_{t+1}/\partial c_{t+1,k}}.
    \label{eq:cefol_general_chain}
\end{equation}

Simultaneously, rearranging the stationarity condition \eqref{eq:cefol_general_stationarity_condition} at $t+1$ yields
    \begin{equation}
    \begin{aligned}
        \left.
        \frac{\partial w}{\partial C}
        \right|_{C=\mathcal C(s_{t+1},c_{t+1})}
        \frac{\partial\mathcal C}{\partial c_{t+1,k}}
        ={}&
        -\left.
        \frac{\partial w}{\partial c_{t+1,k}}
        \right|_{C=\mathcal C(s_{t+1},c_{t+1})}
        \\
        &-\sum_{m=1}^{M_g}\lambda_m(s_{t+1})\frac{\partial g_m(s_{t+1},c_{t+1})}{\partial c_{t+1,k}}
        \\
        &-\sum_{r=1}^{M_q}\nu_r(s_{t+1})\frac{\partial q_r(s_{t+1},c_{t+1})}{\partial c_{t+1,k}} .
    \end{aligned}
        \label{eq:cefol_general_foc_elimination}
    \end{equation}
    Multiplying \eqref{eq:cefol_general_chain} by $\partial w/\partial C$ and using \eqref{eq:cefol_general_foc_elimination} to replace $(\partial w/\partial C)(\partial\mathcal C/\partial c_{t+1,k})$ eliminates the derivative of the certainty-equivalent value from the envelope, yielding
    \begin{equation}
    \begin{aligned}
        \frac{\partial V(s_{t+1})}{\partial s_{t+1,j}}
        ={}&
        \left.
        \frac{\partial w}{\partial s_{t+1,j}}
        \right|_{C=\mathcal C(s_{t+1},c_{t+1})}
        \\
        &+
        \bigg(
        -\left.
        \frac{\partial w}{\partial c_{t+1,k}}
        \right|_{C=\mathcal C(s_{t+1},c_{t+1})}
        \\
        &\qquad
        -\sum_{m=1}^{M_g}\lambda_m(s_{t+1})\frac{\partial g_m(s_{t+1},c_{t+1})}{\partial c_{t+1,k}}
        \\
        &\qquad
        -\sum_{r=1}^{M_q}\nu_r(s_{t+1})\frac{\partial q_r(s_{t+1},c_{t+1})}{\partial c_{t+1,k}}
        \bigg)
        \frac{\partial y_{t+1}/\partial s_{t+1,j}}{\partial y_{t+1}/\partial c_{t+1,k}}
        \\
        &+
        \sum_{m=1}^{M_g}\lambda_m(s_{t+1})\frac{\partial g_m(s_{t+1},c_{t+1})}{\partial s_{t+1,j}}
        +
        \sum_{r=1}^{M_q}\nu_r(s_{t+1})\frac{\partial q_r(s_{t+1},c_{t+1})}{\partial s_{t+1,j}} .
    \end{aligned}
        \label{eq:cefol_envelope_eliminated}
    \end{equation}
    Thus, under the structural condition, the envelope expresses $\partial V(s_{t+1})/\partial s_{t+1,j}$ entirely in terms of $\partial w/\partial s_{t+1,j}$, $\partial w/\partial c_{t+1,k}$, and constraint multipliers, with the derivative of the certainty-equivalent value eliminated. Substituting \eqref{eq:cefol_envelope_eliminated} into \eqref{eq:cefol_general_foc_integrand} then replaces $\partial V(s_{t+1})/\partial s_{t+1}$ in $\mathbb{E}_t[F_{t+1,k}]$ by $\partial w/\partial s_{t+1,j}$, $\partial w/\partial c_{t+1,k}$, and constraint multipliers.

    \subsection{Computational Challenges with Recursive Utility}
    \label{subsec:cefol_computational_challenges}
    
    This subsection expands the five computational challenges summarized in the Introduction. These challenges explain why recursive utility requires a separate treatment of the state-control certainty-equivalent value and why policy and multiplier updates in CEFOL are more complicated than in standard expected-utility models.
    
    The first challenge is that recursive utility generally has no explicit expression in terms of future states and controls. With expected utility, once a policy is fixed, the value function can sometimes be represented as a discounted stream of future utilities generated by that policy. This representation often reduces the certainty-equivalent value to a standard conditional expectation. With recursive utility, this simplification is generally unavailable. The value function is defined implicitly through the Bellman equation, and the current value depends on the next period value function through the certainty-equivalent transformation.

    The second challenge is that the Bellman equation contains a nonlinear certainty-equivalent value. In the notation of \eqref{eq:cefol_state_control_ce}, the certainty-equivalent value is
    \[
    \mathcal C(s_t,c_t)
    =
    f^{-1}
    \left(
    \mathbb E_t
    \left[
    f\left(V(s_{t+1})\right)
    \right]
    \right).
    \]
    This quantity is not only nonlinear in the next period value function, but also a function of both the current state and the current control. In high-dimensional models, directly approximating this state-control function is difficult because the relevant domain includes both simulated states and candidate controls used during optimization.
    
    The third challenge is that nonlinear transformations and conditional expectations do not commute. With expected utility, the certainty-equivalent value is linear in the conditional expectation, so a sample average of simulated next-period values provides a natural Monte Carlo approximation. With recursive utility, a sample average of \(V(s_{t+1})\) does not recover the certainty-equivalent value. Even if
    \[
    \widehat G_t
    =
    \frac{1}{N_z}
    \sum_{j=1}^{N_z}
    f\left(V(s_{t+1}^{(j)})\right)
    \]
    is an unbiased estimator of
    \(\mathbb E_t[f(V(s_{t+1}))]\), the transformed quantity
    \(f^{-1}(\widehat G_t)\) is generally not an unbiased estimator of
    \(\mathcal C(s_t,c_t)\), because \(f^{-1}\) is nonlinear. This finite-sample bias can become more important when \(f\) is strongly curved or when the conditional distribution of next period values is dispersed.
    
    Fourth, policy and multiplier updates are more difficult under recursive utility because the model-specific first-order residuals contain the state-control certainty-equivalent value and may, in a direct differentiated formulation, also contain derivatives of the certainty-equivalent value and future value function. Therefore, during policy updates, the algorithm must evaluate certainty-equivalent values at candidate controls, not only along the policy path. The model-specific first-order residual is evaluated while updating the policy and multiplier networks. These updates may consider controls generated by the current policy, controls generated by exploratory simulation, or nearby candidate controls used in optimization. A value network evaluated only at future states therefore does not provide the state-control certainty-equivalent value needed to evaluate the Bellman target and the model-specific first-order residuals.
    
    As shown in Section~\ref{subsec:cefol_general_foc_kkt}, direct differentiation of the Bellman equation can generate a transformed conditional expectation involving \(f'(V(s_{t+1}))\), derivatives of the transition function, and the future value gradient \(\partial V(s_{t+1})/\partial s_{t+1}\). This representation is computationally demanding because it may require differentiating through the future value and policy networks. However, when the structural condition in \eqref{eq:cefol_reduced_transition} holds, the envelope theorem and the next-period stationarity condition can eliminate the derivative of the certainty-equivalent value from the envelope and express the future value gradient using current derivatives of the time aggregator, transition rule, and constraint multipliers. In such cases, the model-specific first-order condition can be implemented through a simplified Euler or KKT residual without directly approximating \(\partial V(s_{t+1})/\partial s_{t+1}\).

    The fifth challenge is that value approximation, certainty-equivalent approximation, and policy updates are tightly coupled. The certainty-equivalent target depends on the value approximation, the first-order residuals depend on the learned certainty-equivalent value, and policy updates change the distribution of simulated states and candidate controls. Errors in one component can therefore feed into the others. This feedback is more pronounced in models with strong risk sensitivity, high curvature, stochastic volatility, or occasionally binding constraints.
    
    CEFOL addresses these difficulties by learning the state-control certainty-equivalent value as a separate component. Once \(\mathcal C(s_t,c_t)\) is represented by \(\mathcal C(s_t,c_t;\theta_{\mathcal C})\), the Bellman target and the model-specific first-order residuals can both be evaluated at candidate state-control pairs. When needed, derivatives of the learned certainty-equivalent function with respect to controls can be obtained by automatic differentiation. When the structural condition in Section~\ref{subsec:cefol_general_foc_kkt} is available, CEFOL instead evaluates the corresponding simplified Euler or KKT residual, thereby avoiding direct use of future value gradients in the policy--multiplier update. The certainty-equivalent and Bellman losses use forward simulation of next-period states, while additional derivatives enter only through the model-specific residuals used to update the policy and multiplier networks. CEFOL therefore separates certainty-equivalent approximation, Bellman consistency, and policy and multiplier updates. When constraints are explicitly imposed, the corresponding KKT terms are included in the policy and multiplier update.

	\section{The Certainty-Equivalent First-Order Learning (CEFOL) Algorithm}
	\label{sec:algorithm}
	
	This section presents the CEFOL algorithm. Section~\ref{subsec:cefol_network_framework} introduces the alternative neural-network architectures that can be used in CEFOL, including the baseline four-network architecture, the five-network decomposition, and the compact three-network architecture. Sections~\ref{subsec:cefol_ce_bellman_losses} and~\ref{subsec:cefol_foc_kkt_losses} define the certainty-equivalent, Bellman, and first-order/KKT losses. Sections~\ref{subsec:cefol_algorithmic_enhancements} and~\ref{subsec:cefol_training_procedure} describe stabilization devices and training procedures.
	
	\subsection{Neural Network Framework}
	\label{subsec:cefol_network_framework}
	
	CEFOL can be implemented with several neural-network architectures. We consider three related architectures. The terms four-network, five-network, and three-network refer to functional blocks in the CEFOL learning scheme rather than to the literal number of subnetworks used in every implementation. For example, when a model has several control components, such as consumption and labor or consumption and portfolio weights, these components may be represented by separate policy subnetworks but are treated as one policy-network block in the framework. Section~\ref{subsubsec:cefol_four_network} presents the baseline four-network architecture with a value network, a policy network, a multiplier network, and a certainty-equivalent network. Section~\ref{subsubsec:cefol_five_network} presents the five-network architecture, which decomposes the certainty-equivalent network into the conditional expectation of next-period value and a nonlinear-difference component. Section~\ref{subsubsec:cefol_three_network} presents the compact three-network architecture without a separate value network. These architectures differ only in how the value and certainty-equivalent terms are represented; in all cases, the relevant losses are minimized sequentially rather than through one simultaneous optimization step.
	
	\subsubsection{Four-network architecture}
	\label{subsubsec:cefol_four_network}
	
	The baseline CEFOL architecture uses one value network, one policy network, one multiplier network, and one certainty-equivalent network. Throughout this section, $s_t$ denotes the current state, $c_t$ denotes the current control, $w$ denotes the time aggregator, and $\mathcal C(s_t,c_t)$ denotes the state-control certainty-equivalent value. The four-network architecture is written as
	\begin{align}
		V(s_t)
		&\equiv
		V(s_t;\theta_V),
		\label{eq:cefol_four_value_network}
		\\
		c_t
		&\equiv
		c(s_t;\theta_c),
		\label{eq:cefol_four_policy_network}
		\\
		m(s_t)
		&\equiv
		m(s_t;\theta_m),
		\label{eq:cefol_four_multiplier_network}
		\\
		\mathcal C(s_t,c_t)
		&\equiv
		\mathcal C(s_t,c_t;\theta_{\mathcal C}).
		\label{eq:cefol_four_ce_network}
	\end{align}
	The value network approximates the value function, the policy network represents the decision rule, the multiplier network represents the equality and inequality multipliers entering the KKT system, and the certainty-equivalent network represents the nonlinear certainty-equivalent value in \eqref{eq:cefol_state_control_ce}.
	
	The multiplier network output is decomposed as
	\begin{equation}
		m(s_t;\theta_m)
		=
		\left(\lambda(s_t),\nu(s_t)\right).
		\label{eq:cefol_multiplier_decomposition}
	\end{equation}
	Here
	\begin{equation}
		\lambda(s_t)
		=
		\left(\lambda_1(s_t),\ldots,\lambda_{M_g}(s_t)\right)^{\top}
		\label{eq:cefol_inequality_multiplier_vector}
	\end{equation}
	denotes the vector of inequality-constraint multipliers, and
	\begin{equation}
		\nu(s_t)
		=
		\left(\nu_1(s_t),\ldots,\nu_{M_q}(s_t)\right)^{\top}
		\label{eq:cefol_equality_multiplier_vector}
	\end{equation}
	denotes the vector of equality-constraint multipliers. The inequality multipliers must be nonnegative. In implementation, nonnegativity can be imposed directly by choosing the activation function of the $\lambda$ output layer, such as softplus, exponential, or a squared-output transformation. The equality multipliers are unrestricted.

    \subsubsection{Five-network architecture}
    \label{subsubsec:cefol_five_network}
    
    The five-network architecture keeps the value, policy, and multiplier networks, but decomposes the certainty-equivalent value into the conditional expectation of next-period value and a nonlinear-difference component. Define
    \begin{align}
    	V_e(s_t,c_t)
    	&=
    	\mathbb E_t\left[V(s_{t+1})\mid s_t,c_t\right],
    	\label{eq:cefol_expected_future_definition}
    	\\
    	D(s_t,c_t)
    	&=
    	V_e(s_t,c_t)-\mathcal C(s_t,c_t).
    	\label{eq:cefol_difference_definition}
    \end{align}
    
    The motivation for introducing the nonlinear-difference component is that the conditional expectation of next-period value and the certainty-equivalent value capture different quantities. The network \(V_e(s_t,c_t)\) learns the standard conditional expectation of next-period value, whereas \(\mathcal C(s_t,c_t)\) contains the nonlinear transformation induced by recursive utility. Their difference \(D(s_t,c_t)\) therefore measures the nonlinear risk-adjustment, or Jensen-type gap, between the expected-value component and the certainty-equivalent value. When the transformation \(f\) is close to linear, this gap is small; when \(f\) is strongly curved, the gap can be important. Learning \(V_e\) and \(D\) separately can therefore make the representation more interpretable and can help the network isolate the nonlinear certainty-equivalent adjustment rather than forcing a single network to learn the full transformed value directly.
    
    The functions \(V_e(s_t,c_t)\) and \(D(s_t,c_t)\) are represented by two neural networks, \(V_e(s_t,c_t;\theta_e)\) and \(D(s_t,c_t;\theta_D)\). The five-network architecture is written as
    \begin{align}
    	V(s_t)
    	&\equiv
    	V(s_t;\theta_V),
    	\label{eq:cefol_five_value_network}
    	\\
    	c_t
    	&\equiv
    	c(s_t;\theta_c),
    	\label{eq:cefol_five_policy_network}
    	\\
    	m(s_t)
    	&\equiv
    	m(s_t;\theta_m),
    	\label{eq:cefol_five_multiplier_network}
    	\\
    	V_e(s_t,c_t)
    	&\equiv
    	V_e(s_t,c_t;\theta_e),
    	\label{eq:cefol_five_expected_network}
    	\\
    	D(s_t,c_t)
    	&\equiv
    	D(s_t,c_t;\theta_D).
    	\label{eq:cefol_five_difference_network}
    \end{align}
    
    Equivalently, the certainty-equivalent value is represented by
    \begin{equation}
    	\mathcal C(s_t,c_t)
    	\equiv
    	\mathcal C(s_t,c_t;\theta_e,\theta_D)
    	=
    	V_e(s_t,c_t;\theta_e)-D(s_t,c_t;\theta_D).
    	\label{eq:cefol_five_ce_decomposition}
    \end{equation}
    
      The sign of \(D\) depends on the curvature and monotonicity of the transformation \(f\). Let \(X_{t+1}=V(s_{t+1})\). Since \(V_e(s_t,c_t)=\mathbb E_t[X_{t+1}]\) and \(\mathcal C(s_t,c_t)=f^{-1}(\mathbb E_t[f(X_{t+1})])\), Jensen's inequality determines whether the expected-value component lies above or below the certainty-equivalent value. If \(f\) is increasing and concave, or decreasing and convex, then \(V_e(s_t,c_t)\geq \mathcal C(s_t,c_t)\), so \(D(s_t,c_t)\geq0\) under the decomposition in \eqref{eq:cefol_five_ce_decomposition}. In this case, the output layer of the \(D\)-network can impose non-negativity if desired. If \(f\) is increasing and convex, or decreasing and concave, the ordering is reversed, and \(D\) should either be treated as a signed adjustment or the sign convention in \eqref{eq:cefol_five_ce_decomposition} should be reversed. When no global ordering is imposed, an unrestricted output for \(D\) is the conservative choice. The first-order residuals are constructed exactly as in the four-network architecture, except that every occurrence of \(\mathcal C(s_t,c_t;\theta_{\mathcal C})\) is replaced by \(\mathcal C(s_t,c_t;\theta_e,\theta_D)\).

	\subsubsection{Three-network architecture without a separate value update}
	\label{subsubsec:cefol_three_network}
	
	The compact three-network architecture removes the explicit value network and represents current value implicitly through the policy, multiplier, and certainty-equivalent networks:
	\begin{align}
		c_t
		&\equiv
		c(s_t;\theta_c),
		\label{eq:cefol_three_policy_network}
		\\
		m(s_t)
		&\equiv
		m(s_t;\theta_m),
		\label{eq:cefol_three_multiplier_network}
		\\
		\mathcal C(s_t,c_t)
		&\equiv
		\mathcal C(s_t,c_t;\theta_{\mathcal C}).
		\label{eq:cefol_three_ce_network}
	\end{align}
	Instead of introducing a separate value network, this architecture uses
	\begin{equation}
		V^{I}(s_t;\theta_c,\theta_{\mathcal C})
		\equiv
		w\left(
		s_t,
		c(s_t;\theta_c),
		\mathcal C(s_t,c(s_t;\theta_c);\theta_{\mathcal C})
		\right)
		\label{eq:cefol_three_implicit_value}
	\end{equation}
	to represent the value function. When next period value evaluations are needed in the certainty-equivalent target or in model-specific first-order residuals, $V(s_t;\theta_V)$ is replaced by $V^{I}(s_t;\theta_c,\theta_{\mathcal C})$.
	
	\subsection{Certainty-Equivalent and Bellman Residuals}
	\label{subsec:cefol_ce_bellman_losses}
	
	In the four-network architecture, the certainty-equivalent network is trained by minimizing a transformed certainty-equivalent loss. The target identity is
	\begin{equation}
		f\left(\mathcal C(s_t,c_t;\theta_{\mathcal C})\right)
		=
		\mathbb{E}_t\left[f\left(V(s_{t+1};\theta_V^-)\right)\right],
		\label{eq:cefol_ce_identity}
	\end{equation}
	where $\theta_V^-$ denotes the target value-network parameters used to construct the certainty-equivalent target. Given a mini-batch of current state-control pairs and $N_z$ future shock draws, construct
	\begin{equation}
		s_{t+1}^{(i,j)}
		=
		\psi\left(s_t^{(i)},c_t^{(i)},z_{t+1}^{(i,j)}\right),
		\qquad
		j=1,\ldots,N_z.
		\label{eq:cefol_future_states}
	\end{equation}
	Using the target value network, define the transformed certainty-equivalent target
	\begin{equation}
		\widehat G_t^{(i)}
		=
		\frac{1}{N_z}
		\sum_{j=1}^{N_z}
		f\left(V(s_{t+1}^{(i,j)};\theta_V^-)\right).
		\label{eq:cefol_ce_mc_target}
	\end{equation}
	The certainty-equivalent loss is
	\begin{equation}
		\widehat{\mathcal L}_{C}(\theta_{\mathcal C})
		=
		\frac{1}{N}
		\sum_{i=1}^{N}
		\left[
		f\left(\mathcal C(s_t^{(i)},c_t^{(i)};\theta_{\mathcal C})\right)
		-
		\widehat G_t^{(i)}
		\right]^2.
		\label{eq:cefol_ce_loss}
	\end{equation}

	Given the certainty-equivalent network, the Bellman target can be written as
	\begin{equation}
		y_t^{(i)}
		=
		w\left(
		s_t^{(i)},
		c_t^{(i)},
		\mathcal C(s_t^{(i)},c_t^{(i)};\theta_{\mathcal C})
		\right).
		\label{eq:cefol_bellman_target}
	\end{equation}
	When an explicit value network is used, the value network is trained by minimizing
	\begin{equation}
		\widehat{\mathcal L}_{V}(\theta_V)
		=
		\frac{1}{N}
		\sum_{i=1}^{N}
		\left[
		V(s_t^{(i)};\theta_V)-y_t^{(i)}
		\right]^2.
		\label{eq:cefol_value_loss}
	\end{equation}
	The target-network update is described in Section~\ref{subsubsec:cefol_target_network}.
	
	In the five-network architecture, the conditional-expectation network is trained by minimizing
	\begin{equation}
		\widehat{\mathcal L}_{e}(\theta_e)
		=
		\frac{1}{N}
		\sum_{i=1}^{N}
		\left[
		V_e(s_t^{(i)},c_t^{(i)};\theta_e)
		-
		\frac{1}{N_z}
		\sum_{j=1}^{N_z}
		V(s_{t+1}^{(i,j)};\theta_V^-)
		\right]^2,
		\label{eq:cefol_expected_value_loss}
	\end{equation}
	and the nonlinear-difference network is trained by minimizing
	\begin{equation}
		\widehat{\mathcal L}_{D}(\theta_D)
		=
		\frac{1}{N}
		\sum_{i=1}^{N}
		\left[
		f\left(V_e(s_t^{(i)},c_t^{(i)};\theta_e)-D(s_t^{(i)},c_t^{(i)};\theta_D)\right)
		-
		\widehat G_t^{(i)}
		\right]^2.
		\label{eq:cefol_difference_loss}
	\end{equation}
	The Bellman target in this architecture is obtained by replacing $\mathcal C(s_t,c_t;\theta_{\mathcal C})$ in \eqref{eq:cefol_bellman_target} with $\mathcal C(s_t,c_t;\theta_e,\theta_D)$.
	
	In the three-network architecture without a separate value update, the explicit value-network output is replaced by the implicit value $V^{I}(s_t;\theta_c,\theta_{\mathcal C})$ in \eqref{eq:cefol_three_implicit_value}. The certainty-equivalent target is then constructed as
	\begin{equation}
		\widehat G_t^{I,(i)}
		=
		\frac{1}{N_z}
		\sum_{j=1}^{N_z}
		f\left(V^{I}(s_{t+1}^{(i,j)};\theta_c,\theta_{\mathcal C})\right),
		\label{eq:cefol_three_ce_target}
	\end{equation}
	The corresponding certainty-equivalent loss has the same transformed form as \eqref{eq:cefol_ce_loss}, with $\widehat G_t^{(i)}$ replaced by $\widehat G_t^{I,(i)}$. There is no separate value loss $\widehat{\mathcal L}_V$ in this compact architecture.
	
	\subsection{First-Order and KKT Residuals}
	\label{subsec:cefol_foc_kkt_losses}
	
	The main objective of policy-learning is the system of conditional first-order conditions. For each current state $s_t^{(i)}$, draw two conditionally independent future shocks
	\begin{equation}
		z_{t+1}^{(i,1)},
		\qquad
		z_{t+1}^{(i,2)}.
		\label{eq:cefol_two_shock_draws}
	\end{equation}
	Construct the corresponding future states and future controls:
	\begin{align}
		s_{t+1}^{(i,\ell)}
		&=
		\psi\left(s_t^{(i)},c_t^{(i)},z_{t+1}^{(i,\ell)}\right),
		\label{eq:cefol_future_state_two_draws}
		\\
		c_{t+1}^{(i,\ell)}
		&=
		c\left(s_{t+1}^{(i,\ell)};\theta_c\right),
		\qquad
		\ell=1,2.
		\label{eq:cefol_future_control_two_draws}
	\end{align}
	For the general FOC/KKT formulation, define the stochastic stationarity residual
	\begin{align}
		R_{k,t}^{\mathrm{stat},(i,\ell)}
		&=
		F_{t+1,k}^{(i,\ell)}
		+
		\sum_{m=1}^{M_g}
		\lambda_m(s_t^{(i)})
		\frac{\partial g_m(s_t^{(i)},c_t^{(i)})}{\partial c_{t,k}}
		\nonumber\\
		&\quad+
		\sum_{r=1}^{M_q}
		\nu_r(s_t^{(i)})
		\frac{\partial q_r(s_t^{(i)},c_t^{(i)})}{\partial c_{t,k}},
		\qquad
		k=1,\ldots,n_c.
		\label{eq:cefol_stationarity_residual_sample}
	\end{align}
	The independent-draw stationarity loss is
	\begin{equation}
		\widehat{\mathcal L}_{S}(\theta_c,\theta_m)
		=
		\frac{1}{N}
		\sum_{i=1}^{N}
		\sum_{k=1}^{n_c}
		v_{S,k}
		R_{k,t}^{\mathrm{stat},(i,1)}
		R_{k,t}^{\mathrm{stat},(i,2)}.
		\label{eq:cefol_stationarity_loss}
	\end{equation}
	Because the two residuals use conditionally independent future shocks, their product estimates the squared conditional mean of the stationarity residual without explicitly computing the inner conditional expectation. The underlying conditional-moment identity is
	\[
	\mathbb{E}_t\left[R_{t+1}^{(1)}R_{t+1}^{(2)}\right]
	=
	\mathbb{E}_t\left[R_{t+1}^{(1)}\right]
	\mathbb{E}_t\left[R_{t+1}^{(2)}\right]
	=
	\left(\mathbb{E}_t[R_{t+1}]\right)^2,
	\]
	for conditionally independent residual evaluations. The estimator follows the conditional-independence idea used in simulation-based residual learning by \citet{maliar2021deep}, but here it is used only for the stationarity component of the first-order/KKT loss. Although the sample product need not be nonnegative for a finite mini-batch, its expectation is the squared conditional mean residual.
	
	The inequality constraints are imposed by a Fischer--Burmeister loss:
	\begin{equation}
		\widehat{\mathcal L}_{FB}(\theta_c,\theta_m)
		=
		\frac{1}{N}
		\sum_{i=1}^{N}
		\sum_{m=1}^{M_g}
		v_{FB,m}
		\left[
		\Phi^{FB}\left(g_m(s_t^{(i)},c_t^{(i)}),\lambda_m(s_t^{(i)})\right)
		\right]^2,
		\label{eq:cefol_fb_loss}
	\end{equation}
	where
	\begin{equation}
		\Phi^{FB}(a,\lambda)
		=
		a+\lambda-\sqrt{a^2+\lambda^2}.
		\label{eq:cefol_fischer_burmeister}
	\end{equation}
	Under the convention $g_m(s_t,c_t)\geq0$ and $\lambda_m(s_t)\geq0$, $\Phi^{FB}(g_m,\lambda_m)=0$ is equivalent to feasibility, nonnegativity, and complementarity. The equality constraints are imposed by
	\begin{equation}
		\widehat{\mathcal L}_{EQ}(\theta_c)
		=
		\frac{1}{N}
		\sum_{i=1}^{N}
		\sum_{r=1}^{M_q}
		v_{EQ,r}
		\left[q_r(s_t^{(i)},c_t^{(i)})\right]^2.
		\label{eq:cefol_eq_loss}
	\end{equation}
	Here $v_{S,k}$, $v_{FB,m}$, and $v_{EQ,r}$ are positive normalization weights that put residual components on comparable scales. The general first-order/KKT loss is
	\begin{equation}
		\widehat{\mathcal L}_{FOC}(\theta_c,\theta_m)
		=
		\lambda_S\widehat{\mathcal L}_{S}(\theta_c,\theta_m)
		+
		\lambda_{FB}\widehat{\mathcal L}_{FB}(\theta_c,\theta_m)
		+
		\lambda_{EQ}\widehat{\mathcal L}_{EQ}(\theta_c).
		\label{eq:cefol_foc_loss}
	\end{equation}
	The loss combines stationarity, complementarity, and equality-constraint residuals and provides the policy-learning objective in CEFOL.

	\subsection{Algorithmic Enhancements for Stability and Efficiency}
	\label{subsec:cefol_algorithmic_enhancements}
	
	CEFOL uses four implementation refinements for numerical stability and simulation efficiency. These refinements target distinct sources of difficulty in recursive-utility learning. Exploratory control perturbation broadens the simulated state-control region and reduces dependence on early policy-network outputs. The value-network target stabilizes bootstrapped certainty-equivalent and value updates when recursive targets depend on learned future values. The independent sample-mean estimator reduces simulation noise in the conditional first-order residual loss by averaging conditionally independent residual evaluations before forming the product estimator. Delayed first-order/KKT updates prevent the policy and multiplier networks from reacting too frequently to still-inaccurate value and certainty-equivalent approximations. Section~\ref{subsubsec:cefol_exploration} introduces exploratory control perturbation during path simulation. Section~\ref{subsubsec:cefol_target_network} describes the value-network target used for stabilized value updates. Section~\ref{subsubsec:cefol_independent_sample_mean} introduces the independent sample-mean estimator for the first-order residual loss. Section~\ref{subsubsec:cefol_delayed_updates} presents delayed first-order/KKT updates.
	
	\subsubsection{Exploratory Control Perturbation During Path Simulation}
	\label{subsubsec:cefol_exploration}
	
	Simulated paths generated only by the current policy network may cover too narrow a region of the feasible control space, especially in early training. This can lead to premature convergence, inaccurate certainty-equivalent estimates, and poor approximation of first-order residuals away from the current policy path. We therefore introduce exploratory perturbations during path simulation.
	
	Let the policy network generate the baseline control
	\begin{equation}
		c_t=c(s_t;\theta_c).
		\label{eq:cefol_baseline_control}
	\end{equation}
	During path simulation, we use a perturbed control
	\begin{equation}
		\widehat c_t
		=
		\Pi_{\mathcal A(s_t)}\left[c(s_t;\theta_c)+\zeta_k\eta_t\right],
		\label{eq:cefol_perturbed_control}
	\end{equation}
	where $\Pi_{\mathcal A(s_t)}$ projects the perturbed control back to the feasible set, $\eta_t$ is a random exploration shock, and $\zeta_k\geq0$ controls the magnitude of exploration. The perturbation scale can be gradually reduced over training so that early iterations explore broadly while later iterations focus on the neighborhood of the learned policy.
	
	The simulated state evolves according to
	\begin{equation}
		s_{t+1}=\psi(s_t,\widehat c_t,z_{t+1}).
		\label{eq:cefol_perturbed_transition}
	\end{equation}
	The resulting simulated states and perturbed controls form the training sample for the current iteration. When updating the networks, the current unperturbed control $c(s_t;\theta_c)$ is used to construct first-order/KKT and Bellman residuals, while the perturbed paths improve the diversity of the sampled state distribution.
	
	\subsubsection{Target Network for Stabilized Value Updates}
	\label{subsubsec:cefol_target_network}
	
	Value learning can be unstable when the same value network is used both to construct and fit the target. With recursive utility, small errors in next period value estimates can be amplified by the certainty-equivalent transformation and propagated back into the current value update.
	
	To stabilize bootstrapped targets, CEFOL uses a target network only for the value function in the baseline four-network architecture. Let $V(s;\theta_V)$ be the current value network and $V(s;\theta_V^-)$ be the target value network. The target value network is used when constructing Monte Carlo targets for the certainty-equivalent network:
	\begin{equation}
		\widehat G_t^{(i)}
		=
		\frac{1}{N_z}
		\sum_{j=1}^{N_z}
		f\left(V(s_{t+1}^{(i,j)};\theta_V^-)\right).
		\label{eq:cefol_target_value_ce_target}
	\end{equation}
	The value network itself is updated using the Bellman target constructed from the certainty-equivalent network. After each value-network update, the target value network is softly updated by
	\begin{equation}
		\theta_V^-
		\leftarrow
		\tau\theta_V+(1-\tau)\theta_V^-,
		\qquad
		0<\tau<1.
		\label{eq:cefol_target_update}
	\end{equation}
	No separate target networks are introduced for the policy network, multiplier network, or certainty-equivalent network in the baseline implementation.

	\subsubsection{Independent Sample-Mean Estimator for First-Order Residuals}
	\label{subsubsec:cefol_independent_sample_mean}
	
	The stationarity loss in Section~\ref{subsec:cefol_foc_kkt_losses} is based on the conditional moment restriction
	\[
	\mathbb E_t\left[R^{\mathrm{stat}}_{t+1,k}\right]=0.
	\]
	The baseline estimator uses two conditionally independent future shock draws to estimate the squared conditional mean of the residual. When more future shock draws are available, CEFOL can reduce simulation variance by replacing each single residual draw with an independent sample mean.
	
	Suppose \(N_z\) is even and let \(J_z=N_z/2\). For each current state-control pair \((s_t^{(i)},c_t^{(i)})\), compute the stationarity residuals
	\[
	R_{k,t}^{\mathrm{stat},(i,j)},
	\qquad
	j=1,\ldots,N_z,
	\qquad
	k=1,\ldots,n_c .
	\]
	Split the \(N_z\) future draws into two conditionally independent groups and define
	\[
	\overline R_{k,t}^{\mathrm{stat},(i,1)}
	=
	\frac{1}{J_z}
	\sum_{j=1}^{J_z}
	R_{k,t}^{\mathrm{stat},(i,j)},
	\qquad
	\overline R_{k,t}^{\mathrm{stat},(i,2)}
	=
	\frac{1}{J_z}
	\sum_{j=J_z+1}^{N_z}
	R_{k,t}^{\mathrm{stat},(i,j)}.
	\]
	The stationarity loss then uses the product of the two independent sample means:
	\[
	\widehat{\mathcal L}_{S}(\theta_c,\theta_m)
	=
	\frac{1}{N}
	\sum_{i=1}^{N}
	\sum_{k=1}^{n_c}
	v_{S,k}
	\overline R_{k,t}^{\mathrm{stat},(i,1)}
	\overline R_{k,t}^{\mathrm{stat},(i,2)}.
	\]
	
	Because the two groups are conditionally independent, this estimator preserves the same conditional-moment target as the two-draw estimator. The original two-draw estimator is the special case \(J_z=1\). For \(J_z>1\), each independent copy of the residual is averaged over multiple future shocks before the product is formed. This uses all \(N_z\) simulated future draws and reduces Monte Carlo noise in the first-order/KKT update.
	
	\subsubsection{Delayed First-Order/KKT Updates}
	\label{subsubsec:cefol_delayed_updates}
	
	The policy and multiplier updates use first-order/KKT residuals that depend on the current value and certainty-equivalent approximations. When these approximations are still inaccurate, frequent first-order/KKT updates can produce unstable residual estimates. CEFOL therefore updates the policy and multiplier parameters less frequently than the value and certainty-equivalent networks.
	
	Let $d\geq1$ denote the first-order/KKT update-delay parameter. At iteration $k$, the certainty-equivalent and value networks follow their own update steps, while the policy and multiplier parameter updates are activated only when
	\begin{equation}
		k\equiv0\pmod d.
		\label{eq:cefol_update_delay_condition}
	\end{equation}
	Thus, when the update is activated, the policy and multiplier parameters are updated as
	\begin{align}
		\theta_c
		&\leftarrow
		\theta_c
		-
		\alpha_c
		\nabla_{\theta_c}
		\widehat{\mathcal L}_{FOC}(\theta_c,\theta_m),
		\nonumber\\
		\theta_m
		&\leftarrow
		\theta_m
		-
		\alpha_m
		\nabla_{\theta_m}
		\widehat{\mathcal L}_{FOC}(\theta_c,\theta_m),
		\qquad
		\text{only if } k\equiv0\pmod d.
		\label{eq:cefol_delayed_policy_multiplier_update}
	\end{align}
	During the intervening iterations, $\theta_c$ and $\theta_m$ are held fixed while $\theta_{\mathcal C}$ and $\theta_V$ continue to be updated. Setting $d=1$ recovers the baseline update frequency.
	
	\subsection{Training Procedure}
	\label{subsec:cefol_training_procedure}
	
	CEFOL cycles through certainty-equivalent estimation, first-order/KKT updating, and, when an explicit value network is used, value-function fitting. The precise sequence depends on whether the implementation uses the four-network, five-network, or three-network architecture. Section~\ref{subsubsec:cefol_train_four} presents the four-network training procedure in full detail. Section~\ref{subsubsec:cefol_train_five} presents the five-network procedure, which replaces the single certainty-equivalent network with a conditional-expectation network and a nonlinear-difference network. Section~\ref{subsubsec:cefol_train_three} presents the three-network procedure without a separate value update. The complete procedure alternates between these updates until convergence.
	
	\subsubsection{Four-Network Training Procedure}
	\label{subsubsec:cefol_train_four}
	
	The four-network version updates the value network $V(s_t;\theta_V)$, policy network $c(s_t;\theta_c)$, multiplier network $m(s_t;\theta_m)$, and certainty-equivalent network $\mathcal C(s_t,c_t;\theta_{\mathcal C})$. At iteration $k$, it proceeds as follows.
	\begin{enumerate}
		\item \textbf{Simulate state paths and form a mini-batch.} Generate simulated paths using the exploratory control rule in \eqref{eq:cefol_perturbed_control} and the perturbed transition in \eqref{eq:cefol_perturbed_transition}. Take a mini-batch of current states from the simulated paths:
		\begin{equation}
			\{s_t^{(i)}\}_{i=1}^{N}.
			\label{eq:cefol_train4_sample_states}
		\end{equation}
		For each sampled state, compute the current policy and multipliers by
		\begin{align}
			c_t^{(i)}
			&=c(s_t^{(i)};\theta_c),
			\label{eq:cefol_train4_current_policy}
			\\
			m_t^{(i)}
			&=m(s_t^{(i)};\theta_m)
			=
			\left(\lambda(s_t^{(i)}),\nu(s_t^{(i)})\right).
			\label{eq:cefol_train4_current_multiplier}
		\end{align}
		When the problem has no equality constraints or no inequality constraints, the corresponding multiplier component is omitted.
		
		\item \textbf{Draw future shocks and construct future states.} For each current state-control pair $(s_t^{(i)},c_t^{(i)})$, draw future shocks
		\begin{equation}
			\{z_{t+1}^{(i,j)}\}_{j=1}^{N_z}.
			\label{eq:cefol_train4_future_shocks}
		\end{equation}
		Construct future states and future controls by
		\begin{align}
			s_{t+1}^{(i,j)}
			&=
			\psi\left(s_t^{(i)},c_t^{(i)},z_{t+1}^{(i,j)}\right),
			\qquad j=1,\ldots,N_z,
			\label{eq:cefol_train4_future_states}
			\\
			c_{t+1}^{(i,j)}
			&=
			c(s_{t+1}^{(i,j)};\theta_c).
			\label{eq:cefol_train4_future_controls}
		\end{align}
		For the first-order/KKT update, the $N_z$ future draws can be split into two conditionally independent groups and used in the independent sample-mean estimator in Section~\ref{subsubsec:cefol_independent_sample_mean}. When $N_z=2$, this reduces to the original two-draw product estimator. For the certainty-equivalent target used in the certainty-equivalent update, all $N_z$ future draws can also be used.
		
		\item \textbf{Construct the certainty-equivalent target and loss.} Using the target value network, compute the transformed certainty-equivalent target
		\begin{equation}
			\widehat G_t^{(i)}
			=
			\frac{1}{N_z}
			\sum_{j=1}^{N_z}
			f\left(V(s_{t+1}^{(i,j)};\theta_V^-)\right),
			\label{eq:cefol_train4_ce_target}
		\end{equation}
		which is the neural network representation of \eqref{eq:cefol_ce_mc_target}. The certainty-equivalent loss is
		\begin{equation}
			\widehat{\mathcal L}_{C}(\theta_{\mathcal C})
			=
			\frac{1}{N}
			\sum_{i=1}^{N}
			\left[
			f\left(\mathcal C(s_t^{(i)},c_t^{(i)};\theta_{\mathcal C})\right)
			-
			\widehat G_t^{(i)}
			\right]^2.
			\label{eq:cefol_train4_ce_loss}
		\end{equation}

		\item \textbf{Construct the Bellman target and value loss.} Given the current policy and the learned certainty-equivalent value, form the Bellman target
		\begin{equation}
			y_t^{(i)}
			=
			w\left(s_t^{(i)},c_t^{(i)},\mathcal C(s_t^{(i)},c_t^{(i)};\theta_{\mathcal C})\right),
			\label{eq:cefol_train4_bellman_target}
		\end{equation}
		which corresponds to \eqref{eq:cefol_bellman_target}. The value loss is
		\begin{equation}
			\widehat{\mathcal L}_{V}(\theta_V)
			=
			\frac{1}{N}
			\sum_{i=1}^{N}
			\left[
			V(s_t^{(i)};\theta_V)-y_t^{(i)}
			\right]^2.
			\label{eq:cefol_train4_value_loss}
		\end{equation}

		\item \textbf{Construct the first-order stationarity residual.} For the $k$-th control, compute the model-specific stochastic first-order integrand
		\begin{equation}
			F_{t+1,k}^{(i,j)}
			=
			F_{k}\left(s_t^{(i)},c_t^{(i)},z_{t+1}^{(i,j)};
			V(\cdot;\theta_V),c(\cdot;\theta_c),\mathcal C(\cdot,\cdot;\theta_{\mathcal C}),w,f,\psi\right).
			\label{eq:cefol_train4_foc_integrand}
		\end{equation}
		The corresponding stationarity residual is
		\begin{align}
			R_{k,t}^{\mathrm{stat},(i,j)}
			&=
			F_{t+1,k}^{(i,j)}
			+
			\sum_{m=1}^{M_g}
			\lambda_m(s_t^{(i)})
			\frac{\partial g_m(s_t^{(i)},c_t^{(i)})}{\partial c_{t,k}}
			\nonumber\\
			&\quad+
			\sum_{r=1}^{M_q}
			\nu_r(s_t^{(i)})
			\frac{\partial q_r(s_t^{(i)},c_t^{(i)})}{\partial c_{t,k}}.
			\label{eq:cefol_train4_stationarity_residual}
		\end{align}
		For the first-order/KKT update, suppose $N_z$ is even and let $J_z=N_z/2$. Define two conditionally independent sample means of the stationarity residuals by
		\begin{align}
			\overline R_{k,t}^{\mathrm{stat},(i,1)}
			&=
			\frac{1}{J_z}
			\sum_{j=1}^{J_z}
			R_{k,t}^{\mathrm{stat},(i,j)},
			\nonumber\\
			\overline R_{k,t}^{\mathrm{stat},(i,2)}
			&=
			\frac{1}{J_z}
			\sum_{j=J_z+1}^{N_z}
			R_{k,t}^{\mathrm{stat},(i,j)}.
		\end{align}
		The stationarity loss is
		\begin{equation}
			\widehat{\mathcal L}_{S}(\theta_c,\theta_m)
			=
			\frac{1}{N}
			\sum_{i=1}^{N}
			\sum_{k=1}^{n_c}
			v_{S,k}
			\overline R_{k,t}^{\mathrm{stat},(i,1)}
			\overline R_{k,t}^{\mathrm{stat},(i,2)}.
			\label{eq:cefol_train4_stationarity_loss}
		\end{equation}
		Equation \eqref{eq:cefol_train4_stationarity_loss} uses the product of two conditionally independent sample means. When $N_z=2$, it reduces to the original product of two residual evaluations.
		
		\item \textbf{Construct the KKT constraint losses.} For inequality constraints $g_m(s_t,c_t)\geq0$, compute the Fischer--Burmeister loss
		\begin{equation}
			\widehat{\mathcal L}_{FB}(\theta_c,\theta_m)
			=
			\frac{1}{N}
			\sum_{i=1}^{N}
			\sum_{m=1}^{M_g}
			v_{FB,m}
			\left[
			\Phi^{FB}\left(g_m(s_t^{(i)},c_t^{(i)}),\lambda_m(s_t^{(i)})\right)
			\right]^2,
			\label{eq:cefol_train4_fb_loss}
		\end{equation}
		where $\Phi^{FB}$ is defined in \eqref{eq:cefol_fischer_burmeister}. For equality constraints, compute
		\begin{equation}
			\widehat{\mathcal L}_{EQ}(\theta_c)
			=
			\frac{1}{N}
			\sum_{i=1}^{N}
			\sum_{r=1}^{M_q}
			v_{EQ,r}
			\left[q_r(s_t^{(i)},c_t^{(i)})\right]^2.
			\label{eq:cefol_train4_eq_loss}
		\end{equation}
		The four-network first-order/KKT loss is
		\begin{equation}
			\widehat{\mathcal L}_{FOC}(\theta_c,\theta_m)
			=
			\lambda_S\widehat{\mathcal L}_{S}(\theta_c,\theta_m)
			+
			\lambda_{FB}\widehat{\mathcal L}_{FB}(\theta_c,\theta_m)
			+
			\lambda_{EQ}\widehat{\mathcal L}_{EQ}(\theta_c).
			\label{eq:cefol_train4_foc_loss}
		\end{equation}
		If a model has no equality constraints or no inequality constraints, the corresponding component is omitted.
		
		\item \textbf{Update the policy and multiplier networks.} When delayed updates are used, this first-order/KKT step is performed only if $k\equiv0\pmod d$. The policy and multiplier parameters are updated by
		\begin{align}
			\theta_c
			&\leftarrow
			\theta_c
			-
			\alpha_c
			\nabla_{\theta_c}
			\widehat{\mathcal L}_{FOC}(\theta_c,\theta_m),
			\nonumber\\
			\theta_m
			&\leftarrow
			\theta_m
			-
			\alpha_m
			\nabla_{\theta_m}
			\widehat{\mathcal L}_{FOC}(\theta_c,\theta_m).
			\label{eq:cefol_train4_policy_multiplier_update}
		\end{align}
		The value and certainty-equivalent networks are held fixed during this first-order/KKT step.
		
		\item \textbf{Update the certainty-equivalent network.} Update the certainty-equivalent network by minimizing the transformed certainty-equivalent loss:
		\begin{equation}
			\theta_{\mathcal C}
			\leftarrow
			\theta_{\mathcal C}
			-
			\alpha_C
			\nabla_{\theta_{\mathcal C}}
			\widehat{\mathcal L}_{C}(\theta_{\mathcal C}).
			\label{eq:cefol_train4_ce_update}
		\end{equation}
		
		\item \textbf{Update the value network and its target network.} Update the value network by minimizing the Bellman loss:
		\begin{equation}
			\theta_V
			\leftarrow
			\theta_V
			-
			\alpha_V
			\nabla_{\theta_V}
			\widehat{\mathcal L}_{V}(\theta_V).
			\label{eq:cefol_train4_value_update}
		\end{equation}
		After the value update, softly update the target value network:
		\begin{equation}
			\theta_V^-
			\leftarrow
			\tau\theta_V+(1-\tau)\theta_V^-.
			\label{eq:cefol_train4_target_update}
		\end{equation}
	\end{enumerate}
	The four-network procedure separates the learning problem into three complementary updates: first-order/KKT updating for policy and multipliers, certainty-equivalent learning, and value-function fitting.
	
	\begin{algorithm}[H]
		\caption{CEFOL Algorithm with Four-Network Architecture}
		\label{alg:cefol_four_network}
		\begin{algorithmic}[1]
			\State Initialize network parameters $\theta_V$, $\theta_V^-$, $\theta_c$, $\theta_m$, and $\theta_{\mathcal C}$ for \eqref{eq:cefol_four_value_network}--\eqref{eq:cefol_four_ce_network}.
			\State Initialize target update rate $\tau$, perturbation scale $\zeta_k$, and delay parameter $d$.
			\For{$k=1,2,\ldots,K$}
			\State Simulate state paths using \eqref{eq:cefol_perturbed_control}--\eqref{eq:cefol_perturbed_transition}; form $\{s_t^{(i)}\}_{i=1}^N$ as in \eqref{eq:cefol_train4_sample_states}.
			\State Compute current controls and multipliers using \eqref{eq:cefol_train4_current_policy}--\eqref{eq:cefol_train4_current_multiplier}.
			\State Draw shocks and construct future states and controls using \eqref{eq:cefol_train4_future_shocks}--\eqref{eq:cefol_train4_future_controls}.
			\State Compute $\widehat{\mathcal L}_{C}(\theta_{\mathcal C})$ from \eqref{eq:cefol_train4_ce_target}--\eqref{eq:cefol_train4_ce_loss}.
			\State Compute $\widehat{\mathcal L}_{V}(\theta_V)$ from \eqref{eq:cefol_train4_bellman_target}--\eqref{eq:cefol_train4_value_loss}.
			\State Compute $\widehat{\mathcal L}_{FOC}(\theta_c,\theta_m)$ from \eqref{eq:cefol_train4_foc_integrand}--\eqref{eq:cefol_train4_foc_loss}, using the independent sample-mean stationarity loss in \eqref{eq:cefol_train4_stationarity_loss}.
			\If{$k\equiv0\pmod d$}
			\State Update $\theta_c$ and $\theta_m$ using \eqref{eq:cefol_train4_policy_multiplier_update}.
			\EndIf
			\State Update $\theta_{\mathcal C}$ using \eqref{eq:cefol_train4_ce_update}.
			\State Update $\theta_V$ using \eqref{eq:cefol_train4_value_update}.
			\State Soft-update the target value-network parameters $\theta_V^-$ using \eqref{eq:cefol_train4_target_update}.
			\If{convergence criterion is satisfied}
			\State \Return $(\theta_V,\theta_c,\theta_m,\theta_{\mathcal C})$.
			\EndIf
			\EndFor
		\end{algorithmic}
	\end{algorithm}
	
	\subsubsection{Five-Network Training Procedure}
	\label{subsubsec:cefol_train_five}
	
	The five-network version uses the same simulation step, independent sample-mean estimator for the stationarity residual, delayed first-order/KKT update, value update, and target update as the four-network version. The only substantive change is that the single certainty-equivalent network $\mathcal C(s_t,c_t;\theta_{\mathcal C})$ is replaced by the decomposition $\mathcal C(s_t,c_t;\theta_e,\theta_D)=V_e(s_t,c_t;\theta_e)-D(s_t,c_t;\theta_D)$ in \eqref{eq:cefol_five_ce_decomposition}. Thus, the Bellman target becomes
	\begin{equation}
		y_{5,t}^{(i)}
		=
		w\left(s_t^{(i)},c_t^{(i)},V_e(s_t^{(i)},c_t^{(i)};\theta_e)-D(s_t^{(i)},c_t^{(i)};\theta_D)\right),
		\label{eq:cefol_train5_bellman_target}
	\end{equation}
	and the value loss is
	\begin{equation}
		\widehat{\mathcal L}_{V}(\theta_V)
		=
		\frac{1}{N}
		\sum_{i=1}^{N}
		\left[V(s_t^{(i)};\theta_V)-y_{5,t}^{(i)}\right]^2.
		\label{eq:cefol_train5_value_loss}
	\end{equation}
	The conditional-expectation network is trained by minimizing
	\begin{equation}
		\widehat{\mathcal L}_{e}(\theta_e)
		=
		\frac{1}{N}
		\sum_{i=1}^{N}
		\left[
		V_e(s_t^{(i)},c_t^{(i)};\theta_e)
		-
		\frac{1}{N_z}
		\sum_{j=1}^{N_z}
		V(s_{t+1}^{(i,j)};\theta_V^-)
		\right]^2,
		\label{eq:cefol_train5_expected_loss}
	\end{equation}
	and the nonlinear-difference network is trained by minimizing
	\begin{equation}
		\widehat{\mathcal L}_{D}(\theta_D)
		=
		\frac{1}{N}
		\sum_{i=1}^{N}
		\left[
		f\left(V_e(s_t^{(i)},c_t^{(i)};\theta_e)-D(s_t^{(i)},c_t^{(i)};\theta_D)\right)
		-
		\widehat G_t^{(i)}
		\right]^2.
		\label{eq:cefol_train5_difference_loss}
	\end{equation}
	The five-network first-order/KKT loss is obtained from \eqref{eq:cefol_train4_foc_loss} by replacing $\mathcal C(s_t,c_t;\theta_{\mathcal C})$ with $\mathcal C(s_t,c_t;\theta_e,\theta_D)$ in the model-specific integrand:
	\begin{equation}
		\widehat{\mathcal L}_{FOC}(\theta_c,\theta_m)
		=
		\lambda_S\widehat{\mathcal L}_{S}(\theta_c,\theta_m)
		+
		\lambda_{FB}\widehat{\mathcal L}_{FB}(\theta_c,\theta_m)
		+
		\lambda_{EQ}\widehat{\mathcal L}_{EQ}(\theta_c).
		\label{eq:cefol_train5_foc_loss}
	\end{equation}
	The certainty-equivalent updates are
	\begin{align}
		\theta_e
		&\leftarrow
		\theta_e-
		\alpha_e\nabla_{\theta_e}\widehat{\mathcal L}_{e}(\theta_e),
		\label{eq:cefol_train5_expected_update}
		\\
		\theta_D
		&\leftarrow
		\theta_D-
		\alpha_D\nabla_{\theta_D}\widehat{\mathcal L}_{D}(\theta_D).
		\label{eq:cefol_train5_difference_update}
	\end{align}
	The policy-parameter, multiplier-parameter, value, and target-network updates have the same form as \eqref{eq:cefol_train4_policy_multiplier_update}, \eqref{eq:cefol_train4_value_update}, and \eqref{eq:cefol_train4_target_update}, with the losses replaced by \eqref{eq:cefol_train5_foc_loss} and \eqref{eq:cefol_train5_value_loss}.

	\begin{algorithm}[H]
		\caption{CEFOL Algorithm with Five-Network Architecture}
		\label{alg:cefol_five_network}
		\begin{algorithmic}[1]
			\State Initialize network parameters $\theta_V$, $\theta_V^-$, $\theta_c$, $\theta_m$, $\theta_e$, and $\theta_D$ for \eqref{eq:cefol_five_value_network}--\eqref{eq:cefol_five_difference_network}.
			\State Initialize target update rate $\tau$, perturbation scale $\zeta_k$, and delay parameter $d$.
			\For{$k=1,2,\ldots,K$}
			\State Use the same simulated mini-batch and future-state construction as \eqref{eq:cefol_train4_sample_states}--\eqref{eq:cefol_train4_future_controls}.
			\State Compute $\widehat{\mathcal L}_{e}(\theta_e)$ and $\widehat{\mathcal L}_{D}(\theta_D)$ using \eqref{eq:cefol_train5_expected_loss}--\eqref{eq:cefol_train5_difference_loss}.
			\State Compute $\widehat{\mathcal L}_{V}(\theta_V)$ using \eqref{eq:cefol_train5_value_loss}.
			\State Compute $\widehat{\mathcal L}_{FOC}(\theta_c,\theta_m)$ using \eqref{eq:cefol_train5_foc_loss}, with the same independent sample-mean stationarity estimator as in \eqref{eq:cefol_train4_stationarity_loss}.
			\If{$k\equiv0\pmod d$}
			\State Update $\theta_c$ and $\theta_m$ using the five-network first-order/KKT loss in \eqref{eq:cefol_train5_foc_loss}.
			\EndIf
			\State Update $\theta_e$ using \eqref{eq:cefol_train5_expected_update} and update $\theta_D$ using \eqref{eq:cefol_train5_difference_update}.
			\State Update $\theta_V$ as in \eqref{eq:cefol_train4_value_update}, using the Bellman target in \eqref{eq:cefol_train5_bellman_target}.
			\State Soft-update $\theta_V^-$ as in \eqref{eq:cefol_train4_target_update}.
			\If{convergence criterion is satisfied}
			\State \Return $(\theta_V,\theta_c,\theta_m,\theta_e,\theta_D)$.
			\EndIf
			\EndFor
		\end{algorithmic}
	\end{algorithm}
	
	\subsubsection{Three-Network Training Procedure without a Separate Value Update}
	\label{subsubsec:cefol_train_three}
	
	The three-network version updates only the policy network $c(s_t;\theta_c)$, multiplier network $m(s_t;\theta_m)$, and certainty-equivalent network $\mathcal C(s_t,c_t;\theta_{\mathcal C})$. The current value is not represented by a separate value network. Instead, it is defined by the value mapping $V^{I}(s_t;\theta_c,\theta_{\mathcal C})$ in \eqref{eq:cefol_three_implicit_value}. The transformed certainty-equivalent target is
	\begin{equation}
		\widehat G_t^{I,(i)}
		=
		\frac{1}{N_z}
		\sum_{j=1}^{N_z}
		f\left(V^{I}(s_{t+1}^{(i,j)};\theta_c,\theta_{\mathcal C})\right),
		\label{eq:cefol_train3_ce_target}
	\end{equation}
	The three-network certainty-equivalent loss is
	\begin{equation}
		\widehat{\mathcal L}_{C}^{I}(\theta_{\mathcal C})
		=
		\frac{1}{N}
		\sum_{i=1}^{N}
		\left[
		f\left(\mathcal C(s_t^{(i)},c_t^{(i)};\theta_{\mathcal C})\right)
		-
		\widehat G_t^{I,(i)}
		\right]^2.
		\label{eq:cefol_train3_ce_loss}
	\end{equation}
	There is no separate Bellman loss. The first-order/KKT residual uses the implicit value mapping in the model-specific integrand,
	\begin{equation}
		F_{t+1,k}^{I,(i,j)}
		=
		F_{k}\left(s_t^{(i)},c_t^{(i)},z_{t+1}^{(i,j)};
		V^{I}(\cdot;\theta_c,\theta_{\mathcal C}),c(\cdot;\theta_c),\mathcal C(\cdot,\cdot;\theta_{\mathcal C}),w,f,\psi\right),
		\label{eq:cefol_train3_foc_integrand}
	\end{equation}
	The stationarity component $\widehat{\mathcal L}_{S}^{I}(\theta_c,\theta_m)$ is constructed as in \eqref{eq:cefol_train4_stationarity_loss}, replacing $F_{t+1,k}^{(i,j)}$ with $F_{t+1,k}^{I,(i,j)}$ and using the same two independent sample means.
	The first-order/KKT loss is
	\begin{equation}
		\widehat{\mathcal L}_{FOC}^{I}(\theta_c,\theta_m)
		=
		\lambda_S\widehat{\mathcal L}_{S}^{I}(\theta_c,\theta_m)
		+
		\lambda_{FB}\widehat{\mathcal L}_{FB}^{I}(\theta_c,\theta_m)
		+
		\lambda_{EQ}\widehat{\mathcal L}_{EQ}^{I}(\theta_c).
		\label{eq:cefol_train3_foc_loss}
	\end{equation}
	The certainty-equivalent update and first-order/KKT updates are
	\begin{align}
		\theta_{\mathcal C}
		&\leftarrow
		\theta_{\mathcal C}-
		\alpha_C\nabla_{\theta_{\mathcal C}}\widehat{\mathcal L}_{C}^{I}(\theta_{\mathcal C}),
		\label{eq:cefol_train3_ce_update}
		\\
		\theta_c
		&\leftarrow
		\theta_c-
		\alpha_c\nabla_{\theta_c}\widehat{\mathcal L}_{FOC}^{I}(\theta_c,\theta_m),
		\nonumber\\
		\theta_m
		&\leftarrow
		\theta_m-
		\alpha_m\nabla_{\theta_m}\widehat{\mathcal L}_{FOC}^{I}(\theta_c,\theta_m).
		\label{eq:cefol_train3_policy_multiplier_update}
	\end{align}
	The first-order/KKT update is again delayed until $k\equiv0\pmod d$. Since this variant has no explicit value network, it does not use the value-target-network mechanism in Section~\ref{subsubsec:cefol_target_network}. This version is compact because it removes the value network and its Bellman loss, but it also removes a direct value-consistency diagnostic.
	
	\begin{algorithm}[H]
		\caption{CEFOL Algorithm with Three-Network Architecture without Separate Value Update}
		\label{alg:cefol_three_network}
		\begin{algorithmic}[1]
			\State Initialize network parameters $\theta_c$, $\theta_m$, and $\theta_{\mathcal C}$.
			\State Initialize perturbation scale $\zeta_k$ and delay parameter $d$.
			\For{$k=1,2,\ldots,K$}
			\State Use the same simulated mini-batch and future-state construction as \eqref{eq:cefol_train4_sample_states}--\eqref{eq:cefol_train4_future_controls}.
			\State Compute $\widehat{\mathcal L}_{C}^{I}(\theta_{\mathcal C})$ using \eqref{eq:cefol_train3_ce_target}--\eqref{eq:cefol_train3_ce_loss}.
			\State Compute $\widehat{\mathcal L}_{FOC}^{I}(\theta_c,\theta_m)$ using \eqref{eq:cefol_train3_foc_integrand}--\eqref{eq:cefol_train3_foc_loss}, with the independent sample-mean stationarity estimator.
			\If{$k\equiv0\pmod d$}
			\State Update $\theta_c$ and $\theta_m$ using \eqref{eq:cefol_train3_policy_multiplier_update}.
			\EndIf
			\State Update $\theta_{\mathcal C}$ using \eqref{eq:cefol_train3_ce_update}.
			\If{convergence criterion is satisfied}
			\State \Return $(\theta_c,\theta_m,\theta_{\mathcal C})$.
			\EndIf
			\EndFor
		\end{algorithmic}
	\end{algorithm}
	
	\subsection{Discussion}
	\label{subsec:algorithm_discussion}
	
	CEFOL is a deep learning method for recursive utility that updates the policy and multiplier networks using first-order/KKT residuals. The conditional stationarity residuals discipline the policy and multiplier networks, and the multiplier network learns the inequality and equality multipliers associated with the constraints. The certainty-equivalent network remains essential because recursive utility introduces a nonlinear certainty-equivalent value into both the Bellman equation and the first-order conditions.
	
	The four-network architecture provides the most transparent decomposition: the value network enforces Bellman consistency, the policy network learns the control rule, the multiplier network enforces KKT optimality, and the certainty-equivalent network approximates the nonlinear certainty-equivalent value. The five-network architecture further separates the conditional expectation of next-period value from the nonlinear-difference component. The three-network architecture removes the explicit value update and evaluates next period values through the value mapping $V^{I}$. This decomposition makes the method applicable to recursive utility models with vector controls, multiple constraints, and general stochastic dynamics, while preserving consistency with the CEFOL notation used in the numerical results.

\section{Numerical Results}\label{sec:numerical_results}

This section evaluates the accuracy and consistency of the value and policy functions learned by the deep learning algorithms described in the preceding sections. Assessing solution quality in discrete-time dynamic programming problems requires metrics tailored to the availability of benchmarks.

Two classical diagnostic metrics are used throughout: the Bellman error and the Euler residual. The Bellman error quantifies the discrepancy between the two sides of the recursive Bellman equation, the fundamental optimality condition of discrete-time dynamic programming. The Euler residual evaluates violations of the first-order intertemporal condition derived from the maximization step inside the Bellman equation. Both metrics are identically zero for the exact solution; consequently, small values provide evidence that the learned policy and value functions approximately satisfy the necessary conditions for optimality.

To make the Euler-residual diagnostic comparable across models, we use a common pointwise normalization. When an interior Euler equation can be written as
\begin{equation}
	\mathrm{LHS}_t=\mathrm{RHS}_t,
	\label{eq:app_generic_euler}
\end{equation}
we define the Euler residual as the relative deviation between the two sides,
\begin{equation}
	\mathrm{ER}_t
	=
	\left|
	1-\frac{\mathrm{RHS}_t}{\mathrm{LHS}_t}
	\right|.
	\label{eq:app_generic_euler_residual}
\end{equation}
If the same condition is reported in normalized form with a unit left-hand side, this definition reduces to
\begin{equation}
	\mathrm{ER}_t
	=
	\left|1-\mathrm{RHS}_t\right|.
	\label{eq:app_normalized_euler_residual}
\end{equation}

Computing these diagnostics requires evaluating conditional expectations of future quantities under the learned policy. We adopt a nested simulation approach: for each state along a test trajectory, multiple independent future paths are simulated, the relevant next-period quantities are averaged, and the recursive aggregation operator is applied. This procedure is flexible and model-agnostic, and a sufficiently large number of inner simulations yields precise diagnostic estimates.

We consider three classes of discrete-time dynamic programming problems with recursive utility. In Section~\ref{subsec:rs_consumption_saving_rs}, we study a recursive consumption-saving framework with risk-sensitive preferences, following \citet{hansen2012recursive} and the environment of \citet{maliar2021deep}. In Section~\ref{subsec:rs_consumption_saving_ez}, we consider the same consumption-saving environment with Epstein--Zin preferences \citep{epstein1989substitution}. In Section~\ref{subsec:small_noise_robust_control}, we examine the small-noise robust-control model of \citet{andreasen2012effects}, benchmarking the general nonlinear case against value function iteration (VFI).
Finally, in Section~\ref{subsec:rs_dsge}, we apply the method to a DSGE model with recursive preferences and stochastic volatility, adapted from \citet{caldara2012computing}, which extends the analysis to a macroeconomic setting with a three-dimensional state vector and two continuous controls.

Throughout the figures, CEFOL denotes the direct neural-network value or policy. For value-function plots, CEFOL-td denotes the one-step temporal-difference expansion using the learned policy and the next-period value network, while CEFOL-vc denotes the value implied by the learned certainty-equivalent network. These three value representations are internal consistency checks rather than separate solution methods: close agreement among them indicates that the learned value, policy, and certainty-equivalent networks satisfy the recursive relationships implied by the Bellman equation. For small-noise models, VFI is the grid-based benchmark. 
\subsection{Recursive Consumption-Saving Model: Risk-Sensitive Formulation}
\label{subsec:rs_consumption_saving_rs}

We first consider a risk-sensitive recursive version of the consumption-saving framework studied in \citet{maliar2021deep}. The agent chooses consumption and saving subject to a cash-on-hand constraint and an occasionally binding borrowing limit. This environment features four exogenous driving processes and a borrowing friction that gives rise to a Kuhn--Tucker multiplier, thereby requiring a richer treatment of the Euler equation.

\subsubsection{Problem Setup}
\label{subsubsec:rs_problem_setup}

At the beginning of period \(t\), the household has cash-on-hand \(w_t\). After observing the current exogenous states, it chooses the consumption ratio \(c_t\in(0,1]\). Actual consumption is
\begin{equation*}
\tilde c_t = c_t w_t.
\end{equation*}
The remaining resources,
\begin{equation*}
(1-c_t)w_t,
\end{equation*}
are saved at time \(t\).

Between periods \(t\) and \(t+1\), these savings earn the stochastic gross return \(\bar r\exp(r_{t+1})\), where \(\bar r>0\) is the deterministic component of the gross return and \(r_{t+1}\) is the next-period interest-rate shock. At time \(t+1\), the household also receives labor income
\begin{equation}
Y_{t+1}
=
\exp(p_{t+1})\exp(q_{t+1})
=
\exp(p_{t+1}+q_{t+1}),
\label{eq:rs_income}
\end{equation}
where \(p_{t+1}\) and \(q_{t+1}\) are the permanent and transitory components of labor income, respectively. Therefore, next-period cash-on-hand is
\begin{equation}
w_{t+1}
=
(1-c_t)w_t\bar r\exp(r_{t+1})
+
\exp(p_{t+1}+q_{t+1}).
\label{eq:rs_wealth_transition}
\end{equation}

The exogenous variables \(q_t\), \(p_t\), \(r_t\), and \(\delta_t\) follow independent AR(1) processes:
\begin{align*}
q_{t+1} &= \rho_q q_t+\sigma_q\epsilon^q_{t+1}, \\
p_{t+1} &= \rho_p p_t+\sigma_p\epsilon^p_{t+1}, \\
r_{t+1} &= \rho_r r_t+\sigma_r\epsilon^r_{t+1}, \\
\delta_{t+1} &= \rho_\delta\delta_t+\sigma_\delta\epsilon^\delta_{t+1},
\end{align*}
where \(\epsilon^j_{t+1}\sim\mathcal{N}(0,1)\) are mutually independent for \(j\in\{q,p,r,\delta\}\). The future shock vector is
\begin{equation*}
z_{t+1}
=
\left(
\epsilon^q_{t+1},
\epsilon^p_{t+1},
\epsilon^r_{t+1},
\epsilon^\delta_{t+1}
\right).
\end{equation*}

Accordingly, the state vector is
\begin{equation*}
s_t
=
\left(
w_t,
r_t,
\delta_t,
q_t,
p_t
\right),
\end{equation*}
where \(w_t\) denotes beginning-of-period cash-on-hand, \(r_t\) is the interest-rate shock, \(\delta_t\) is the preference shock, and \(q_t\) and \(p_t\) are the transitory and permanent components of labor income. The state transition can be written compactly as
\begin{equation*}
s_{t+1}
=
\psi(s_t,c_t,z_{t+1})
=
\left(
w_{t+1},
r_{t+1},
\delta_{t+1},
q_{t+1},
p_{t+1}
\right),
\end{equation*}
where \(w_{t+1}\) is given by \eqref{eq:rs_wealth_transition}.

Given this timing, the household maximizes current utility plus a risk-sensitive certainty-equivalent value. The risk-sensitive consumption-saving problem is
\begin{equation}
V(s_t)
=
\max_{c_t}
\left\{
\exp(\delta_t)u(\tilde c_t)
-
\frac{1}{\sigma}
\log
\mathbb{E}_t
\left[
\exp\{-\sigma\beta V(s_{t+1})\}
\right]
\right\},
\label{eq:rs_bellman}
\end{equation}
subject to the primitive feasibility condition
\begin{equation}
0<\tilde c_t\le w_t,
\qquad
\tilde c_t=c_t w_t,
\label{eq:rs_primitive_constraint}
\end{equation}
the wealth transition in \eqref{eq:rs_wealth_transition}, and the exogenous processes specified above.

The lower boundary \(\tilde c_t=0\) is not treated as a separate KKT constraint. With CRRA utility,
\begin{equation*}
u(\tilde c_t)
=
\frac{\tilde c_t^{1-\gamma}-1}{1-\gamma},
\qquad
u'(\tilde c_t)
=
\tilde c_t^{-\gamma},
\qquad
\gamma\neq1,
\end{equation*}
we have
\begin{equation*}
u'(\tilde c_t)\rightarrow\infty
\qquad
\text{as }
\tilde c_t\downarrow0.
\end{equation*}
Thus, zero consumption is ruled out by the marginal utility behavior of the period utility. The logarithmic utility case is obtained as the limit \(\gamma\to1\).

In computation, the policy network uses a sigmoid output layer. Hence,
\begin{equation*}
0<c_t<1,
\qquad
0<\tilde c_t=c_t w_t<w_t,
\end{equation*}
so the primitive feasibility condition \eqref{eq:rs_primitive_constraint} is satisfied by construction. For consistency with the general KKT notation, the upper resource boundary is represented by the normalized slack
\begin{equation*}
g(s_t,c_t)
=
1-c_t
\geq0.
\end{equation*}
There is no equality constraint in this example, so \(M_q=0\) and \(M_g=1\).

The risk-sensitive Bellman equation \eqref{eq:rs_bellman} corresponds to the general certainty-equivalent formulation with
\begin{equation*}
f(x)
=
\exp(-\sigma\beta x),
\qquad
f^{-1}(y)
=
-\frac{1}{\sigma\beta}\log y.
\end{equation*}

We use the four-network architecture introduced in Section~\ref{subsubsec:cefol_four_network}.

\subsubsection{First-Order Condition and KKT Losses}
\label{subsubsec:rs_foc_aio_residuals}

To align this model with the general CEFOL formulation in Section~\ref{subsec:cefol_foc_kkt_losses}, we specialize the FOC/KKT system to the scalar-control, single-inequality-constraint case:
\begin{equation*}
n_c=1,
\qquad
M_g=1,
\qquad
M_q=0,
\qquad
g(s_t,c_t)=1-c_t,
\qquad
\frac{\partial g(s_t,c_t)}{\partial c_t}=-1.
\end{equation*}

For the risk-sensitive transformation \(f(x)=\exp(-\sigma\beta x)\), the general distortion term \(\chi_{t+1}^{f}\) becomes
\begin{equation}
\chi_{t+1}^{f}
=
\frac{
\exp\{-\sigma\beta V(s_{t+1})\}
}{
\mathbb{E}_t
\left[
\exp\{-\sigma\beta V(s_{t+1})\}
\right]
}.
\label{eq:rs_distortion}
\end{equation}
This term satisfies
\begin{equation*}
\mathbb{E}_t\left[\chi_{t+1}^{f}\right]=1.
\end{equation*}

Using the certainty-equivalent network, the denominator in \eqref{eq:rs_distortion} is represented by
\begin{equation*}
\exp
\left\{
-\sigma\beta
\mathcal{C}(s_t,c_t;\theta_{\mathcal C})
\right\}.
\end{equation*}
Accordingly, the network-based approximation to the distortion term is
\begin{equation}
\widehat{\chi}_{t+1}^{f}
=
\frac{
\exp\{-\sigma\beta V(s_{t+1};\theta_V)\}
}{
\exp\{-\sigma\beta
\mathcal{C}(s_t,c_t;\theta_{\mathcal C})\}
}
=
\exp
\left\{
-\sigma\beta
\left[
V(s_{t+1};\theta_V)
-
\mathcal{C}(s_t,c_t;\theta_{\mathcal C})
\right]
\right\}.
\label{eq:rs_hat_distortion}
\end{equation}

The constrained risk-sensitive Bellman problem implies the population stationarity condition
\begin{equation}
\mathbb{E}_t\left[F_{t+1,1}\right]
+
\lambda_t
\frac{\partial g(s_t,c_t)}{\partial c_t}
=
\mathbb{E}_t\left[F_{t+1,1}\right]
-
\lambda_t
=
0,
\label{eq:rs_stationarity_condition}
\end{equation}
where the stochastic stationarity integrand is
\begin{equation}
F_{t+1,1}
=
w_t
\left[
\exp(\delta_t)u'(\tilde c_t)
-
\beta
\chi_{t+1}^{f}
\exp(\delta_{t+1})
u'(\tilde c_{t+1})
\bar r\exp(r_{t+1})
\right].
\label{eq:rs_F_integrand}
\end{equation}
Here,
\begin{equation*}
\tilde c_t=c_t w_t,
\qquad
\tilde c_{t+1}=c_{t+1}w_{t+1}.
\end{equation*}

In implementation, the sampled integrand uses the network-based distortion \(\widehat{\chi}_{t+1}^{f}\):
\begin{equation}
\widehat F_{t+1,1}
=
w_t
\left[
\exp(\delta_t)u'(\tilde c_t)
-
\beta
\widehat{\chi}_{t+1}^{f}
\exp(\delta_{t+1})
u'(\tilde c_{t+1})
\bar r\exp(r_{t+1})
\right].
\label{eq:rs_hat_F_integrand}
\end{equation}

Suppose \(N_z\) is even and let \(J_z=N_z/2\). For each mini-batch state-control pair \(i\), let \(\widehat F_{t+1,1}^{(i,j)}\) denote the realization of \eqref{eq:rs_hat_F_integrand} under future draw \(j=1,\ldots,N_z\). The corresponding stochastic stationarity residual is
\begin{equation}
R_{1,t}^{\mathrm{stat},(i,j)}
=
\widehat F_{t+1,1}^{(i,j)}
-
m(s_t^{(i)};\theta_m).
\label{eq:rs_stationarity_residual}
\end{equation}

Split the \(N_z\) future draws into two conditionally independent groups and define
\begin{equation}
\overline R_{1,t}^{\mathrm{stat},(i,1)}
=
\frac{1}{J_z}
\sum_{j=1}^{J_z}
R_{1,t}^{\mathrm{stat},(i,j)},
\qquad
\overline R_{1,t}^{\mathrm{stat},(i,2)}
=
\frac{1}{J_z}
\sum_{j=J_z+1}^{N_z}
R_{1,t}^{\mathrm{stat},(i,j)}.
\label{eq:rs_split_stationarity_residuals}
\end{equation}
Using these independent sample means, the stationarity loss is the scalar-control specialization of \eqref{eq:cefol_stationarity_loss}:
\begin{equation}
\widehat{\mathcal L}_{S}(\theta_c,\theta_m)
=
\frac{1}{N}
\sum_{i=1}^{N}
\overline R_{1,t}^{\mathrm{stat},(i,1)}
\overline R_{1,t}^{\mathrm{stat},(i,2)}.
\label{eq:rs_stationarity_loss}
\end{equation}

The KKT conditions for the upper resource constraint are
\begin{equation*}
g(s_t,c_t)\geq0,
\qquad
\lambda_t\geq0,
\qquad
g(s_t,c_t)\lambda_t=0.
\end{equation*}
Because this model has one inequality constraint and no equality constraints, the Fischer--Burmeister loss is the one-constraint specialization of \eqref{eq:cefol_fb_loss}:
\begin{equation}
\widehat{\mathcal{L}}_{FB}(\theta_c,\theta_m)
=
\frac{1}{N}
\sum_{i=1}^{N}
v_{FB}
\left[
\Phi^{FB}
\left(
1-c_t^{(i)},
m(s_t^{(i)};\theta_m)
\right)
\right]^2.
\label{eq:rs_fb_loss}
\end{equation}
This loss disciplines feasibility, multiplier nonnegativity, and complementarity for the upper resource constraint. Since \(M_q=0\), there is no equality-constraint term in this model.

The general first-order/KKT loss therefore reduces to
\begin{equation}
\widehat{\mathcal{L}}_{FOC}(\theta_c,\theta_m)
=
\lambda_S
\widehat{\mathcal{L}}_{S}(\theta_c,\theta_m)
+
\lambda_{FB}
\widehat{\mathcal{L}}_{FB}(\theta_c,\theta_m).
\label{eq:rs_foc_loss}
\end{equation}
The policy-network parameters and multiplier-network parameters are updated jointly according to the rule in Section~\ref{subsec:cefol_training_procedure}:
\begin{equation*}
(\theta_c,\theta_m)
\leftarrow
(\theta_c,\theta_m)
-
\alpha_{cm}
\nabla_{\theta_c,\theta_m}
\widehat{\mathcal{L}}_{FOC}(\theta_c,\theta_m).
\end{equation*}
During this policy--multiplier step, the value and certainty-equivalent networks are held fixed. They are then updated separately by their own residual losses, exactly as in Section~\ref{subsec:cefol_training_procedure}.

\subsubsection{Numerical Results}

We report two risk-sensitive recursive-utility calibrations, with robustness parameters $\sigma=1$ and $\sigma=100$. Consistent results are obtained across both cases. Since no external benchmark is available for this model, the validation focuses on internal recursive consistency, Bellman errors, and Euler residuals rather than relative-difference figures. In the value function comparison, the direct value network and the recursively implied value representations are closely aligned, indicating that the learned value and certainty-equivalent value networks satisfy the recursive fixed-point relation. Because the risk-sensitive value function crosses zero, relative Bellman errors are numerically unstable near zero-value regions; we therefore report absolute Bellman errors instead. The absolute Bellman error is generally of order $10^{-3}$. The Euler residual is small over the unconstrained interior region, with medians around $10^{-4}$. The Euler residual is larger near the binding constraint because it measures the interior Euler equation rather than the constrained KKT stationarity condition. In all figures, cash-on-hand $w_t$ is varied over $[0,4]$, while the remaining exogenous state variables are held at their steady-state values.

\paragraph{Results for baseline robustness: $\sigma=1$}

Figure~\ref{fig:rs-value-sigma1} reports the direct value network and its recursive expansions. 

\resultfig{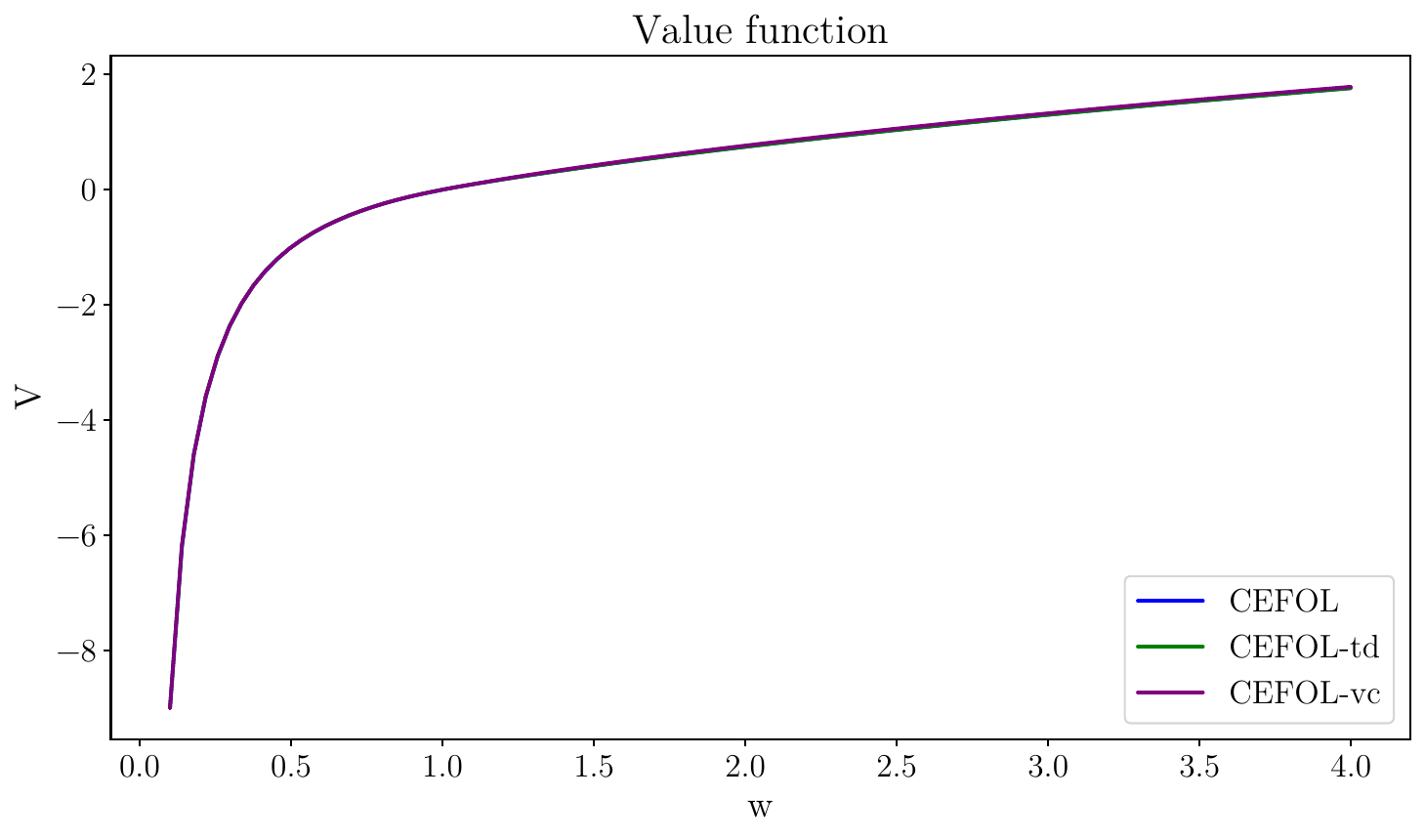}
{Value function and learned recursive expansions for the risk-sensitive consumption-saving model with $\sigma=1$.}
{fig:rs-value-sigma1}

Figure~\ref{fig:rs-policy-sigma1} reports the consumption policy and the upper bound \(\tilde{c}=w\).

\resultfig{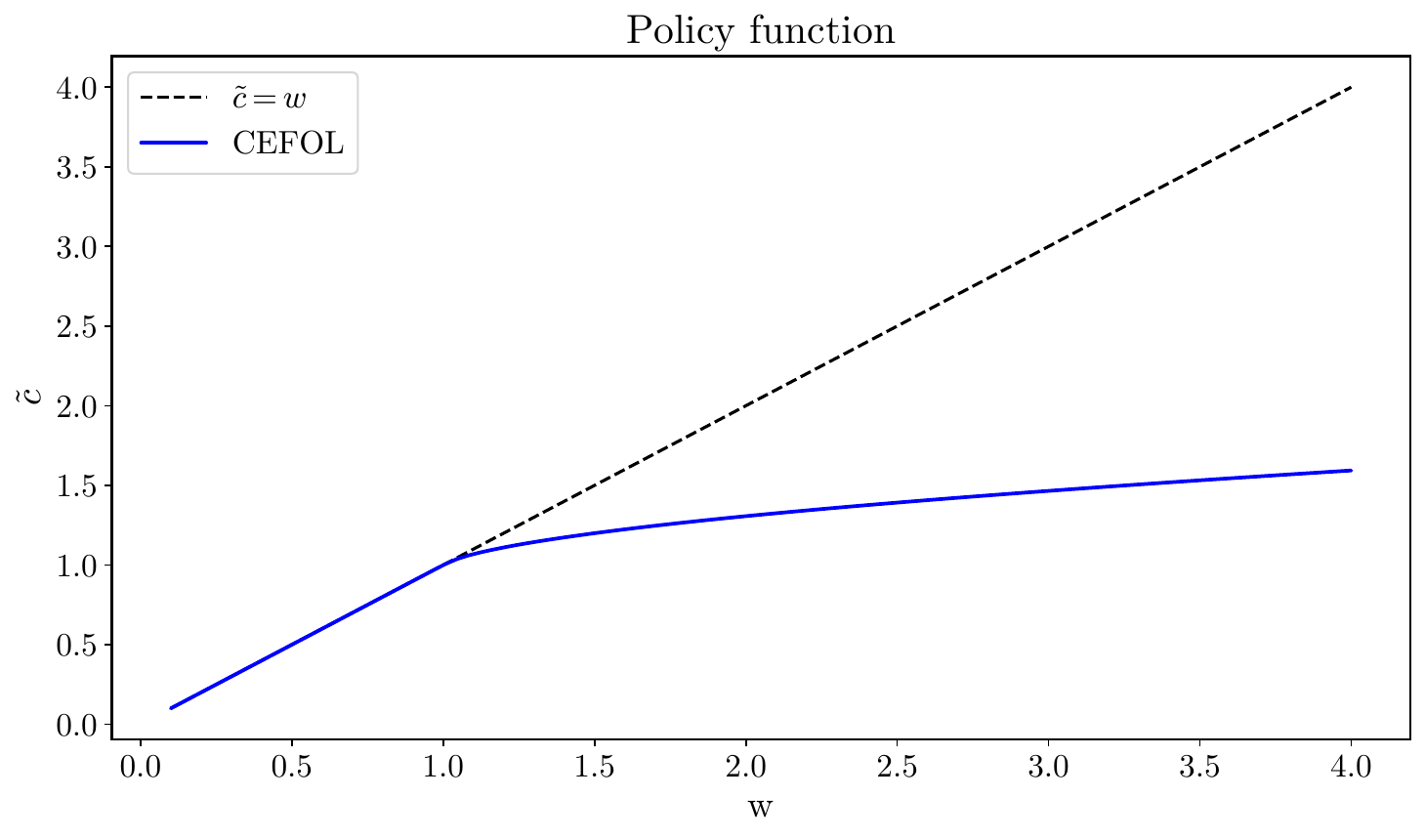}
{Consumption policy for the risk-sensitive consumption-saving model with $\sigma=1$. The dashed line reports the upper bound \(\tilde{c}=w\).}
{fig:rs-policy-sigma1}

Figures~\ref{fig:rs-bellman-sigma1} and \ref{fig:rs-euler-sigma1} report the Bellman error and Euler residual.

\resultfig{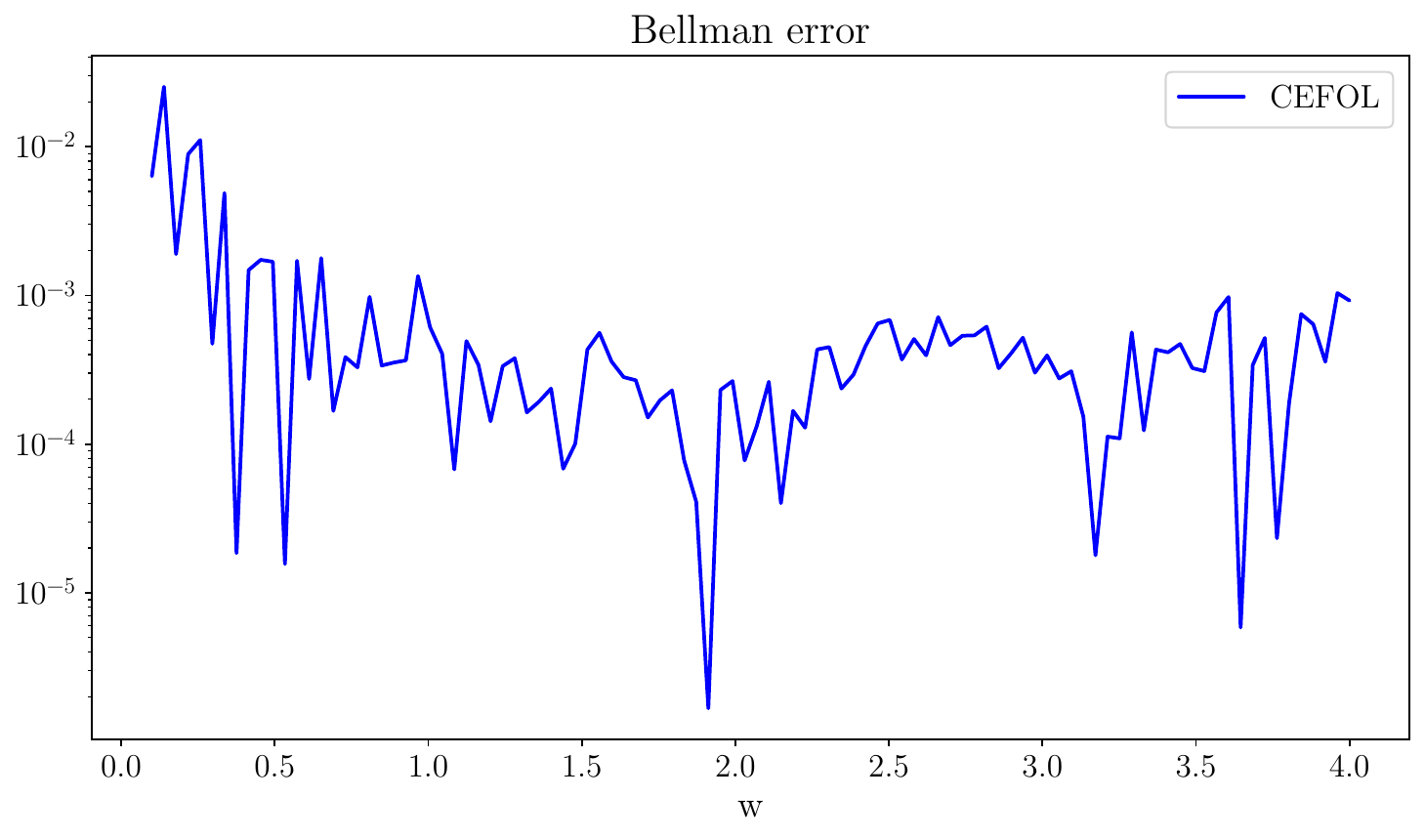}
{Bellman error for the risk-sensitive consumption-saving model with $\sigma=1$. }
{fig:rs-bellman-sigma1}

\resultfig{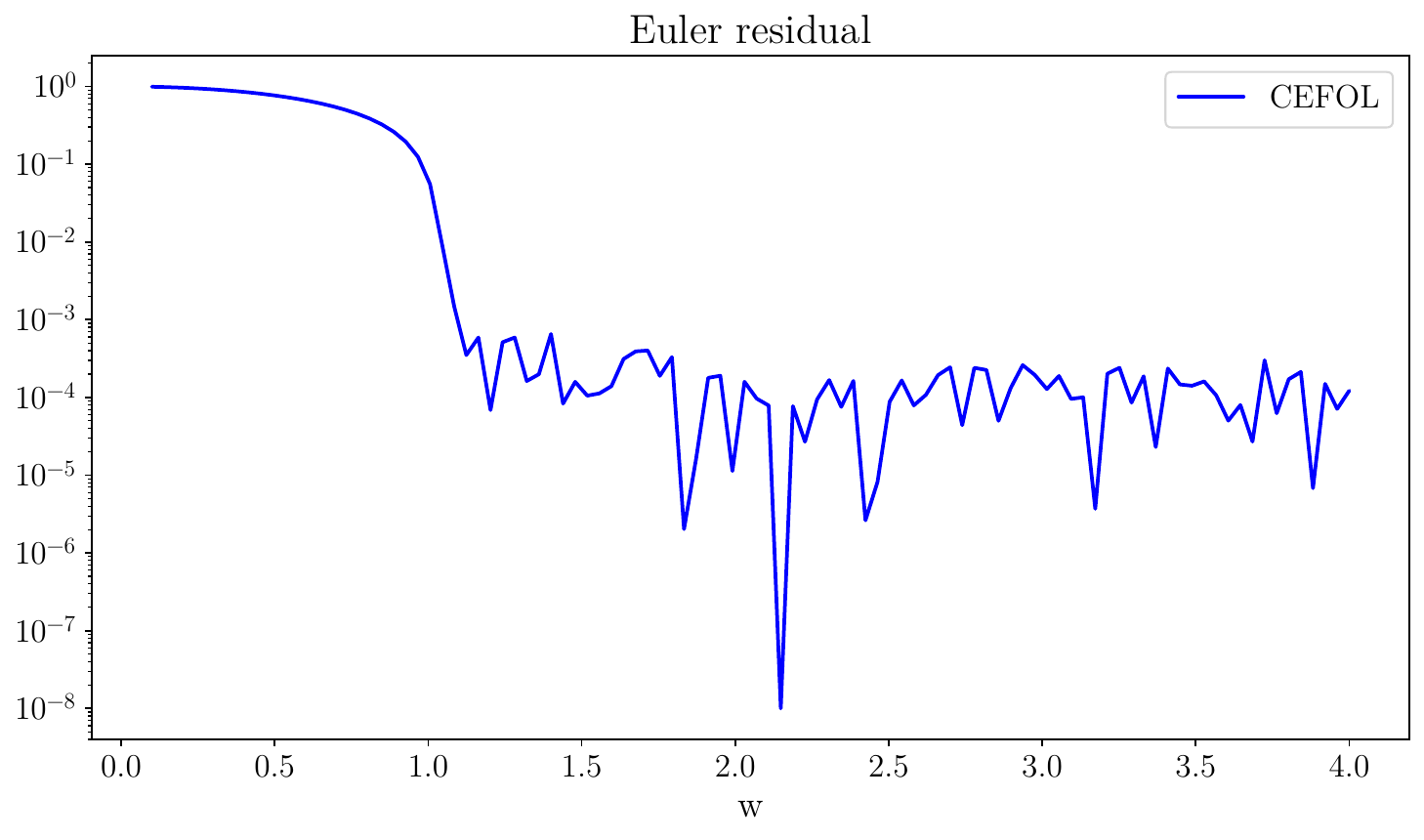}
{Euler residual for the risk-sensitive consumption-saving model with $\sigma=1$.}
{fig:rs-euler-sigma1}

\paragraph{Results for high robustness: $\sigma=100$}

Figure~\ref{fig:rs-value-sigma100} reports the direct value network and its recursive expansions. 

\resultfig{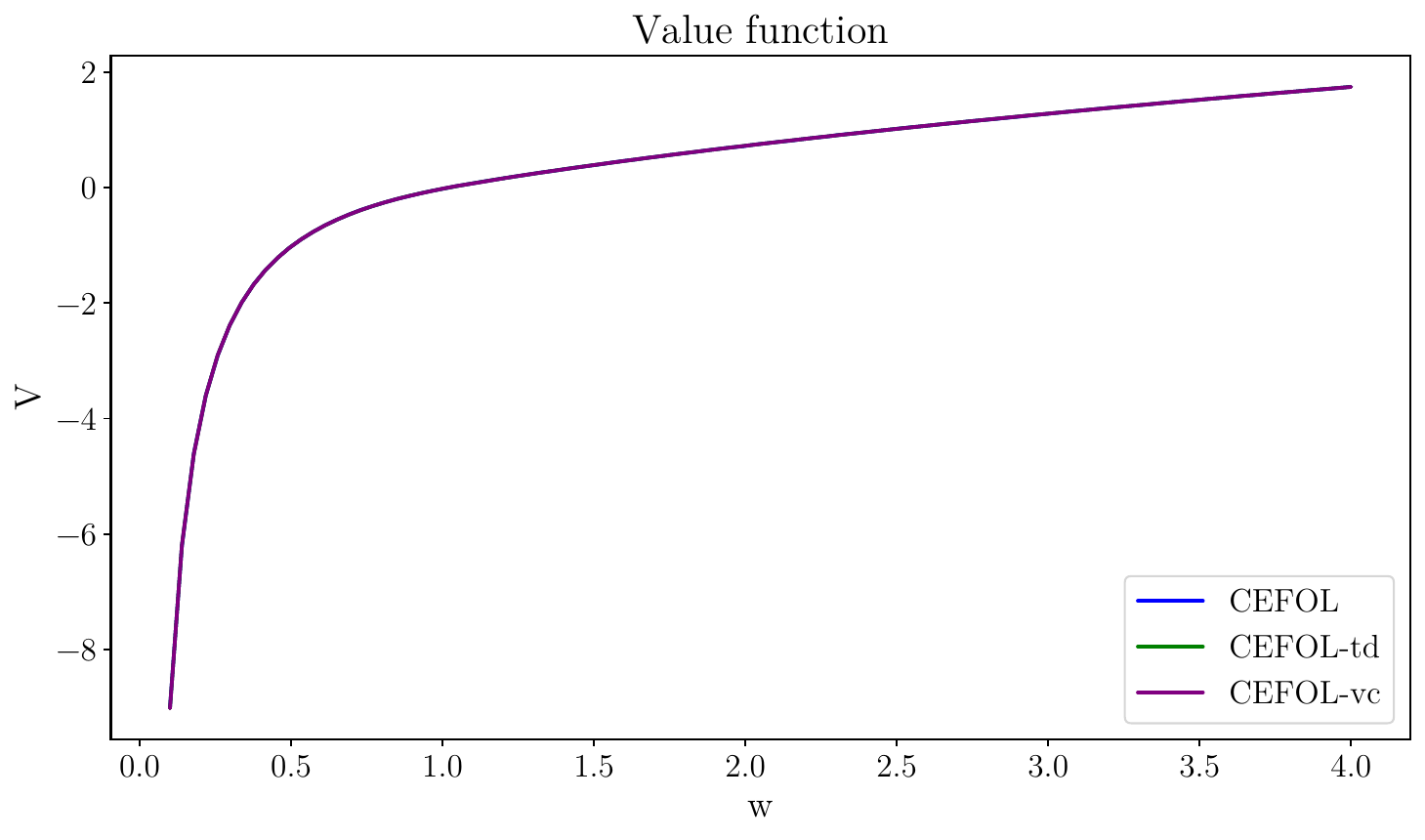}
{Value function and learned recursive expansions for the risk-sensitive consumption-saving model with $\sigma=100$.}
{fig:rs-value-sigma100}

Figure~\ref{fig:rs-policy-sigma100} reports the consumption policy and the resource upper bound $\tilde{c}=w$.

\resultfig{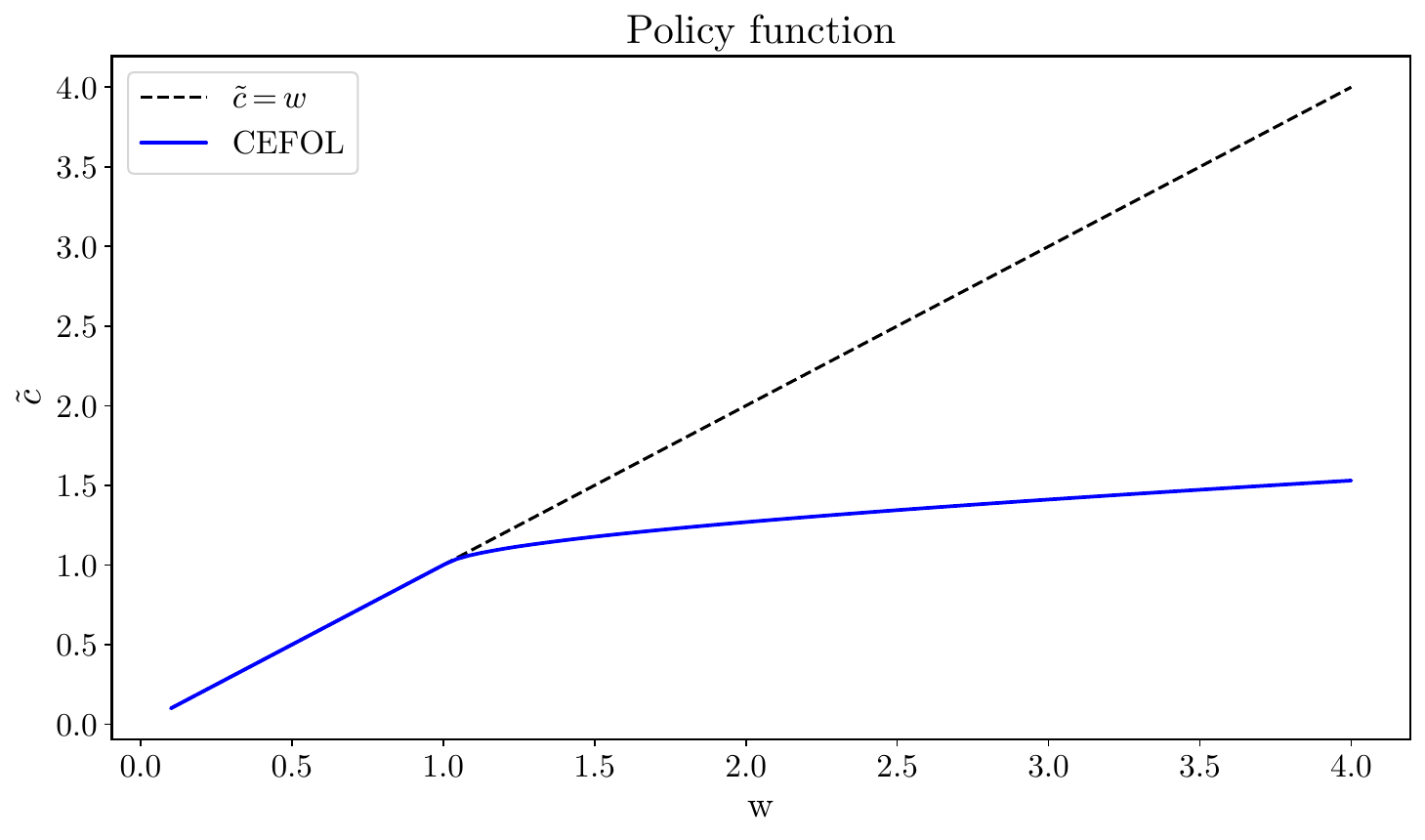}
{Consumption policy for the risk-sensitive consumption-saving model with $\sigma=100$. The dashed line reports $\tilde{c}=w$.}
{fig:rs-policy-sigma100}

Figures~\ref{fig:rs-bellman-sigma100} and \ref{fig:rs-euler-sigma100} report the Bellman error and Euler residual.

\resultfig{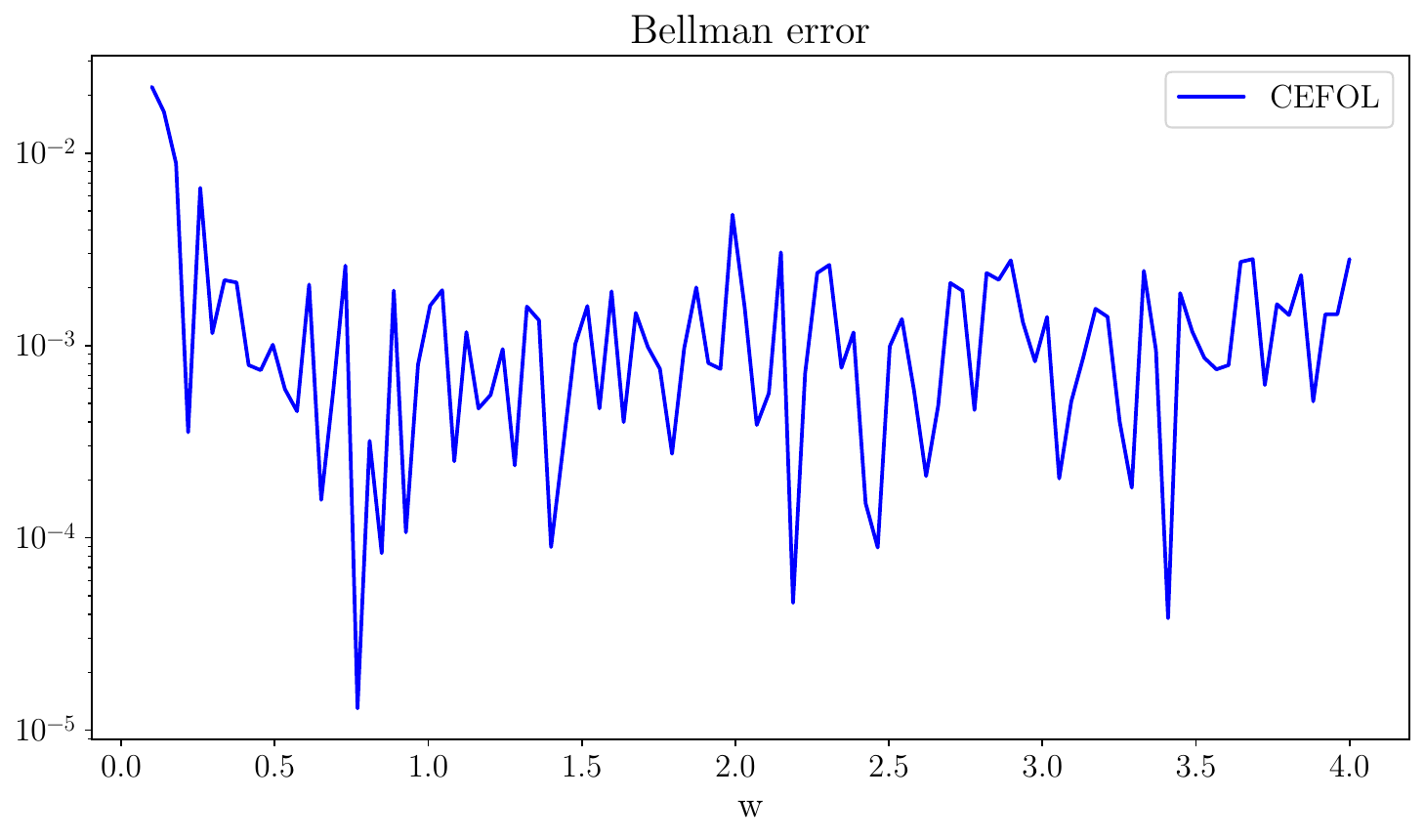}
{Bellman error for the risk-sensitive consumption-saving model with $\sigma=100$. Relative errors are not reported because the value function crosses zero.}
{fig:rs-bellman-sigma100}

\resultfig{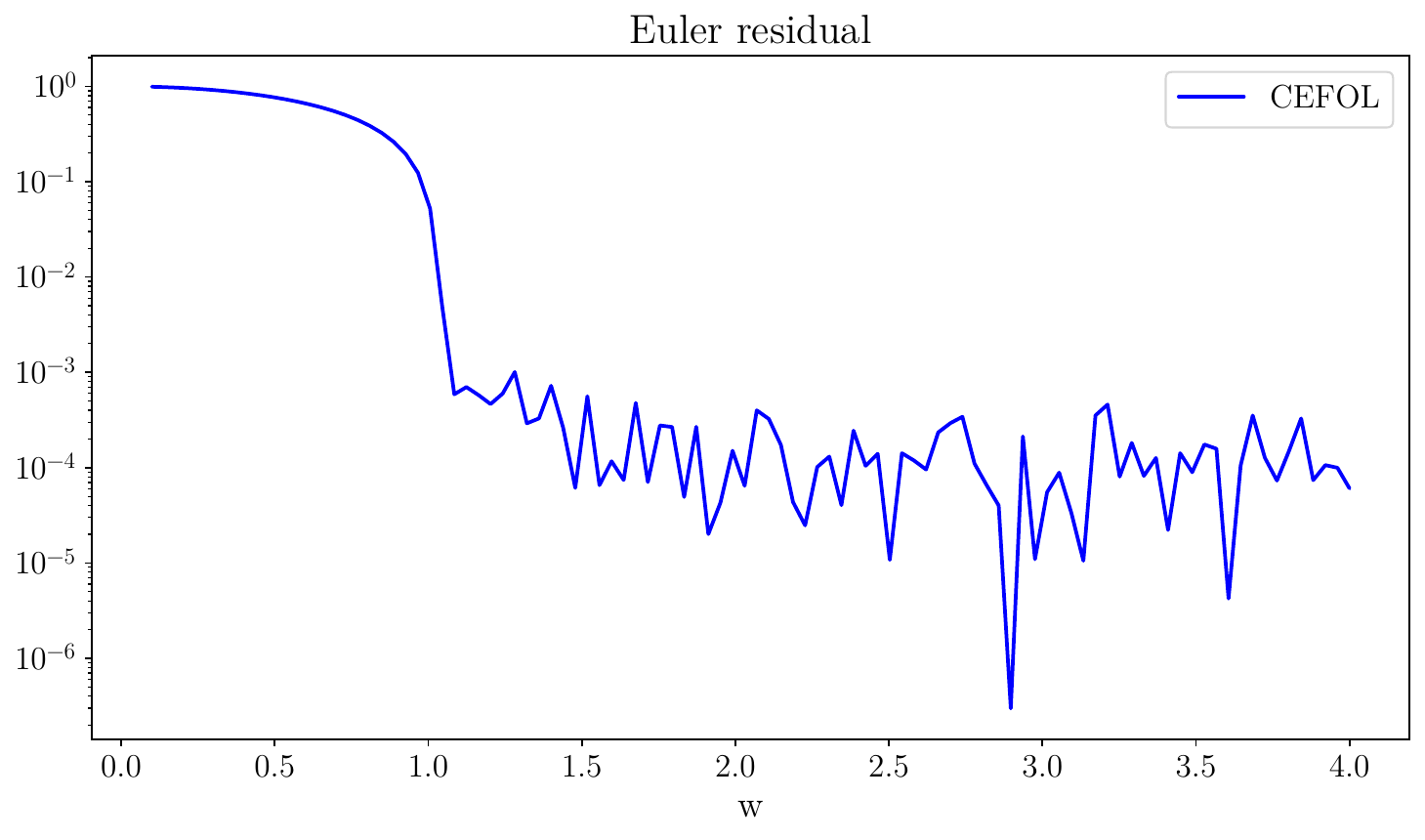}
{Euler residual for the risk-sensitive consumption-saving model with $\sigma=100$.}
{fig:rs-euler-sigma100}

\FloatBarrier

\subsection{Recursive Consumption-Saving Model: Epstein--Zin Formulation}
\label{subsec:rs_consumption_saving_ez}

We consider an Epstein--Zin recursive version of the consumption-saving framework studied in \citet{maliar2021deep}. The agent chooses consumption and saving subject to a cash-on-hand constraint and an occasionally binding borrowing limit. This environment features four exogenous driving processes and a borrowing friction that gives rise to a Kuhn--Tucker multiplier, thereby requiring a richer treatment of the first-order condition.

\subsubsection{Problem Setup}

This model shares the same state vector, income process, control variable, cash-on-hand transition, and borrowing constraint as the risk-sensitive consumption-saving setup in Section~\ref{subsubsec:rs_problem_setup}. The only change is the recursive utility specification. In particular, with \(\tilde c_t=c_t w_t\), the household now solves
\begin{equation*}
V\left(s_t\right)
=
\max_{0<c_t\leq1}
\left\{
(1-\beta)\exp\left(\delta_t\right)\tilde c_t^{1-\rho}
+
\beta
\left(
\mathbb{E}_t
\left[
V\left(s_{t+1}\right)^{1-\gamma}
\right]
\right)^{\frac{1-\rho}{1-\gamma}}
\right\}^{\frac{1}{1-\rho}},
\end{equation*}
where \(\beta\in(0,1)\) is the subjective discount factor, \(\gamma>0\) denotes the coefficient of relative risk aversion, and \(\rho>0\) is the inverse of the elasticity of intertemporal substitution (EIS). The limiting case \(\rho=\gamma\) recovers the standard time-separable expected utility benchmark. As in Section~\ref{subsubsec:rs_problem_setup}, the lower bound \(c_t=0\) is excluded because it implies zero consumption \(\tilde c_t=0\), at which the marginal utility \(\tilde c_t^{-\rho}\to\infty\) and the first-order condition becomes ill-posed.

We use the same four-network architecture introduced in Section~\ref{subsubsec:cefol_four_network}.

\subsubsection{First-Order Condition and KKT Losses}
\label{subsubsec:ez_foc_aio_residuals}

To align this model with the general CEFOL formulation in Section~\ref{subsec:cefol_foc_kkt_losses}, we specialize the FOC/KKT system to the scalar-control, single-inequality-constraint case:
\begin{equation*}
n_c=1,
\qquad
M_g=1,
\qquad
M_q=0,
\qquad
g(s_t,c_t)=1-c_t,
\qquad
\frac{\partial g(s_t,c_t)}{\partial c_t}=-1.
\end{equation*}

For the Epstein--Zin transformation \(f(x)=x^{1-\gamma}\), the general distortion term is
\begin{equation}
\chi_{t+1}^{f}
=
\left(
\frac{
V(s_{t+1})
}{
\mathcal C(s_t,c_t)
}
\right)^{-\gamma}.
\label{eq:ez_distortion}
\end{equation}
Using the value and certainty-equivalent networks, its neural network representation is
\begin{equation}
\widehat{\chi}_{t+1}^{f}
=
\left(
\frac{
V(s_{t+1};\theta_V)
}{
\mathcal C(s_t,c_t;\theta_{\mathcal C})
}
\right)^{-\gamma}.
\label{eq:ez_hat_distortion}
\end{equation}

The constrained Epstein--Zin Bellman problem implies the population stationarity condition
\begin{equation}
\mathbb{E}_t
\left[
F_{t+1,1}
\right]
+
\lambda_t
\frac{\partial g(s_t,c_t)}{\partial c_t}
=
\mathbb{E}_t
\left[
F_{t+1,1}
\right]
-
\lambda_t
=
0,
\label{eq:ez_stationarity_condition}
\end{equation}
where the stochastic stationarity integrand is
\begin{equation}
\begin{aligned}
F_{t+1,1}
={}&
w_t(1-\beta)
\exp(\delta_t)
\tilde c_t^{-\rho}
V(s_t)^\rho
\\
&\times
\Bigg[
1
-
\beta
\chi_{t+1}^{f}
\left(
\frac{
V(s_{t+1})
}{
\mathcal C(s_t,c_t)
}
\right)^\rho
\exp\left(
\delta_{t+1}-\delta_t
\right)
\left(
\frac{
\tilde c_{t+1}
}{
\tilde c_t
}
\right)^{-\rho}
\bar r
\exp(r_{t+1})
\Bigg].
\end{aligned}
\label{eq:ez_foc_integrand}
\end{equation}
Equivalently, the product of the two value ratios in \eqref{eq:ez_foc_integrand} is
\begin{equation*}
\chi_{t+1}^{f}
\left(
\frac{
V(s_{t+1})
}{
\mathcal C(s_t,c_t)
}
\right)^\rho
=
\left(
\frac{
V(s_{t+1})
}{
\mathcal C(s_t,c_t)
}
\right)^{\rho-\gamma}.
\end{equation*}

Here,
\begin{equation*}
\tilde c_t=c_t w_t,
\qquad
\tilde c_{t+1}
=
c(s_{t+1};\theta_c)w_{t+1}.
\end{equation*}

In implementation, the sampled integrand is constructed using the network-based distortion:
\begin{equation}
\begin{aligned}
\widehat F_{t+1,1}
={}&
w_t(1-\beta)
\exp(\delta_t)
\tilde c_t^{-\rho}
V(s_t;\theta_V)^\rho
\\
&\times
\Bigg[
1
-
\beta
\widehat{\chi}_{t+1}^{f}
\left(
\frac{
V(s_{t+1};\theta_V)
}{
\mathcal C(s_t,c_t;\theta_{\mathcal C})
}
\right)^\rho
\exp\left(
\delta_{t+1}-\delta_t
\right)
\left(
\frac{
\tilde c_{t+1}
}{
\tilde c_t
}
\right)^{-\rho}
\bar r
\exp(r_{t+1})
\Bigg].
\end{aligned}
\label{eq:ez_hat_foc_integrand}
\end{equation}

Suppose \(N_z\) is even and let \(J_z=N_z/2\). For each mini-batch state-control pair \(i\), let \(\widehat F_{t+1,1}^{(i,j)}\) denote the realization of \eqref{eq:ez_hat_foc_integrand} under future draw \(j=1,\ldots,N_z\). The corresponding stochastic stationarity residual is
\begin{equation}
R_{1,t}^{\mathrm{stat},(i,j)}
=
\widehat F_{t+1,1}^{(i,j)}
-
m(s_t^{(i)};\theta_m).
\label{eq:ez_stationarity_residual}
\end{equation}

Split the \(N_z\) future draws into two conditionally independent groups and define
\begin{equation}
\overline R_{1,t}^{\mathrm{stat},(i,1)}
=
\frac{1}{J_z}
\sum_{j=1}^{J_z}
R_{1,t}^{\mathrm{stat},(i,j)},
\qquad
\overline R_{1,t}^{\mathrm{stat},(i,2)}
=
\frac{1}{J_z}
\sum_{j=J_z+1}^{N_z}
R_{1,t}^{\mathrm{stat},(i,j)}.
\label{eq:ez_split_stationarity_residuals}
\end{equation}
Using these independent sample means, the stationarity loss is the scalar-control specialization of \eqref{eq:cefol_stationarity_loss}:
\begin{equation}
\widehat{\mathcal L}_{S}(\theta_c,\theta_m)
=
\frac{1}{N}
\sum_{i=1}^{N}
\overline R_{1,t}^{\mathrm{stat},(i,1)}
\overline R_{1,t}^{\mathrm{stat},(i,2)}.
\label{eq:ez_stationarity_loss}
\end{equation}

The KKT conditions for the upper resource constraint are
\begin{equation*}
g(s_t,c_t)\geq0,
\qquad
\lambda_t\geq0,
\qquad
g(s_t,c_t)\lambda_t=0.
\end{equation*}
Because this model has a single inequality constraint and no equality constraints, the Fischer--Burmeister loss is the one-constraint specialization of \eqref{eq:cefol_fb_loss}:
\begin{equation}
\widehat{\mathcal L}_{FB}(\theta_c,\theta_m)
=
\frac{1}{N}
\sum_{i=1}^{N}
\left[
\Phi^{FB}
\left(
1-c_t^{(i)},
m(s_t^{(i)};\theta_m)
\right)
\right]^2.
\label{eq:ez_fb_loss}
\end{equation}
This loss disciplines feasibility, multiplier nonnegativity, and complementarity for the upper resource constraint. Since \(M_q=0\), there is no equality-constraint term in this model.

The general first-order/KKT loss therefore reduces to
\begin{equation}
\widehat{\mathcal L}_{\mathrm{FOC}}(\theta_c,\theta_m)
=
\lambda_S
\widehat{\mathcal L}_{S}(\theta_c,\theta_m)
+
\lambda_{FB}
\widehat{\mathcal L}_{FB}(\theta_c,\theta_m).
\label{eq:ez_foc_loss}
\end{equation}
The policy-network parameters and multiplier-network parameters are updated jointly according to the rule in Section~\ref{subsec:cefol_training_procedure}:
\begin{equation*}
(\theta_c,\theta_m)
\leftarrow
(\theta_c,\theta_m)
-
\alpha_{cm}
\nabla_{\theta_c,\theta_m}
\widehat{\mathcal L}_{\mathrm{FOC}}(\theta_c,\theta_m).
\end{equation*}
During this policy--multiplier step, the value network and certainty-equivalent network are held fixed. They are then updated separately by their own residual losses, exactly as in Section~\ref{subsec:cefol_training_procedure}.

\subsubsection{Numerical Results}

We report five Epstein--Zin consumption-saving calibrations: $(\beta,\gamma,\rho)=(0.9,2.0,0.5)$, $(0.95,2.0,0.5)$, $(0.95,5.0,0.5)$, $(0.95,20.0,0.5)$, and $(0.99,5.0,2.0)$. Consistent results are obtained across all calibrations. In the value function comparison, the three representations---the direct value network, the one-step temporal-difference expansion, and the certainty-equivalent-implied value---are closely aligned. The consumption policy is smooth and identifies regions where the constraint is binding and non-binding. The relative Bellman error remains below $10^{-4}$ at the majority of grid points, with only isolated exceptions slightly above this threshold. The FOC residual stays below $10^{-3}$ over most of the unconstrained region, again with sporadic points marginally exceeding $10^{-3}$. In all figures, cash-on-hand $w_t$ is varied over $[0,4]$, while the remaining exogenous state variables are held at their steady-state values.

\paragraph{Results: $\beta=0.9$, $\gamma=2.0$, $\rho=0.5$}

This baseline calibration combines moderate risk aversion with an EIS of 2.

\resultfig{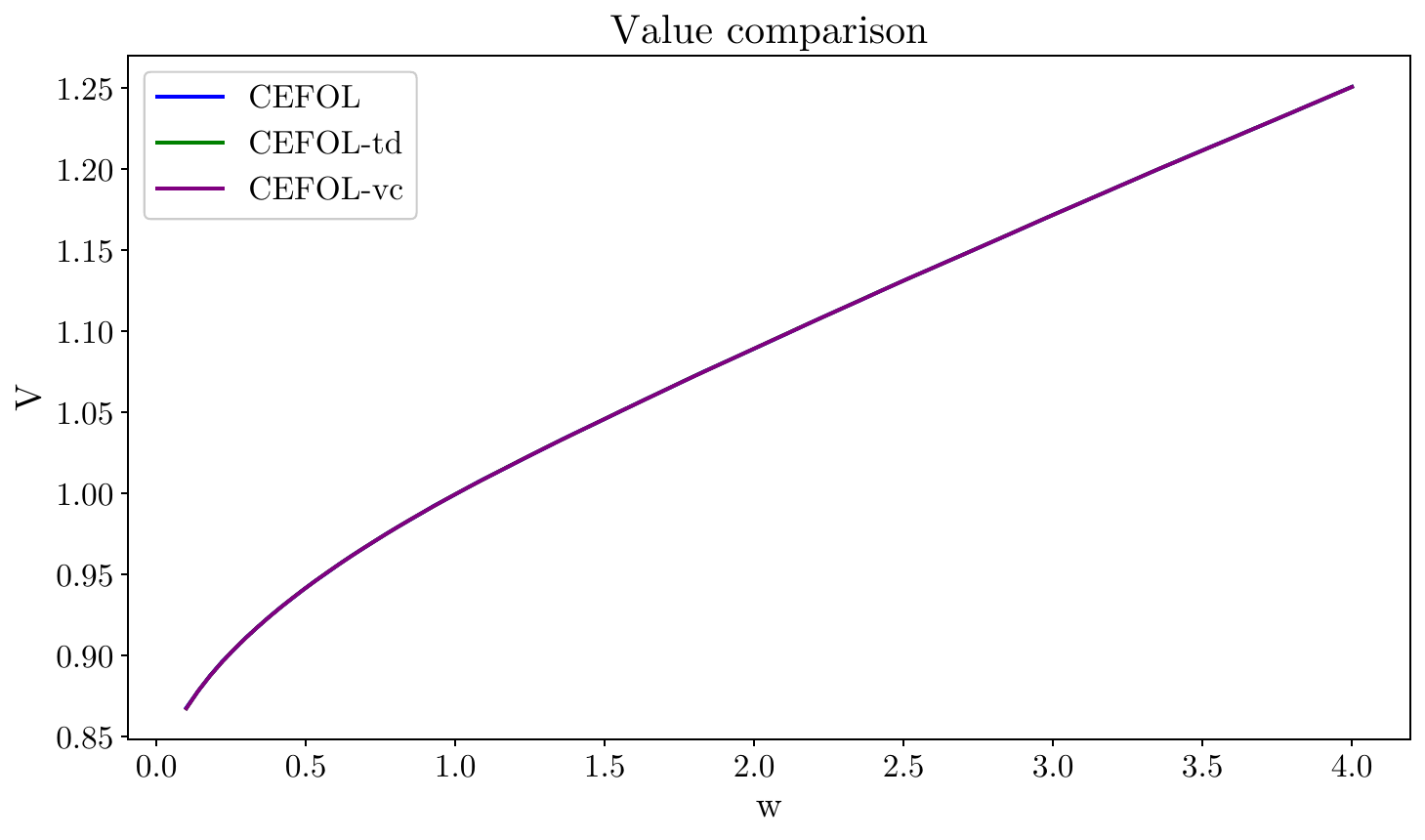}
{Value function comparison for the Epstein--Zin consumption-saving model under $(\beta,\gamma,\rho)=(0.9,2.0,0.5)$. Cash-on-hand $w_{t}$ varies over $[0,4]$, while the remaining exogenous state variables are fixed at their steady-state values.}
{fig:ez-beta09-gamma2-rho05-value}

\resultfig{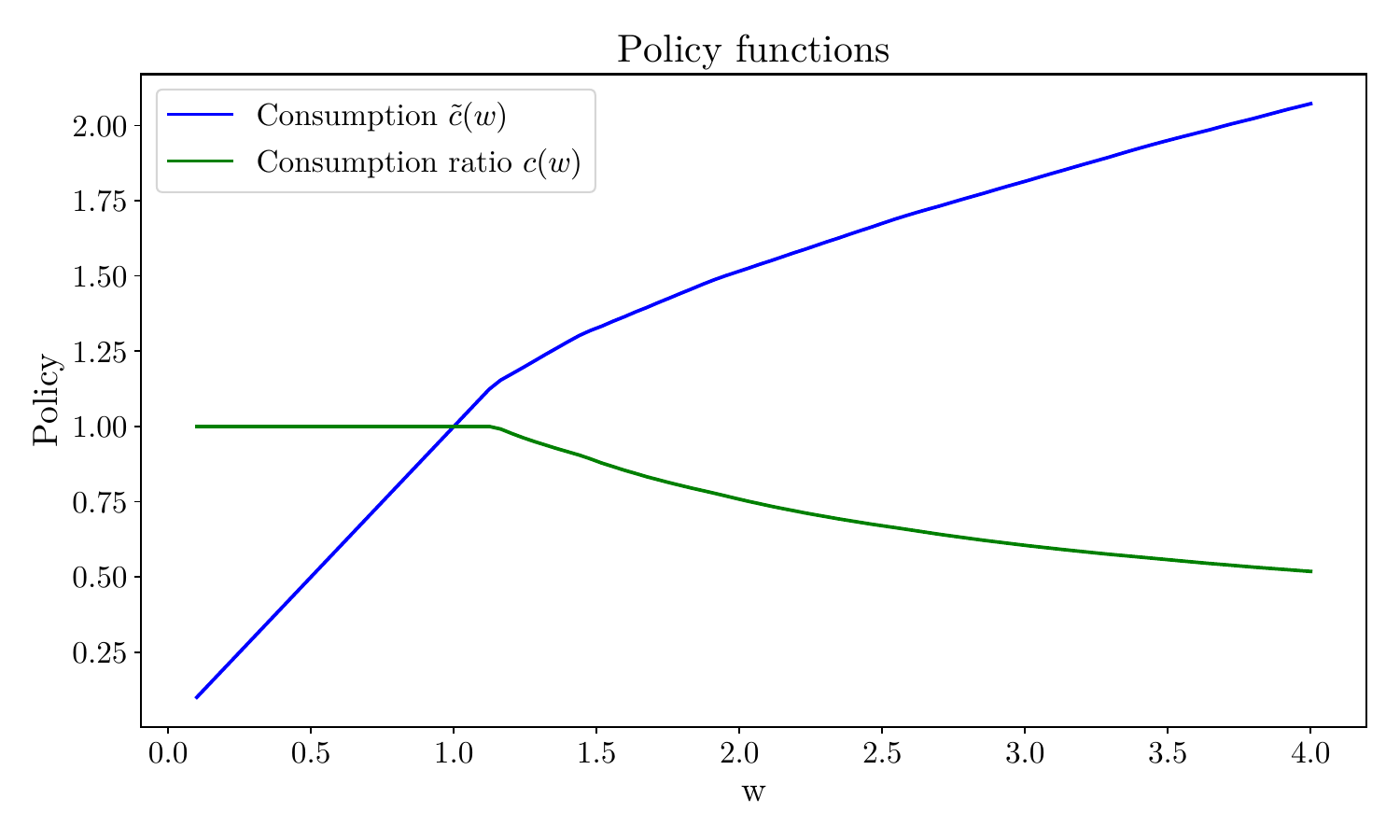}
{Policy function for the Epstein--Zin consumption-saving model under $(\beta,\gamma,\rho)=(0.9,2.0,0.5)$. The figure reports the consumption ratio $c_{t}$ as a function of cash-on-hand $w_{t}$, with the remaining exogenous state variables fixed at their steady-state values.}
{fig:ez-beta09-gamma2-rho05-policy}

\resultfig{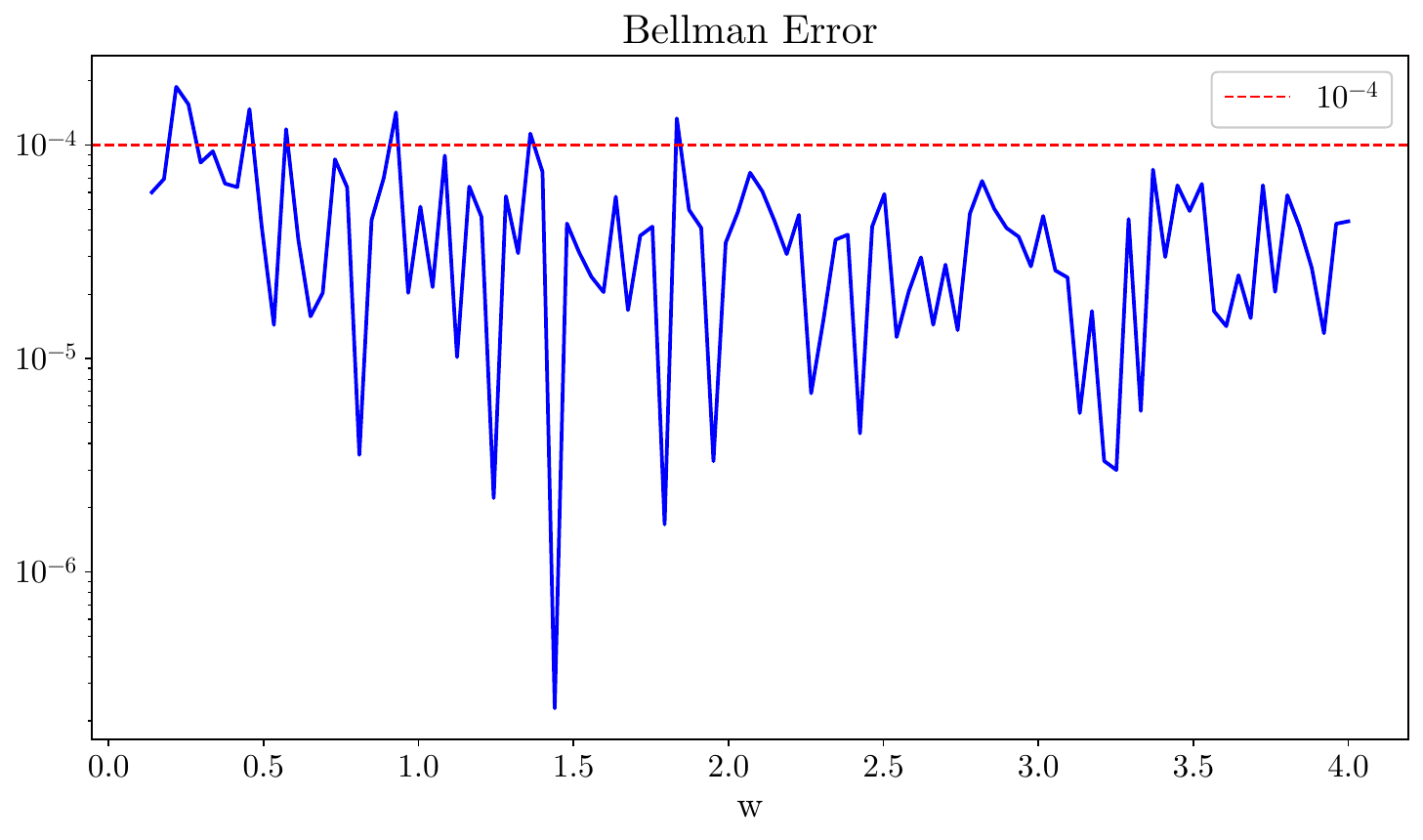}
{Bellman error for the Epstein--Zin consumption-saving model under $(\beta,\gamma,\rho)=(0.9,2.0,0.5)$. Cash-on-hand $w_{t}$ varies over $[0,4]$, while the remaining exogenous state variables are fixed at their steady-state values.}
{fig:ez-beta09-gamma2-rho05-bellman}

\resultfig{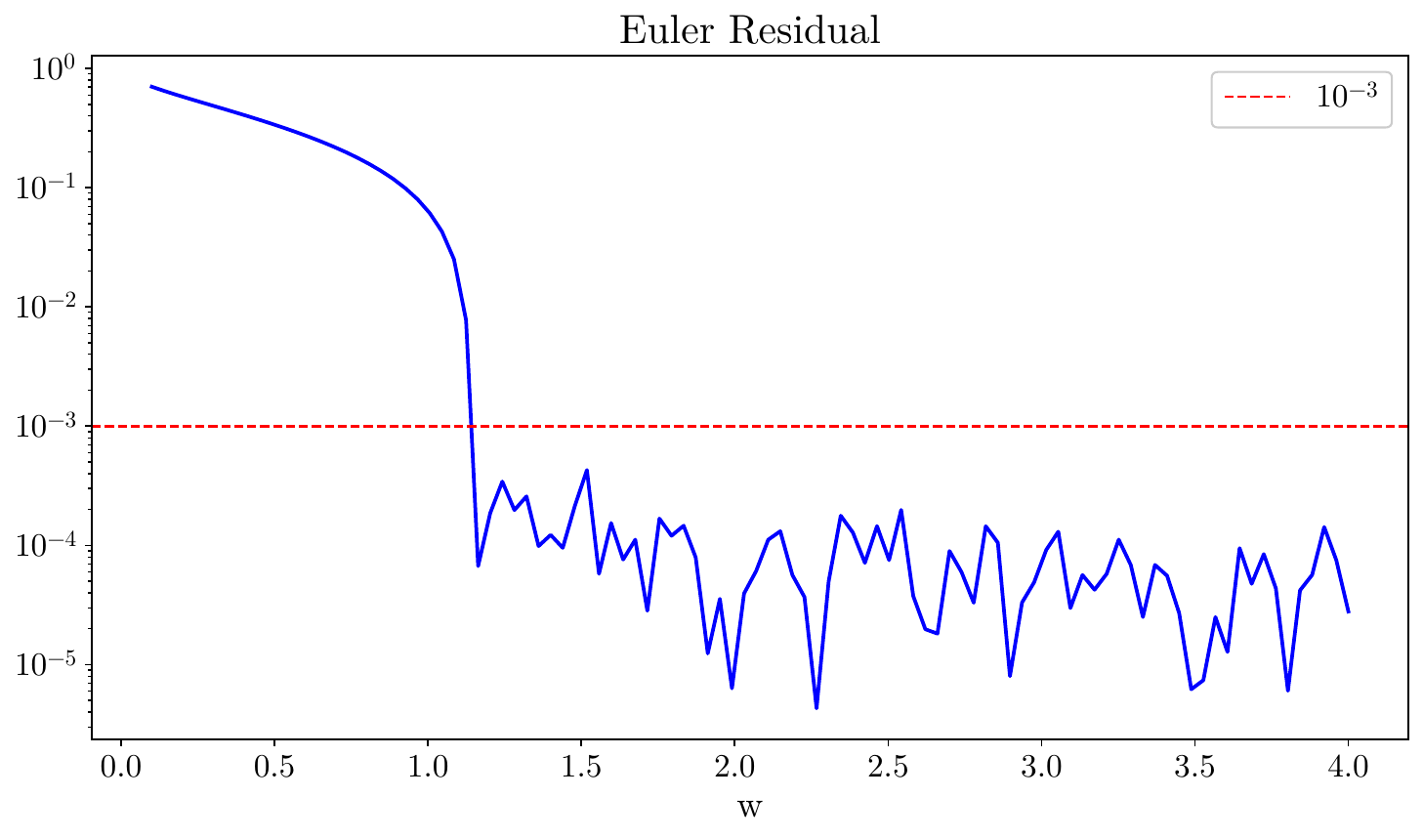}
{FOC residual for the Epstein--Zin consumption-saving model under $(\beta,\gamma,\rho)=(0.9,2.0,0.5)$. Cash-on-hand $w_{t}$ varies over $[0,4]$, while the remaining exogenous state variables are fixed at their steady-state values.}
{fig:ez-beta09-gamma2-rho05-euler}

\paragraph{Results: $\beta=0.95$, $\gamma=2.0$, $\rho=0.5$}

This calibration increases the discount factor to 0.95 while maintaining moderate risk aversion.

\resultfig{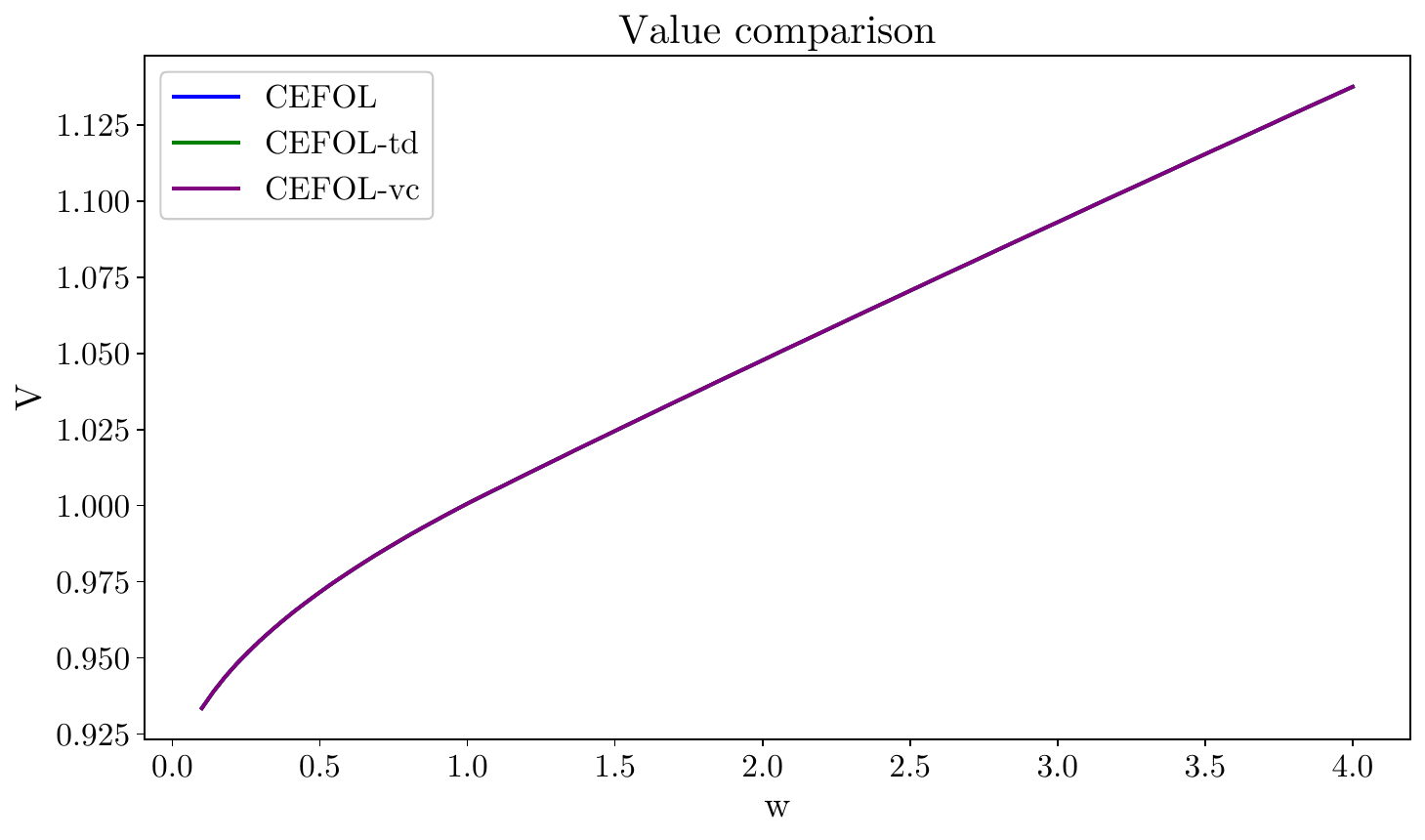}
{Value function comparison for the Epstein--Zin consumption-saving model under $(\beta,\gamma,\rho)=(0.95,2.0,0.5)$. Cash-on-hand $w_{t}$ varies over $[0,4]$, while the remaining exogenous state variables are fixed at their steady-state values.}
{fig:ez-beta095-gamma2-rho05-value}

\resultfig{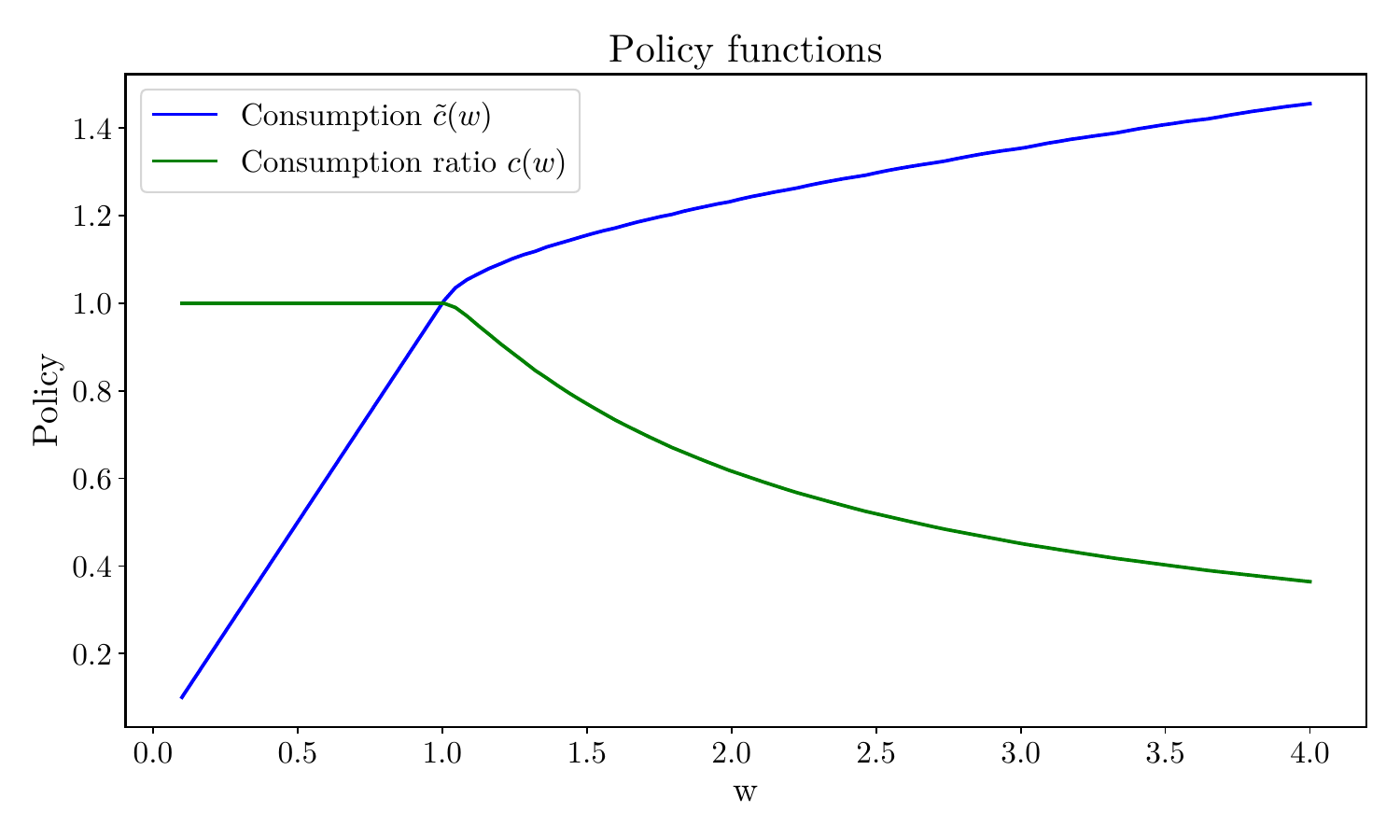}
{Policy function for the Epstein--Zin consumption-saving model under $(\beta,\gamma,\rho)=(0.95,2.0,0.5)$. The figure reports the consumption ratio $c_{t}$ as a function of cash-on-hand $w_{t}$, with the remaining exogenous state variables fixed at their steady-state values.}
{fig:ez-beta095-gamma2-rho05-policy}

\resultfig{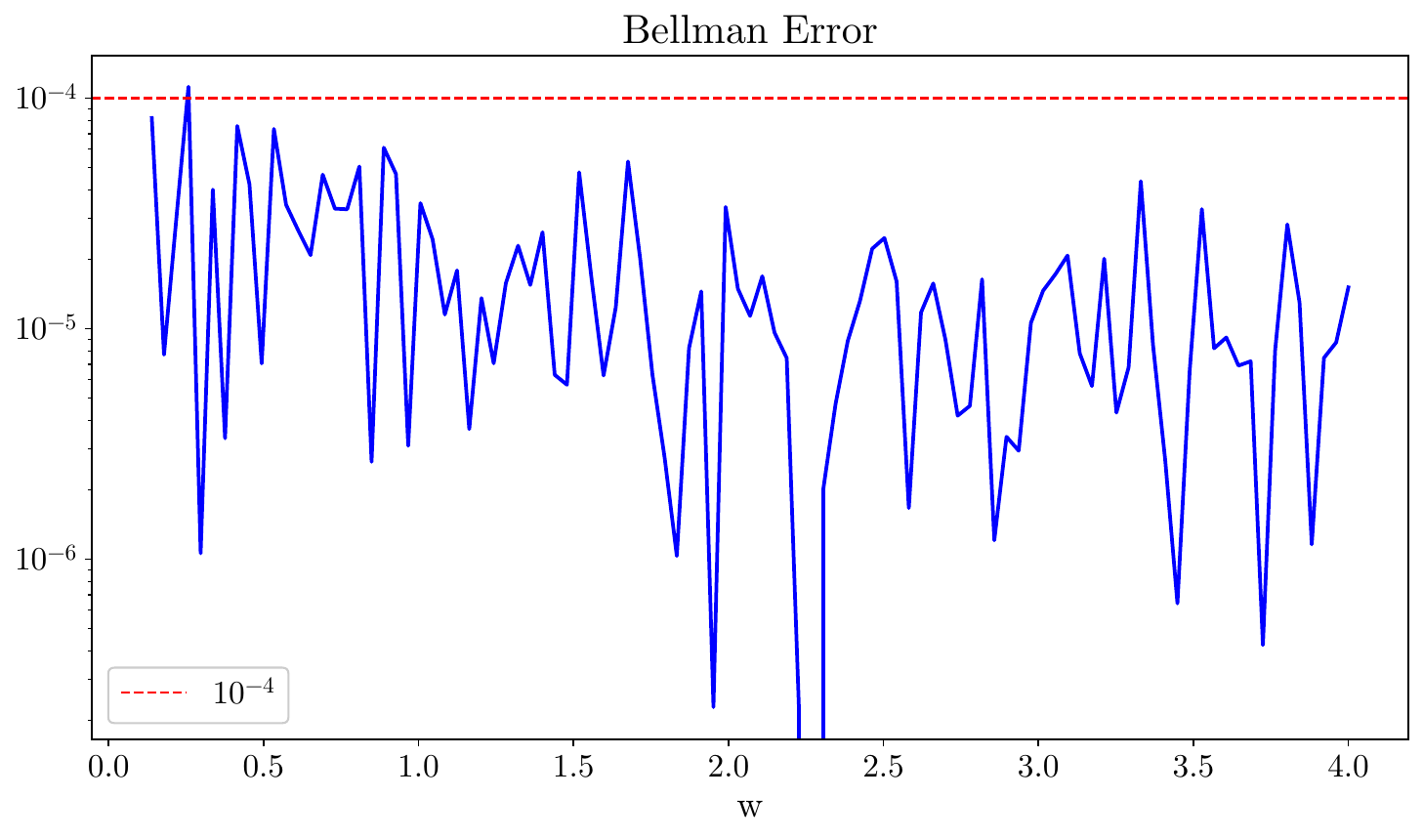}
{Bellman error for the Epstein--Zin consumption-saving model under $(\beta,\gamma,\rho)=(0.95,2.0,0.5)$. Cash-on-hand $w_{t}$ varies over $[0,4]$, while the remaining exogenous state variables are fixed at their steady-state values.}
{fig:ez-beta095-gamma2-rho05-bellman}

\resultfig{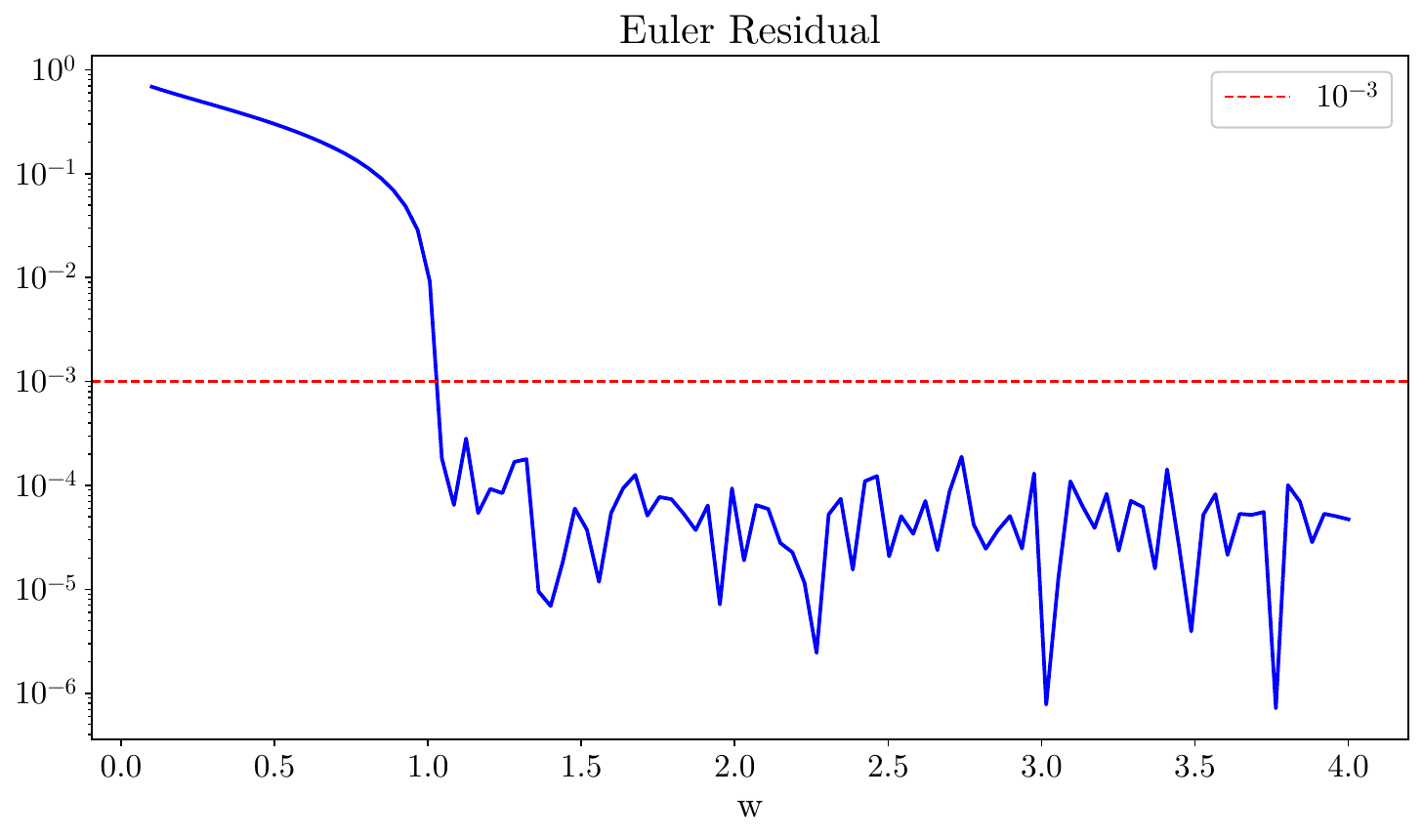}
{FOC residual for the Epstein--Zin consumption-saving model under $(\beta,\gamma,\rho)=(0.95,2.0,0.5)$. Cash-on-hand $w_{t}$ varies over $[0,4]$, while the remaining exogenous state variables are fixed at their steady-state values.}
{fig:ez-beta095-gamma2-rho05-euler}

\paragraph{Results: $\beta=0.95$, $\gamma=5.0$, $\rho=0.5$}

This calibration raises risk aversion to 5.0 under the higher discount factor.

\resultfig{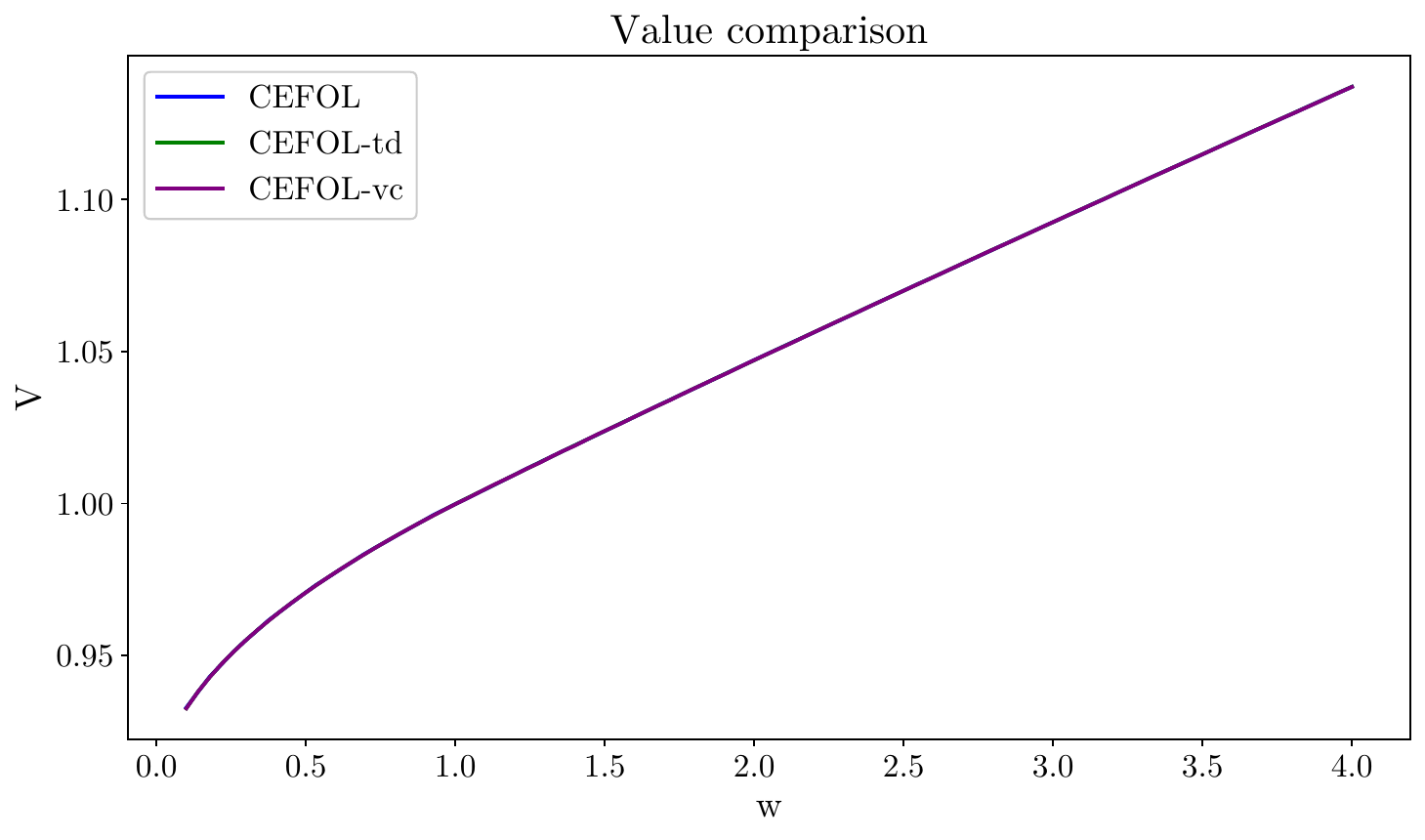}
{Value function comparison for the Epstein--Zin consumption-saving model under $(\beta,\gamma,\rho)=(0.95,5.0,0.5)$. Cash-on-hand $w_{t}$ varies over $[0,4]$, while the remaining exogenous state variables are fixed at their steady-state values.}
{fig:ez-beta095-gamma5-rho05-value}

\resultfig{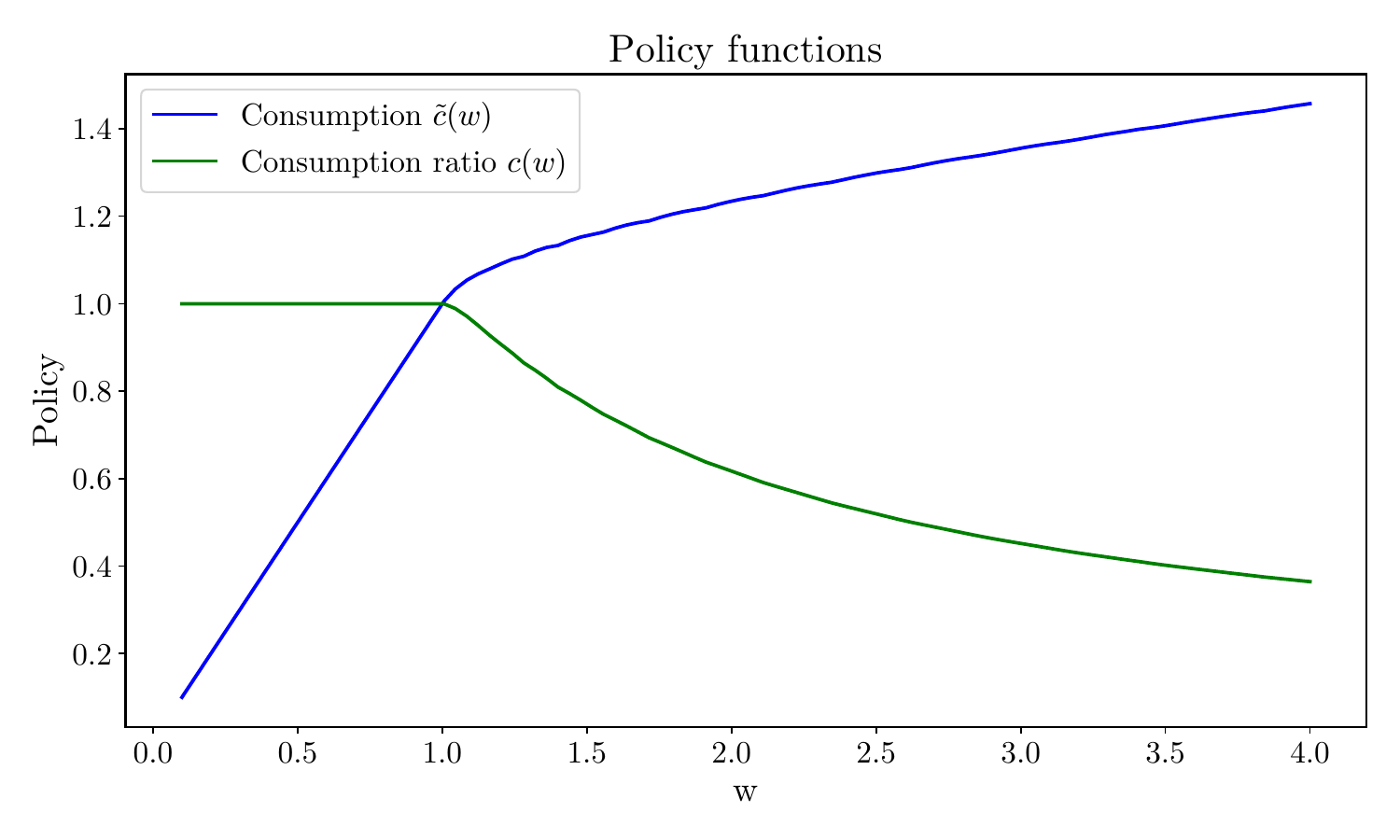}
{Policy function for the Epstein--Zin consumption-saving model under $(\beta,\gamma,\rho)=(0.95,5.0,0.5)$. The figure reports the consumption ratio $c_{t}$ as a function of cash-on-hand $w_{t}$, with the remaining exogenous state variables fixed at their steady-state values.}
{fig:ez-beta095-gamma5-rho05-policy}

\resultfig{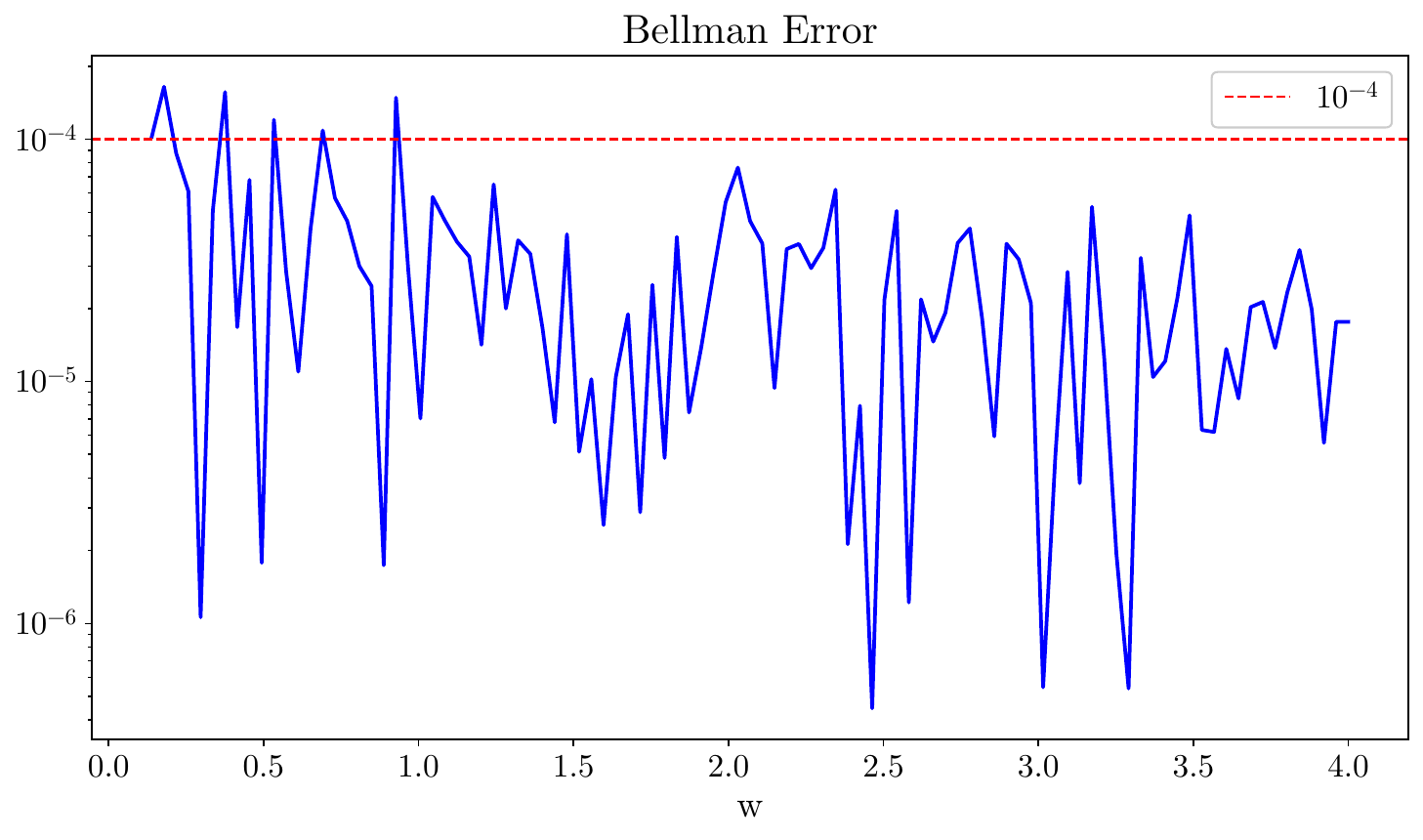}
{Bellman error for the Epstein--Zin consumption-saving model under $(\beta,\gamma,\rho)=(0.95,5.0,0.5)$. Cash-on-hand $w_{t}$ varies over $[0,4]$, while the remaining exogenous state variables are fixed at their steady-state values.}
{fig:ez-beta095-gamma5-rho05-bellman}

\resultfig{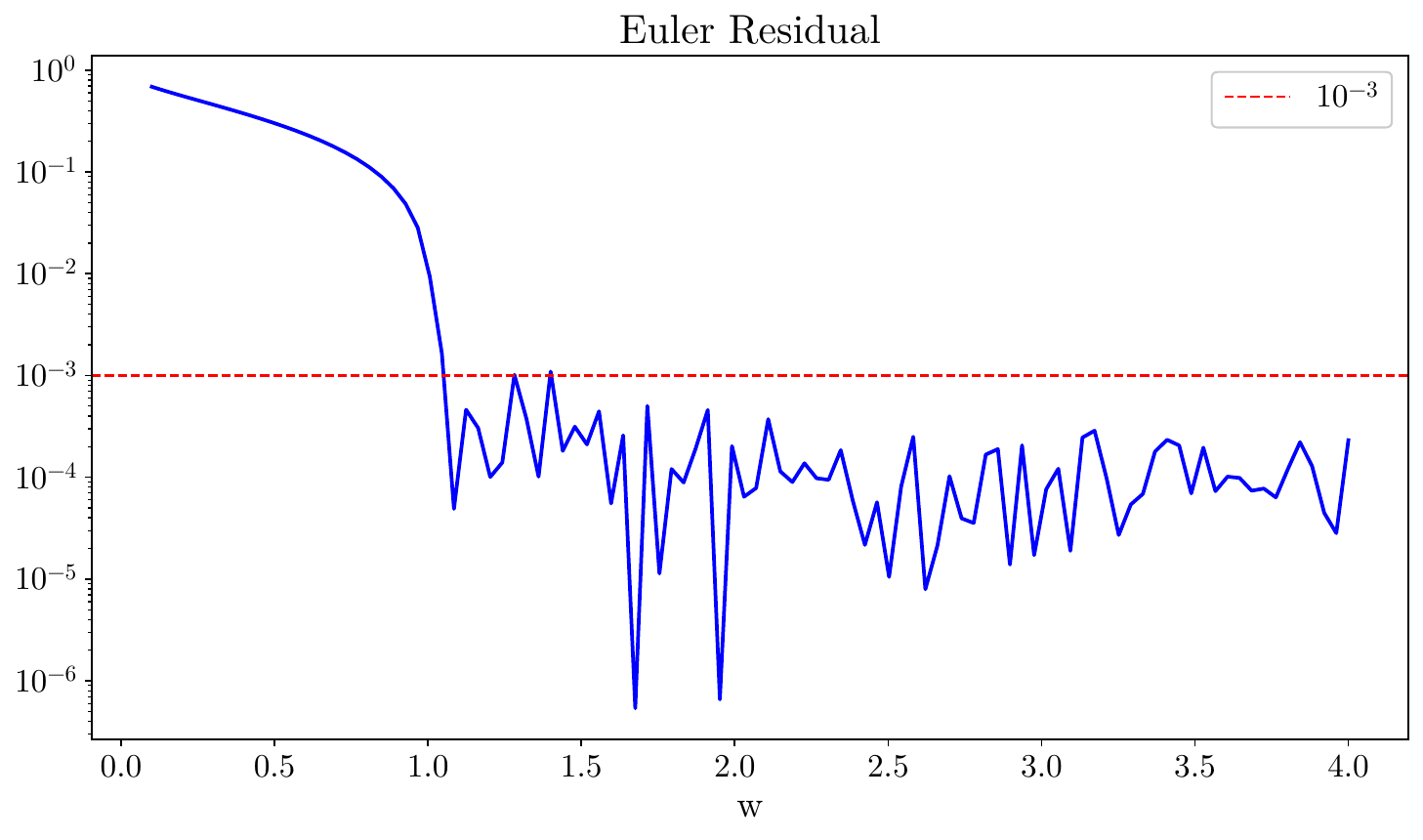}
{FOC residual for the Epstein--Zin consumption-saving model under $(\beta,\gamma,\rho)=(0.95,5.0,0.5)$. Cash-on-hand $w_{t}$ varies over $[0,4]$, while the remaining exogenous state variables are fixed at their steady-state values.}
{fig:ez-beta095-gamma5-rho05-euler}

\paragraph{Results: $\beta=0.95$, $\gamma=20.0$, $\rho=0.5$}

This calibration further raises risk aversion to 20.0, thereby strengthening the recursive nonlinearity.

\resultfig{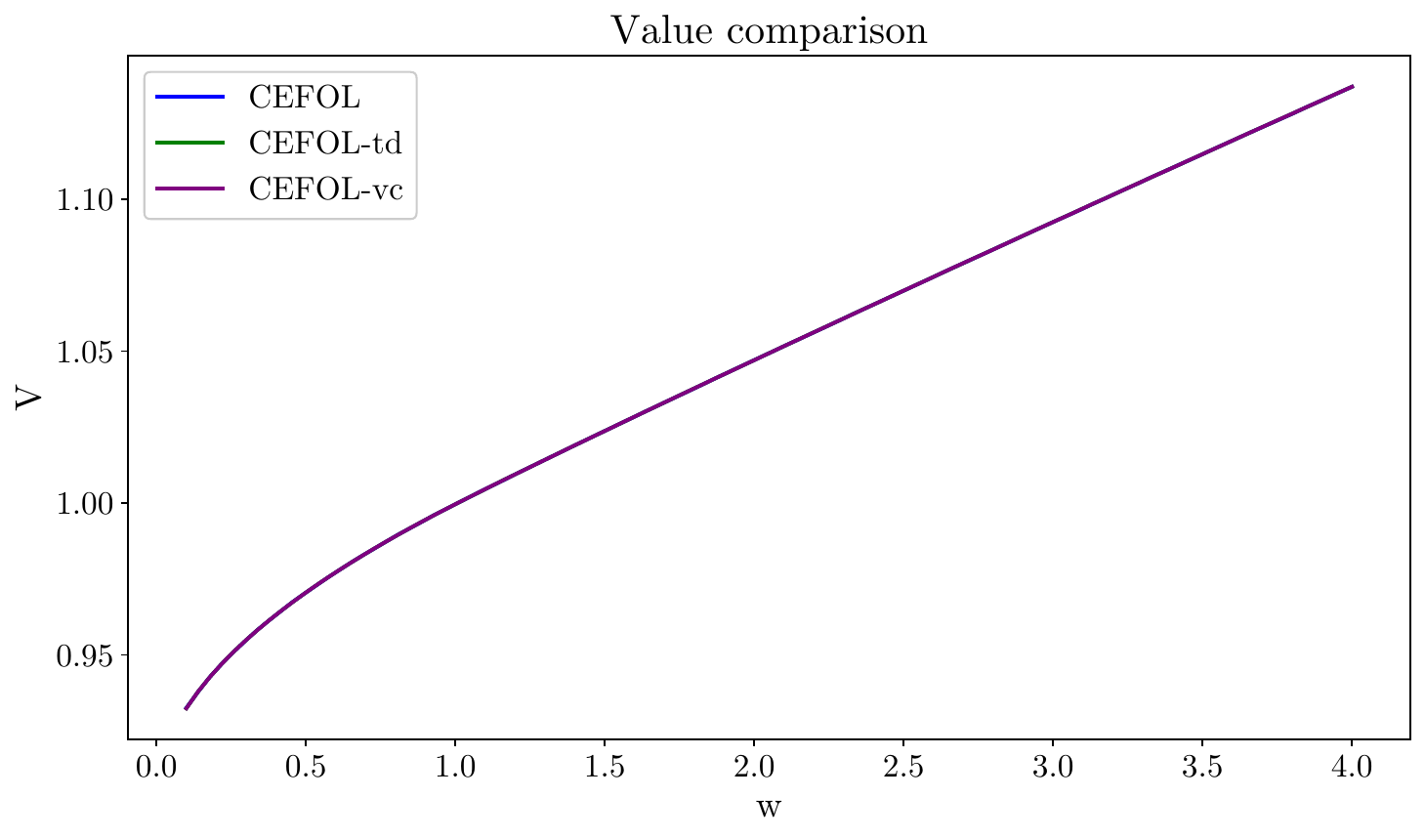}
{Value function comparison for the Epstein--Zin consumption-saving model under $(\beta,\gamma,\rho)=(0.95,20.0,0.5)$. Cash-on-hand $w_{t}$ varies over $[0,4]$, while the remaining exogenous state variables are fixed at their steady-state values.}
{fig:ez-beta095-gamma20-rho05-value}

\resultfig{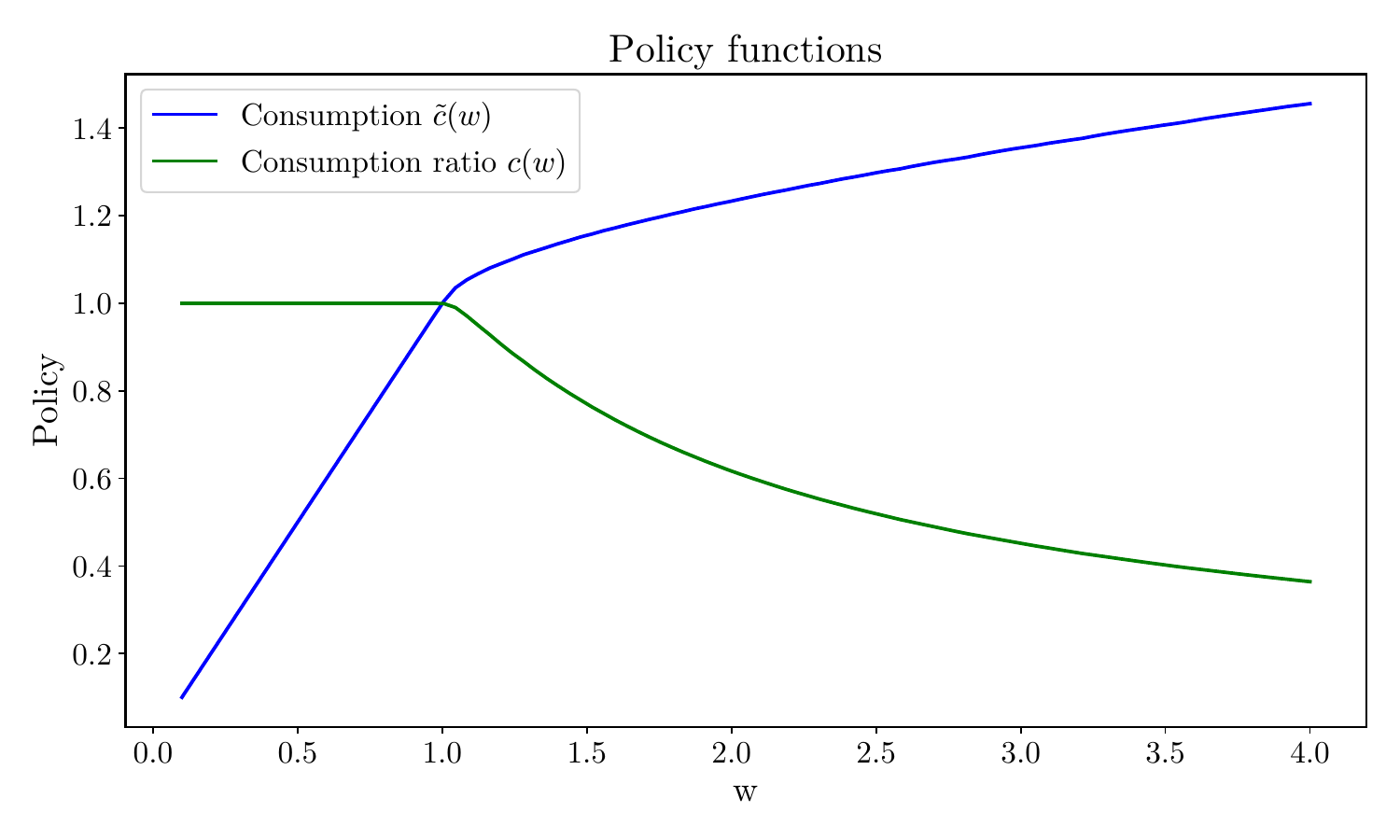}
{Policy function for the Epstein--Zin consumption-saving model under $(\beta,\gamma,\rho)=(0.95,20.0,0.5)$. The figure reports the consumption ratio $c_{t}$ as a function of cash-on-hand $w_{t}$, with the remaining exogenous state variables fixed at their steady-state values.}
{fig:ez-beta095-gamma20-rho05-policy}

\resultfig{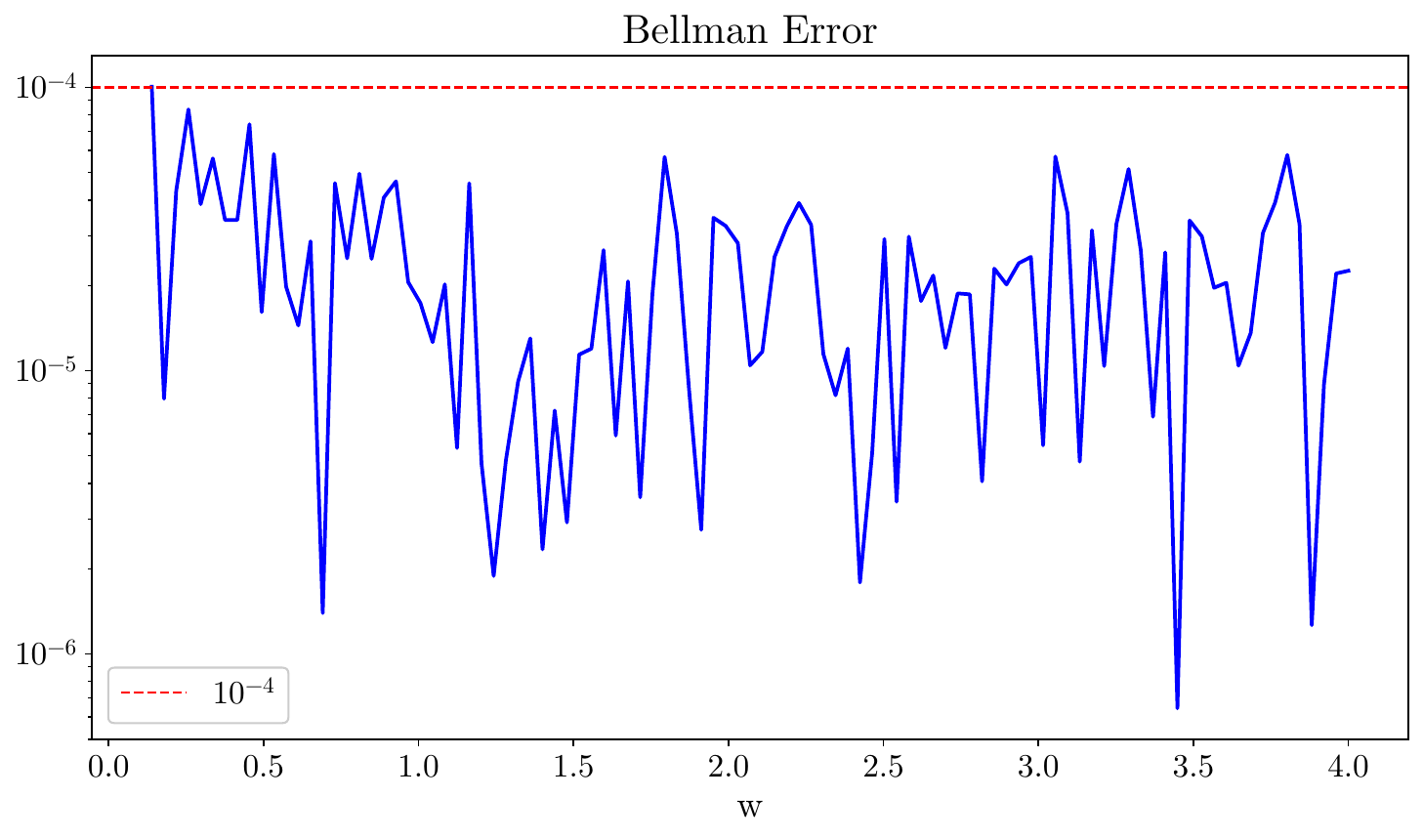}
{Bellman error for the Epstein--Zin consumption-saving model under $(\beta,\gamma,\rho)=(0.95,20.0,0.5)$. Cash-on-hand $w_{t}$ varies over $[0,4]$, while the remaining exogenous state variables are fixed at their steady-state values.}
{fig:ez-beta095-gamma20-rho05-bellman}

\resultfig{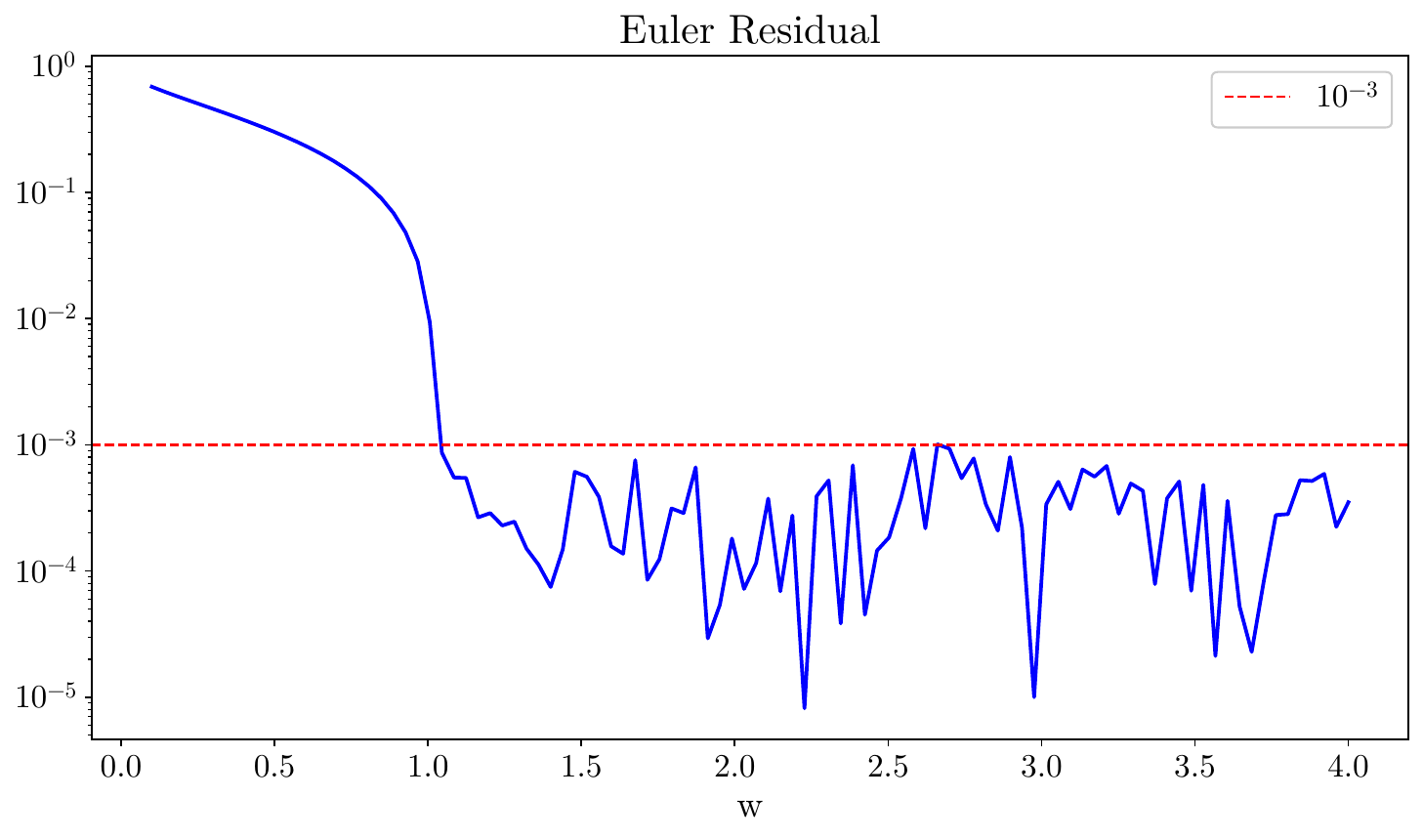}
{FOC residual for the Epstein--Zin consumption-saving model under $(\beta,\gamma,\rho)=(0.95,20.0,0.5)$. Cash-on-hand $w_{t}$ varies over $[0,4]$, while the remaining exogenous state variables are fixed at their steady-state values.}
{fig:ez-beta095-gamma20-rho05-euler}

\paragraph{Results: $\beta=0.99$, $\gamma=5.0$, $\rho=2.0$}

This calibration combines a high discount factor with lower intertemporal substitutability ($\psi = \frac{1}{\rho}=0.5$).

\resultfig{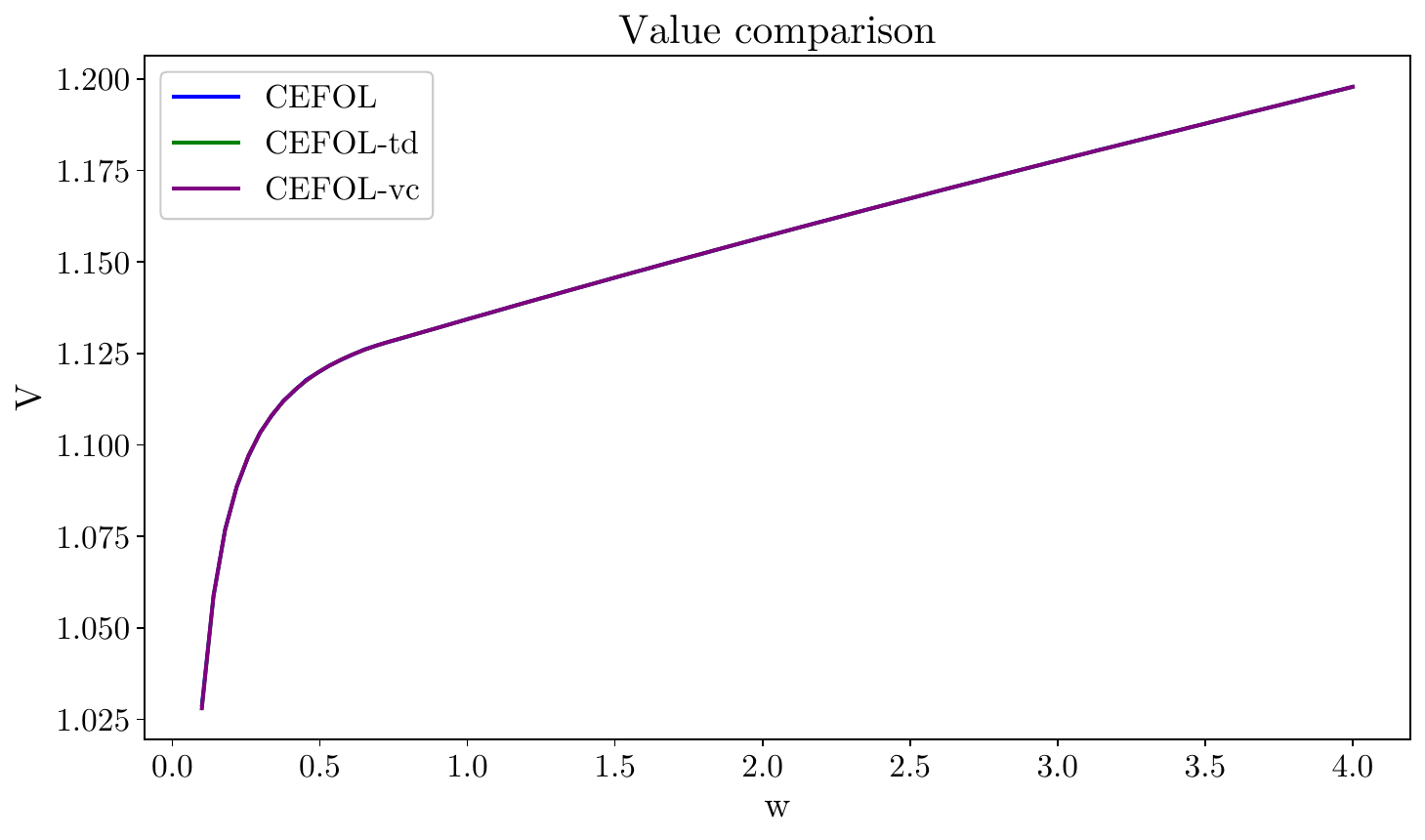}
{Value function comparison for the Epstein--Zin consumption-saving model under $(\beta,\gamma,\rho)=(0.99,5.0,2.0)$. Cash-on-hand $w_{t}$ varies over $[0,4]$, while the remaining exogenous state variables are fixed at their steady-state values.}
{fig:ez-beta099-gamma5-rho20-value}

\resultfig{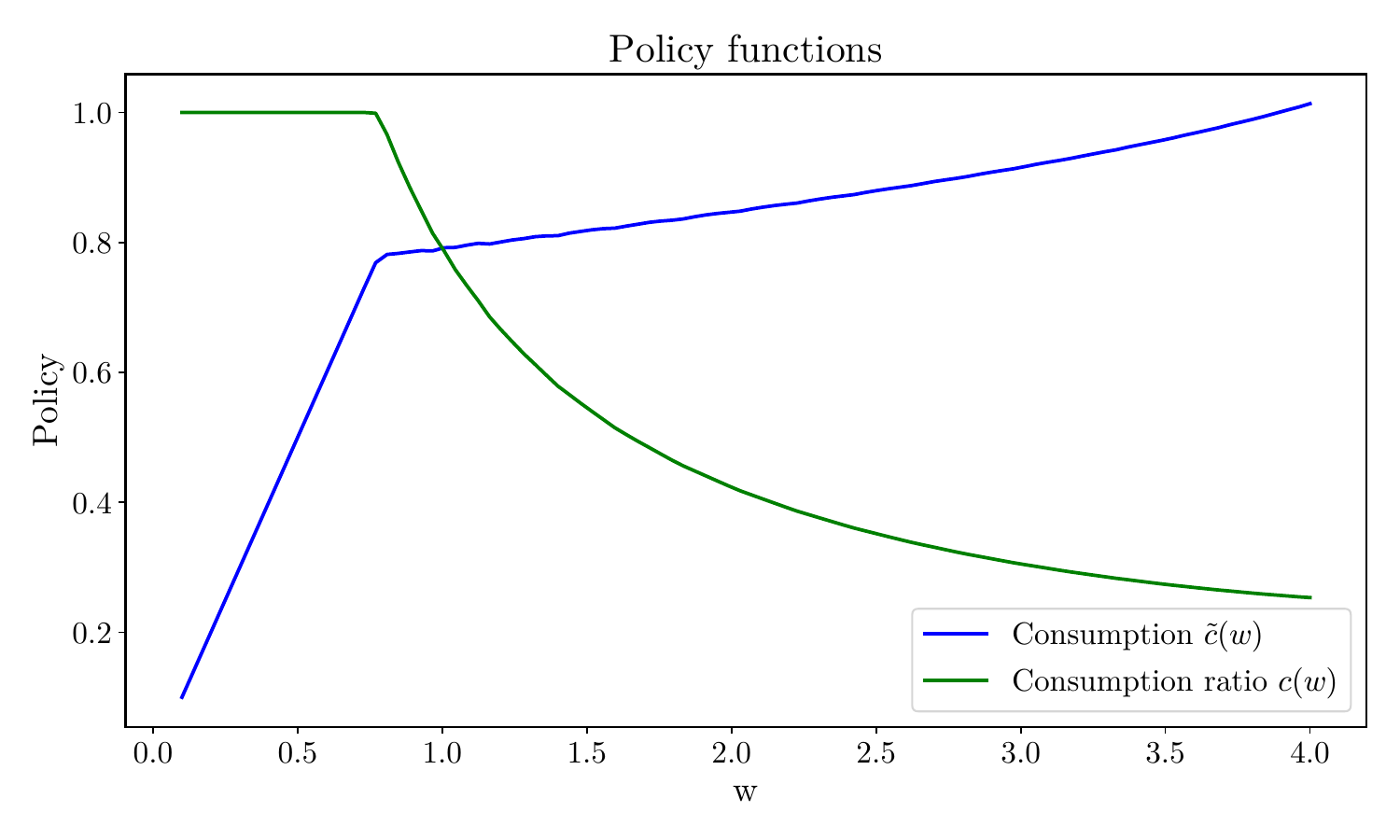}
{Policy function for the Epstein--Zin consumption-saving model under $(\beta,\gamma,\rho)=(0.99,5.0,2.0)$. The figure reports the consumption ratio $c_{t}$ as a function of cash-on-hand $w_{t}$, with the remaining exogenous state variables fixed at their steady-state values.}
{fig:ez-beta099-gamma5-rho20-policy}

\resultfig{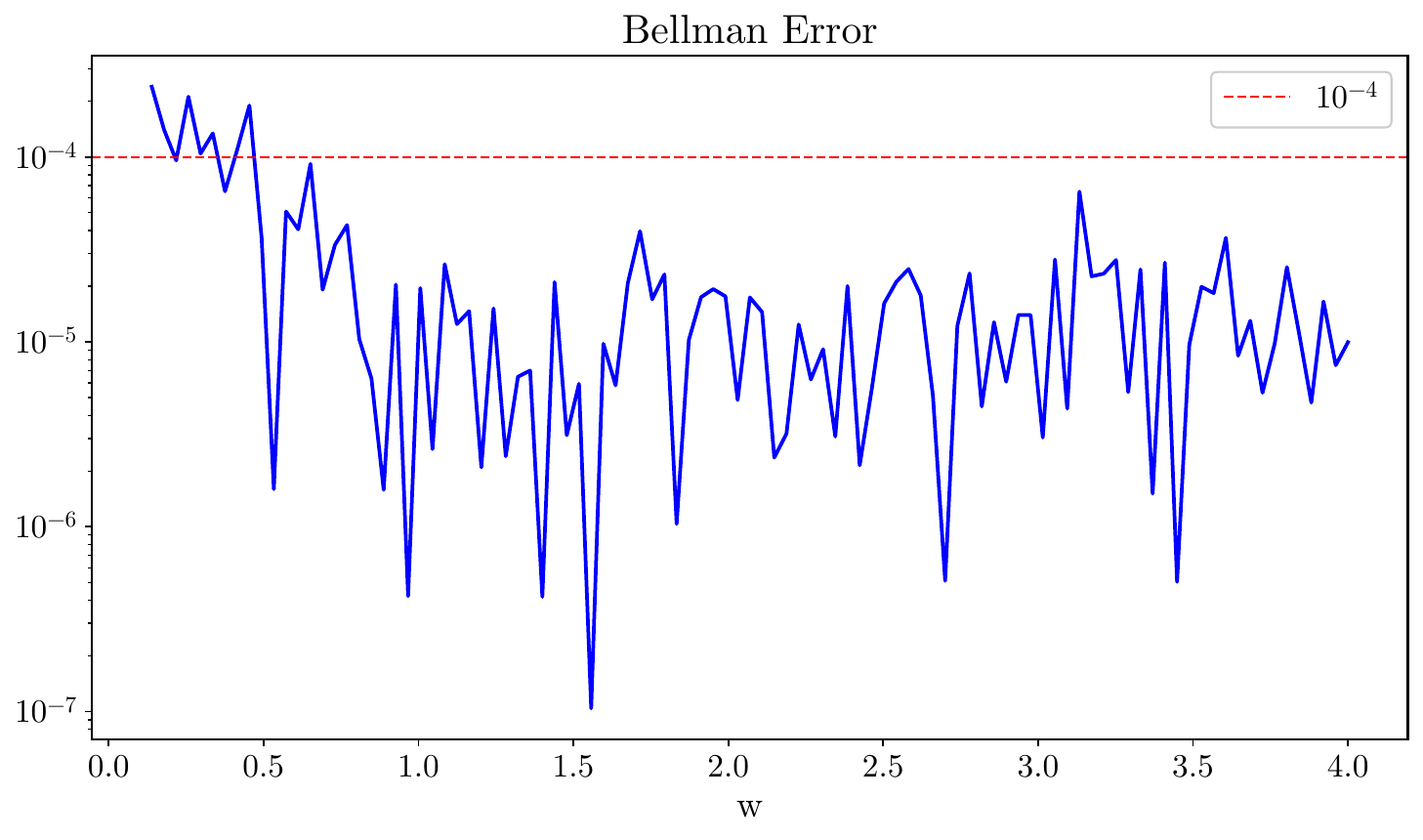}
{Bellman error for the Epstein--Zin consumption-saving model under $(\beta,\gamma,\rho)=(0.99,5.0,2.0)$. Cash-on-hand $w_{t}$ varies over $[0,4]$, while the remaining exogenous state variables are fixed at their steady-state values.}
{fig:ez-beta099-gamma5-rho20-bellman}

\resultfig{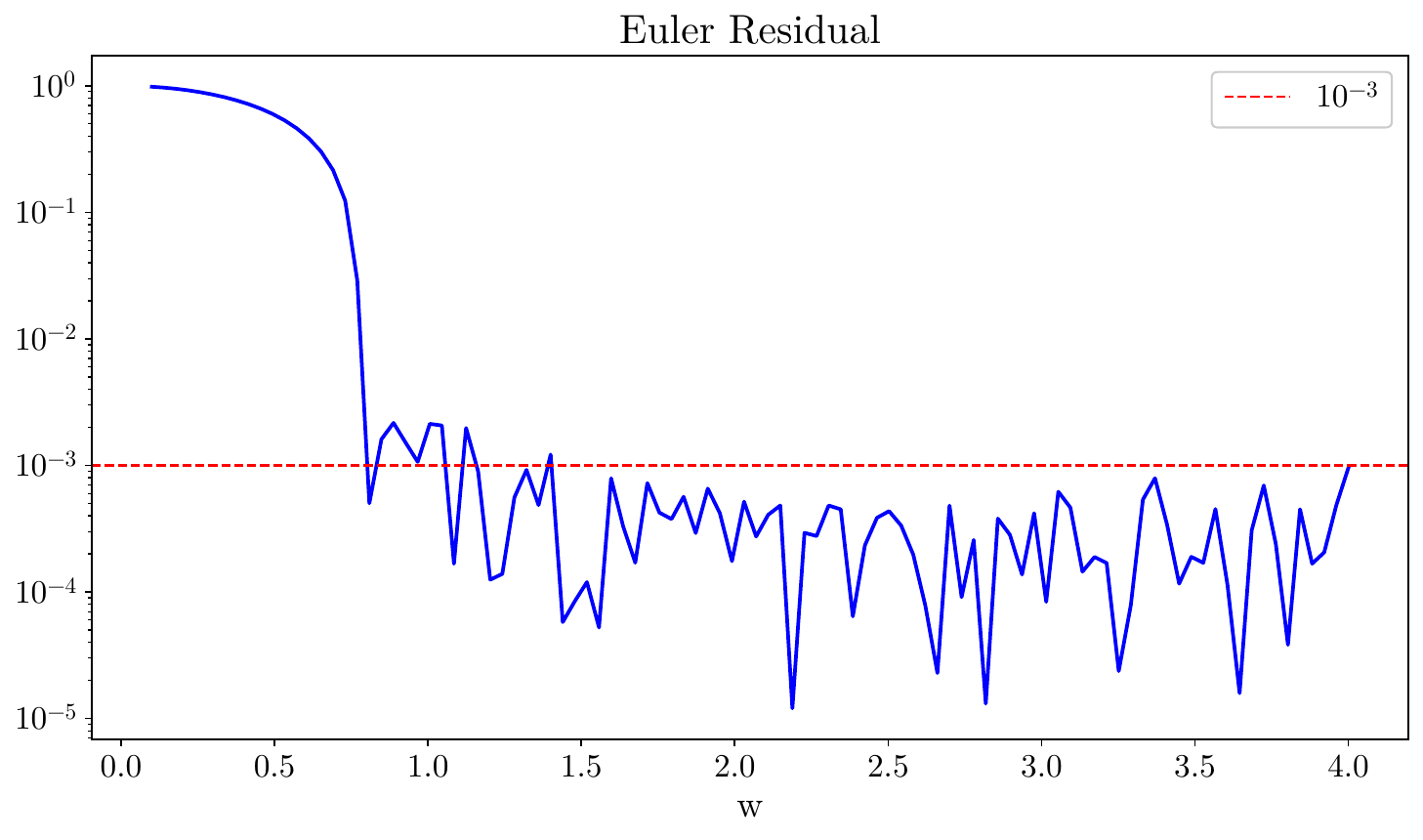}
{FOC residual for the Epstein--Zin consumption-saving model under $(\beta,\gamma,\rho)=(0.99,5.0,2.0)$. Cash-on-hand $w_{t}$ varies over $[0,4]$, while the remaining exogenous state variables are fixed at their steady-state values.}
{fig:ez-beta099-gamma5-rho20-euler}

\subsection{Small-Noise Robust-Control}
\label{subsec:small_noise_robust_control}

The second example is a small-noise robust-control model of consumption and capital accumulation in a stochastic production economy. It considers a representative agent facing a scalar consumption-ratio decision and a single productivity shock. The small-noise feature enters through the productivity process, whose innovation volatility is scaled by \(\sqrt{\epsilon}\). 

\subsubsection{Problem Setup}
\label{subsubsec:sn_problem_setup}

At the beginning of period \(t\), predetermined capital \(k_{t-1}\) and the current productivity shock \(q_t\) determine the total resources available for allocation:
\begin{equation*}
w_t
=
e^{P q_t}k_{t-1}^{\alpha}
+
(1-\delta)k_{t-1}.
\end{equation*}
where $P>0$ is a productivity scale parameter, $\alpha\in(0,1)$ is the capital share, and $\delta\in[0,1]$ is the depreciation rate. Thus, current resources consist of output produced using $k_{t-1}$ and the undepreciated part of existing capital.

The household chooses the consumption ratio \(c_t\). Given \(w_t\), this choice determines actual consumption and next-period capital:
\begin{equation*}
\tilde c_t = c_t w_t,
\qquad
k_t = (1-c_t)w_t,
\qquad
\tilde c_t+k_t = w_t.
\end{equation*}
Hence, the household allocates current resources between consumption and capital carried into the next period.

Between periods \(t\) and \(t+1\), the capital choice \(k_t\) becomes next period's predetermined capital, while productivity evolves according to
\begin{equation*}
q_{t+1}
=
\Omega_0+\Omega_q q_t+\sqrt{\epsilon}\Omega_v\varepsilon_{t+1},
\qquad
\varepsilon_{t+1}\sim\mathcal{N}(0,1).
\end{equation*}
where $\Omega_0$ and $\Omega_q\in(0,1)$ are the intercept and persistence of the productivity process, and $\epsilon>0$ and $\Omega_v>0$ govern the innovation variance. After the realization of $q_{t+1}$, next-period resources are
\begin{equation*}
w_{t+1}
=
e^{P q_{t+1}}k_t^{\alpha}
+
(1-\delta)k_t.
\end{equation*}

Accordingly, the state vector is
\begin{equation*}
s_t = (k_{t-1},q_t).
\end{equation*}
Let \(z_{t+1}=\varepsilon_{t+1}\) denote the next-period innovation. The state transition is
\begin{equation}
s_{t+1}
=
\psi(s_t,c_t,z_{t+1})
=
(k_t,q_{t+1}),
\qquad
k_t=(1-c_t)w_t.
\label{eq:sn_transition}
\end{equation}

Given this timing, the household maximizes current utility plus a risk-sensitive certainty-equivalent value. The small-noise robust-control problem is
\begin{equation}
V(s_t)
=
\max_{c_t}
\left\{
u(\tilde c_t)
-
\frac{1}{\sigma}
\log
\mathbb{E}_t
\left[
\exp\{-\sigma\beta V(s_{t+1})\}
\right]
\right\},
\label{eq:sn_bellman}
\end{equation}
subject to
\begin{equation}
0<\tilde c_t\le w_t,
\qquad
\tilde c_t+k_t=w_t,
\qquad
\tilde c_t=c_tw_t,
\label{eq:sn_primitive_constraint}
\end{equation}
together with the transition law in \eqref{eq:sn_transition} and the productivity process specified above.

As in Section~\ref{subsubsec:rs_problem_setup}, the lower boundary \(\tilde c_t=0\) is not treated as a separate KKT constraint. Once the control is parameterized directly by \(c_t\), the relations \(\tilde c_t=c_tw_t\) and \(k_t=(1-c_t)w_t\) are definitional rather than independent equality constraints. For consistency with the general KKT notation, the upper resource boundary is represented by the normalized slack
\begin{equation*}
g(s_t,c_t) = 1-c_t \geq0.
\end{equation*}
Therefore, \(M_q=0\) and \(M_g=1\).

The logarithmic utility case is obtained as the limit \(\gamma\to1\). The risk-sensitive Bellman equation \eqref{eq:sn_bellman} corresponds to the general certainty-equivalent formulation with
\begin{equation*}
f(x) = \exp(-\sigma\beta x),
\qquad
f^{-1}(y) = -\frac{1}{\sigma\beta}\log y.
\end{equation*}

We use the four-network architecture introduced in Section~\ref{subsubsec:cefol_four_network}.

The general nonlinear calibration is
\begin{equation*}
\beta=0.9,
\qquad
\gamma=0.9,
\qquad
\delta=0.09,
\qquad
\alpha=0.3,
\qquad
P=1,
\end{equation*}
\begin{equation*}
\Omega_0=-0.19996246,
\qquad
\Omega_q=0.5,
\qquad
\Omega_v=0.02,
\qquad
\epsilon=1.
\end{equation*}
There is no analytical solution in the general nonlinear case, so VFI is used as the numerical benchmark.

\subsubsection{First-Order Condition and KKT Losses}
\label{subsubsec:sn_foc_aio_residuals}

To align this model with the general CEFOL formulation in Section~\ref{subsec:cefol_foc_kkt_losses}, we specialize the FOC/KKT system to the scalar-control, single-inequality-constraint case:
\begin{equation*}
n_c=1,
\qquad
M_g=1,
\qquad
M_q=0,
\qquad
g(s_t,c_t)=1-c_t,
\qquad
\frac{\partial g(s_t,c_t)}{\partial c_t}=-1.
\end{equation*}

The gross marginal return on capital from date \(t\) to date \(t+1\) is
\begin{equation}
R_{t+1}
=
\alpha e^{Pq_{t+1}}k_t^{\alpha-1}
+
1-\delta.
\label{eq:sn_return}
\end{equation}

For the risk-sensitive transformation \(f(x)=\exp(-\sigma\beta x)\), the general distortion term is
\begin{equation}
\chi_{t+1}^{f}
=
\frac{
\exp\{-\sigma\beta V(s_{t+1})\}
}{
\mathbb{E}_t
\left[
\exp\{-\sigma\beta V(s_{t+1})\}
\right]
}.
\label{eq:sn_distortion}
\end{equation}
By construction,
\begin{equation*}
\mathbb{E}_t
\left[
\chi_{t+1}^{f}
\right]
=
1.
\end{equation*}

Using the certainty-equivalent network, the denominator in \eqref{eq:sn_distortion} is approximated by
\begin{equation*}
\exp
\left\{
-\sigma\beta
\mathcal{C}(s_t,c_t;\theta_{\mathcal C})
\right\}.
\end{equation*}
Accordingly, the neural network representation is
\begin{equation}
\widehat{\chi}_{t+1}^{f}
=
\frac{
\exp\{-\sigma\beta V(s_{t+1};\theta_V)\}
}{
\exp\{-\sigma\beta
\mathcal{C}(s_t,c_t;\theta_{\mathcal C})\}
}
=
\exp
\left\{
-\sigma\beta
\left[
V(s_{t+1};\theta_V)
-
\mathcal{C}(s_t,c_t;\theta_{\mathcal C})
\right]
\right\}.
\label{eq:sn_hat_distortion}
\end{equation}

The constrained small-noise Bellman problem implies the population stationarity condition
\begin{equation}
\mathbb{E}_t
\left[
F_{t+1,1}
\right]
+
\lambda_t
\frac{\partial g(s_t,c_t)}{\partial c_t}
=
\mathbb{E}_t
\left[
F_{t+1,1}
\right]
-
\lambda_t
=
0,
\label{eq:sn_stationarity_condition}
\end{equation}
where the stochastic stationarity integrand is
\begin{equation}
F_{t+1,1}
=
w_t
\left[
u'(\tilde c_t)
-
\beta
\chi_{t+1}^{f}
u'(\tilde c_{t+1})
R_{t+1}
\right].
\label{eq:sn_F_integrand}
\end{equation}
Here,
\begin{equation*}
\tilde c_t=c_t w_t,
\qquad
k_t=(1-c_t)w_t,
\qquad
\tilde c_{t+1}
=
c(s_{t+1};\theta_c)w_{t+1}.
\end{equation*}

In implementation, the sampled integrand is constructed using the network-based distortion:
\begin{equation}
\widehat F_{t+1,1}
=
w_t
\left[
u'(\tilde c_t)
-
\beta
\widehat{\chi}_{t+1}^{f}
u'(\tilde c_{t+1})
R_{t+1}
\right].
\label{eq:sn_hat_F_integrand}
\end{equation}

Suppose \(N_z\) is even and let \(J_z=N_z/2\). For each mini-batch state-control pair \(i\), let \(\widehat F_{t+1,1}^{(i,j)}\) denote the realization of \eqref{eq:sn_hat_F_integrand} under future draw \(j=1,\ldots,N_z\). The corresponding stochastic stationarity residual is
\begin{equation}
R_{1,t}^{\mathrm{stat},(i,j)}
=
\widehat F_{t+1,1}^{(i,j)}
-
m(s_t^{(i)};\theta_m).
\label{eq:sn_stationarity_residual}
\end{equation}

Split the \(N_z\) future draws into two conditionally independent groups and define
\begin{equation}
\overline R_{1,t}^{\mathrm{stat},(i,1)}
=
\frac{1}{J_z}
\sum_{j=1}^{J_z}
R_{1,t}^{\mathrm{stat},(i,j)},
\qquad
\overline R_{1,t}^{\mathrm{stat},(i,2)}
=
\frac{1}{J_z}
\sum_{j=J_z+1}^{N_z}
R_{1,t}^{\mathrm{stat},(i,j)}.
\label{eq:sn_split_stationarity_residuals}
\end{equation}
Using these independent sample means, the stationarity loss is the scalar-control specialization of \eqref{eq:cefol_stationarity_loss}:
\begin{equation}
\widehat{\mathcal{L}}_{S}(\theta_c,\theta_m)
=
\frac{1}{N}
\sum_{i=1}^{N}
\overline R_{1,t}^{\mathrm{stat},(i,1)}
\overline R_{1,t}^{\mathrm{stat},(i,2)}.
\label{eq:sn_stationarity_loss}
\end{equation}

The KKT conditions for the upper resource constraint are
\begin{equation*}
g(s_t,c_t)\geq0,
\qquad
\lambda_t\geq0,
\qquad
g(s_t,c_t)\lambda_t=0.
\end{equation*}
Because this model has a single inequality constraint and no equality constraints, the Fischer--Burmeister loss is the one-constraint specialization of \eqref{eq:cefol_fb_loss}:
\begin{equation}
\widehat{\mathcal{L}}_{FB}(\theta_c,\theta_m)
=
\frac{1}{N}
\sum_{i=1}^{N}
v_{FB}
\left[
\Phi^{FB}
\left(
1-c_t^{(i)},
m(s_t^{(i)};\theta_m)
\right)
\right]^2.
\label{eq:sn_fb_loss}
\end{equation}
This loss disciplines feasibility, multiplier nonnegativity, and complementarity for the upper resource constraint. Since \(M_q=0\), there is no equality-constraint term in this model.

The general first-order/KKT loss therefore reduces to
\begin{equation}
\widehat{\mathcal{L}}_{FOC}(\theta_c,\theta_m)
=
\lambda_S
\widehat{\mathcal{L}}_{S}(\theta_c,\theta_m)
+
\lambda_{FB}
\widehat{\mathcal{L}}_{FB}(\theta_c,\theta_m).
\label{eq:sn_foc_loss}
\end{equation}
The policy-network parameters and multiplier-network parameters are updated jointly according to the rule in Section~\ref{subsec:cefol_training_procedure}:
\begin{equation*}
(\theta_c,\theta_m)
\leftarrow
(\theta_c,\theta_m)
-
\alpha_{cm}
\nabla_{\theta_c,\theta_m}
\widehat{\mathcal{L}}_{FOC}(\theta_c,\theta_m).
\end{equation*}
During this policy--multiplier step, the value and certainty-equivalent networks are held fixed. They are then updated separately by their own residual losses, exactly as in Section~\ref{subsec:cefol_training_procedure}.

\subsubsection{Numerical Results}

We report two general nonlinear small-noise consumption-saving calibrations, with robustness parameters \(\sigma=1\) and \(\sigma=10\), and evaluate the learned CEFOL solutions against the VFI benchmark. Since VFI is itself a numerical approximation, the reported relative differences should be interpreted as differences from a grid-based benchmark rather than exact errors. The results are qualitatively consistent across both calibrations. In the value-function comparison, the CEFOL and VFI solutions are closely aligned: the relative value difference is generally of order \(10^{-5}\) to \(10^{-4}\) over most of the grid and remains below \(1.0\times10^{-3}\). The policy relative difference is typically around \(10^{-3}\) and remains below \(10^{-2}\). The relative Bellman error is mostly below \(10^{-4}\). The standard Euler residual is small over the unconstrained interior region and remains below \(1.3\times10^{-3}\); its larger values near the upper consumption boundary reflect that it measures the interior Euler equation rather than the constrained KKT stationarity condition. These results indicate that CEFOL remains accurate as the robustness parameter increases from \(\sigma=1\) to \(\sigma=10\). In all figures, capital \(k\) is varied over the plotted grid, while the exogenous productivity state is fixed at its steady-state value.

\paragraph{General nonlinear case with baseline robustness: $\sigma=1$}
Figure~\ref{fig:sn-general-sigma1-value} compares the direct value network, its recursive expansions, and the VFI benchmark. Figure~\ref{fig:sn-general-sigma1-value-diff} reports the relative value difference from VFI. Together these figures assess whether the learned value level and curvature match the grid-based benchmark.

\resultfig{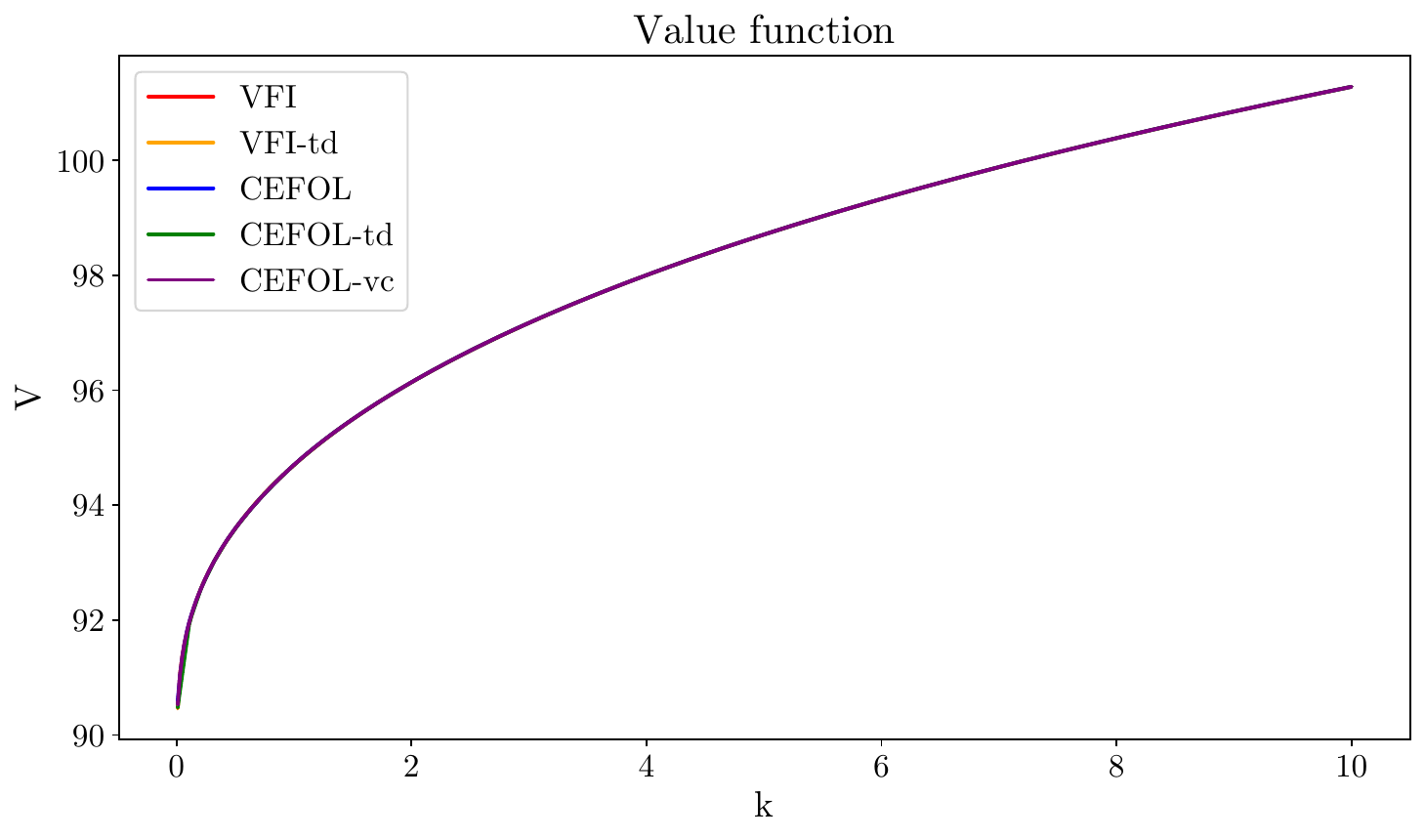}
{Value function and one-step expansions for the general small-noise model with $\sigma=1$.}
{fig:sn-general-sigma1-value}

\resultfig{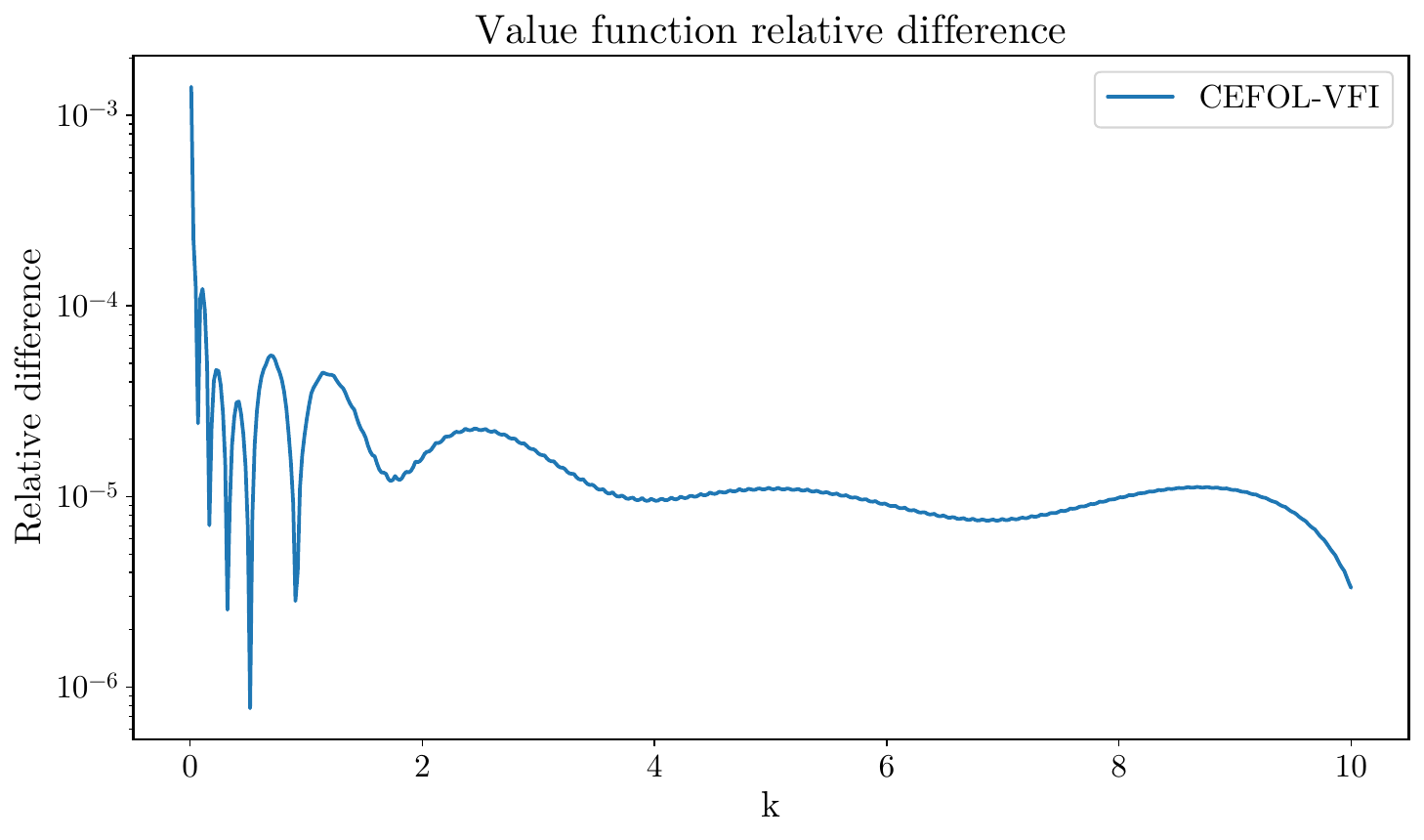}
{Relative difference between the CEFOL value network and the VFI benchmark for the general small-noise model with $\sigma=1$.}
{fig:sn-general-sigma1-value-diff}

Figure~\ref{fig:sn-general-sigma1-policy} reports the consumption policy and the wealth line. Figure~\ref{fig:sn-general-sigma1-policy-diff} reports the relative difference between the learned and VFI consumption policies.

\resultfig{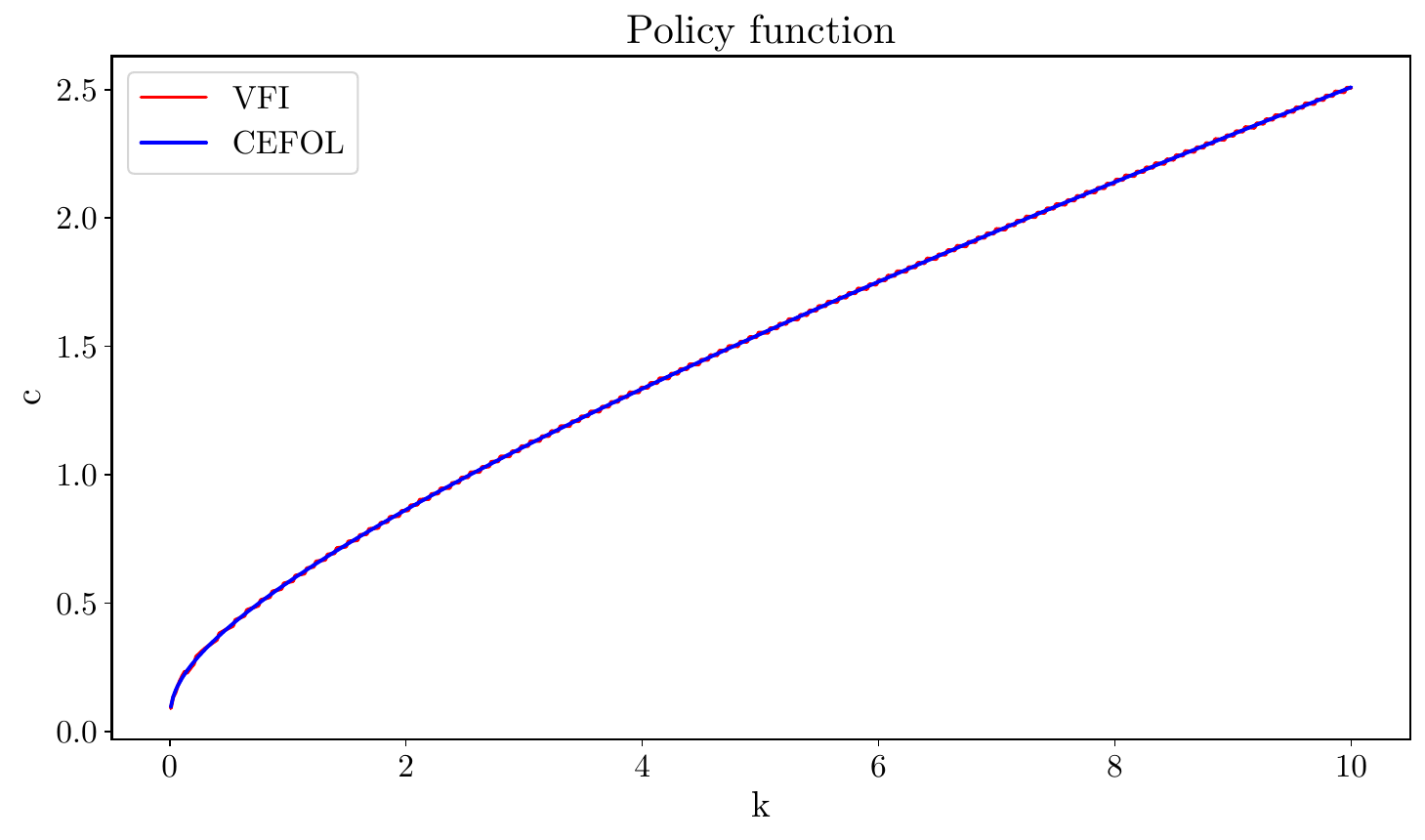}
{Consumption policy for the general small-noise model with $\sigma=1$. The dashed line reports available wealth.}
{fig:sn-general-sigma1-policy}

\resultfig{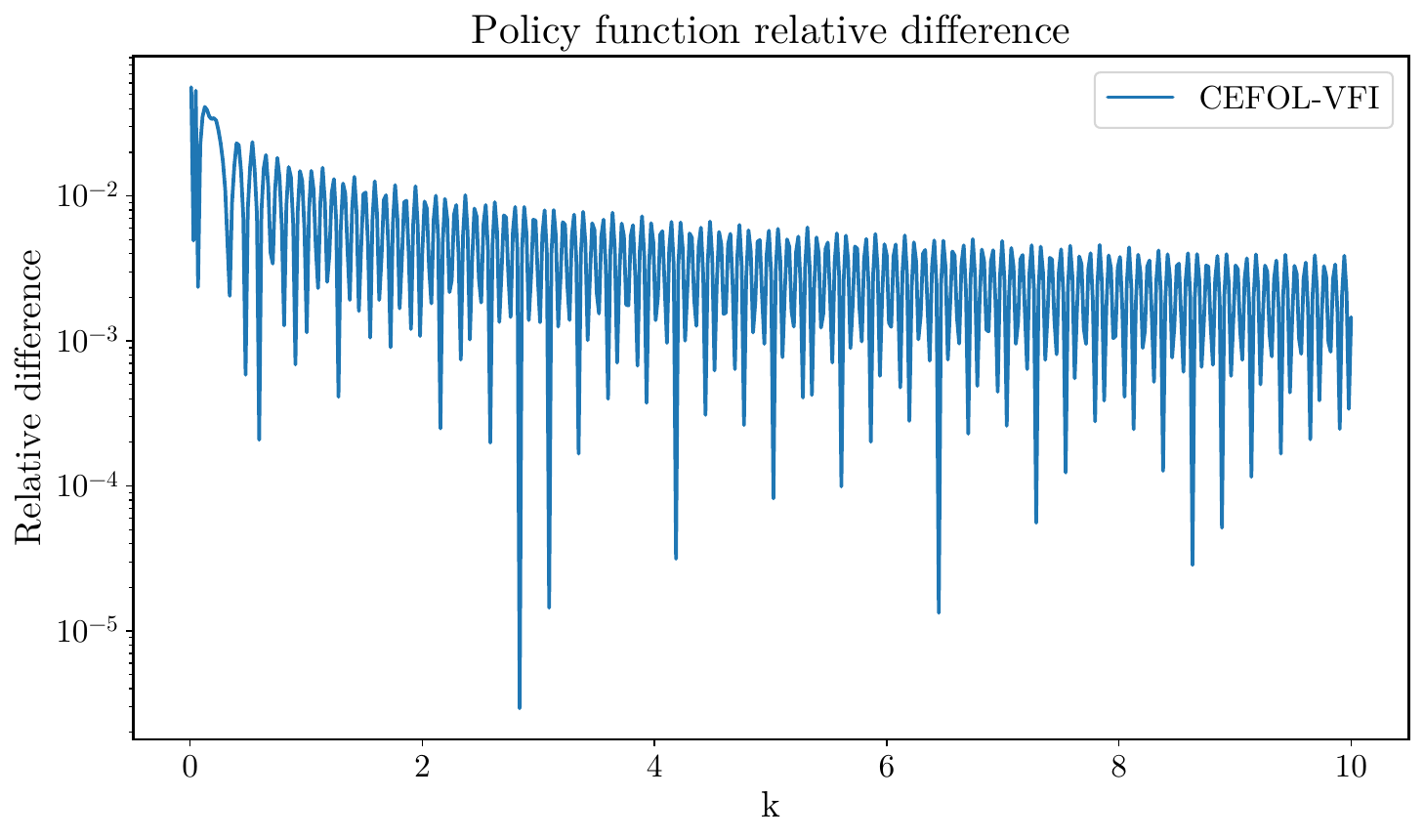}
{Relative difference between the CEFOL consumption policy and the VFI benchmark for the general small-noise model with $\sigma=1$.}
{fig:sn-general-sigma1-policy-diff}

Figures~\ref{fig:sn-general-sigma1-bellman} and \ref{fig:sn-general-sigma1-euler} report fixed-point and Euler-equation diagnostics. Both VFI and CEFOL are shown on the same logarithmic scale.

\resultfig{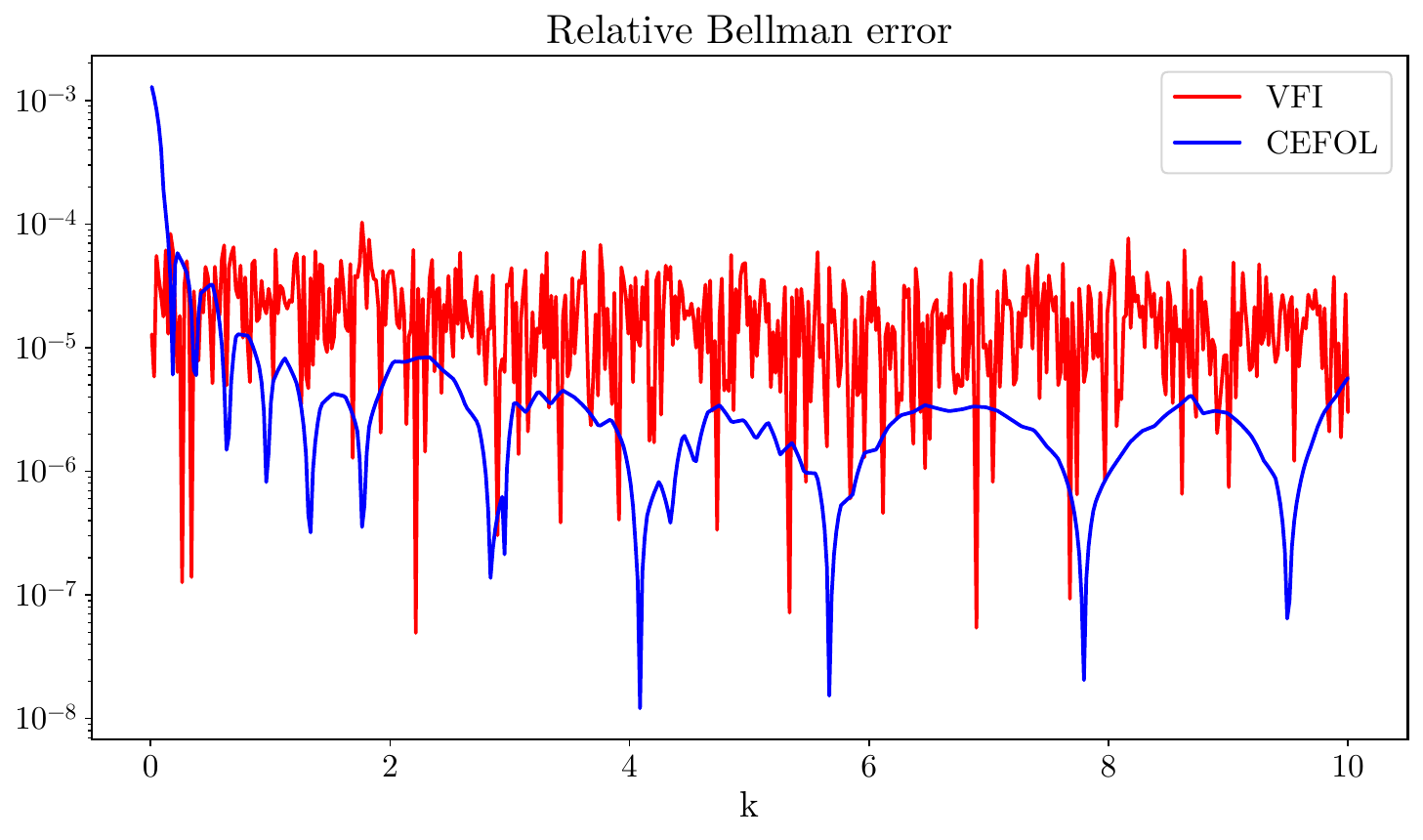}
{Relative Bellman error for the general small-noise model with $\sigma=1$.}
{fig:sn-general-sigma1-bellman}

\resultfig{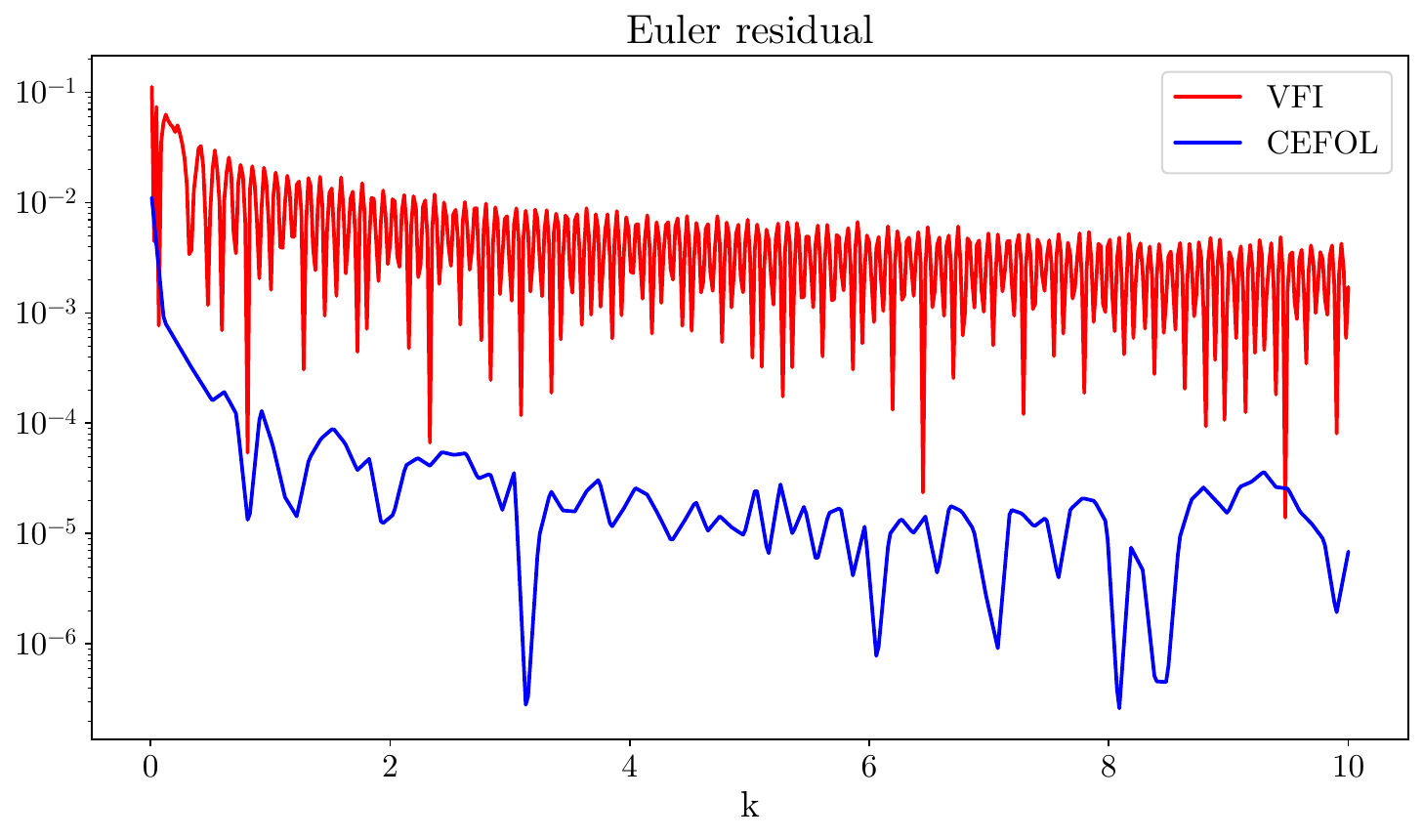}
{Euler residual for the general small-noise model with $\sigma=1$.}
{fig:sn-general-sigma1-euler}

\paragraph{General nonlinear case with moderate robustness: $\sigma=10$}
Figure~\ref{fig:sn-general-sigma10-value} compares the direct value network, its recursive expansions, and the VFI benchmark. Figure~\ref{fig:sn-general-sigma10-value-diff} reports the relative value difference from VFI. Together these figures assess whether the learned value level and curvature match the grid-based benchmark.

\resultfig{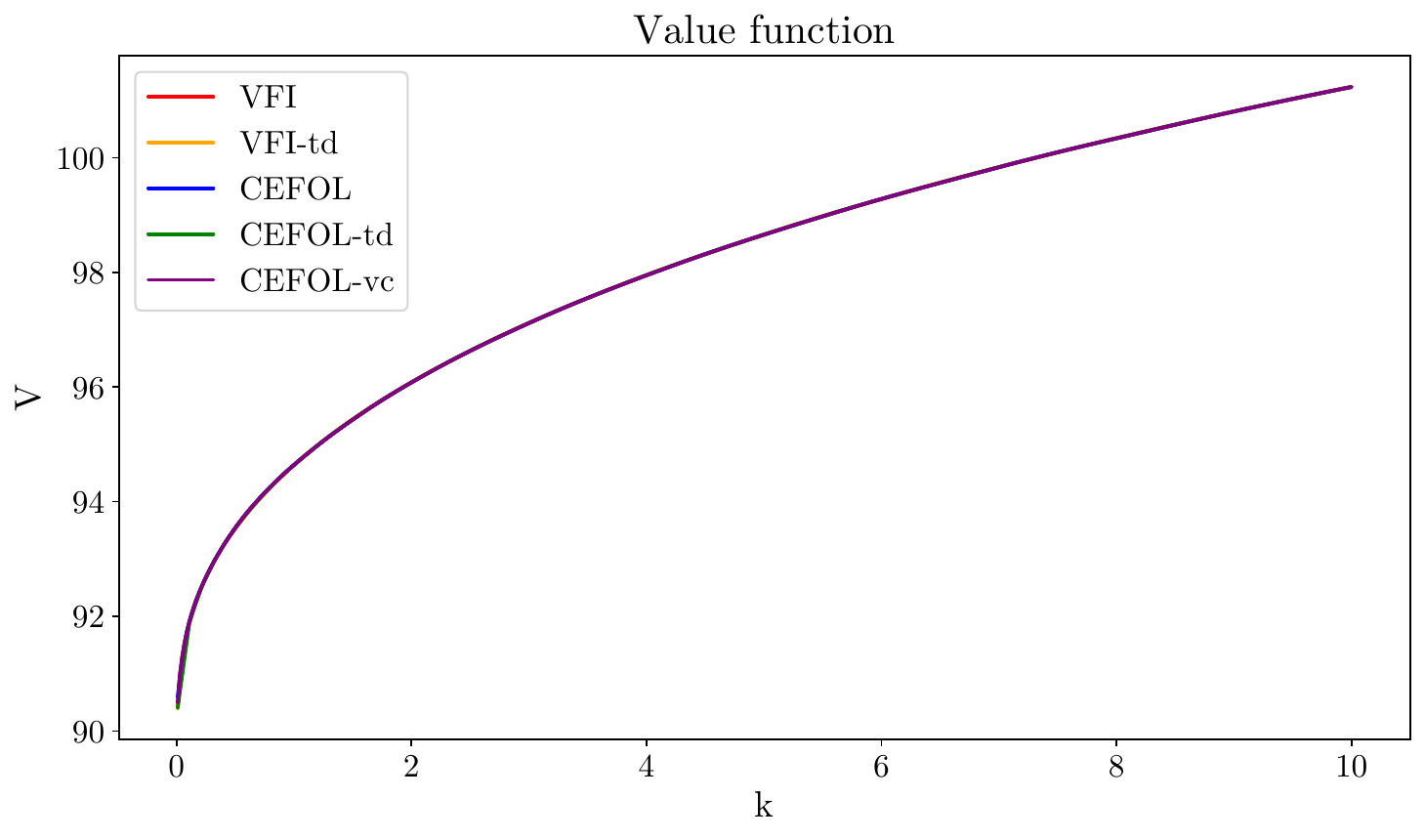}
{Value function and one-step expansions for the general small-noise model with $\sigma=10$.}
{fig:sn-general-sigma10-value}

\resultfig{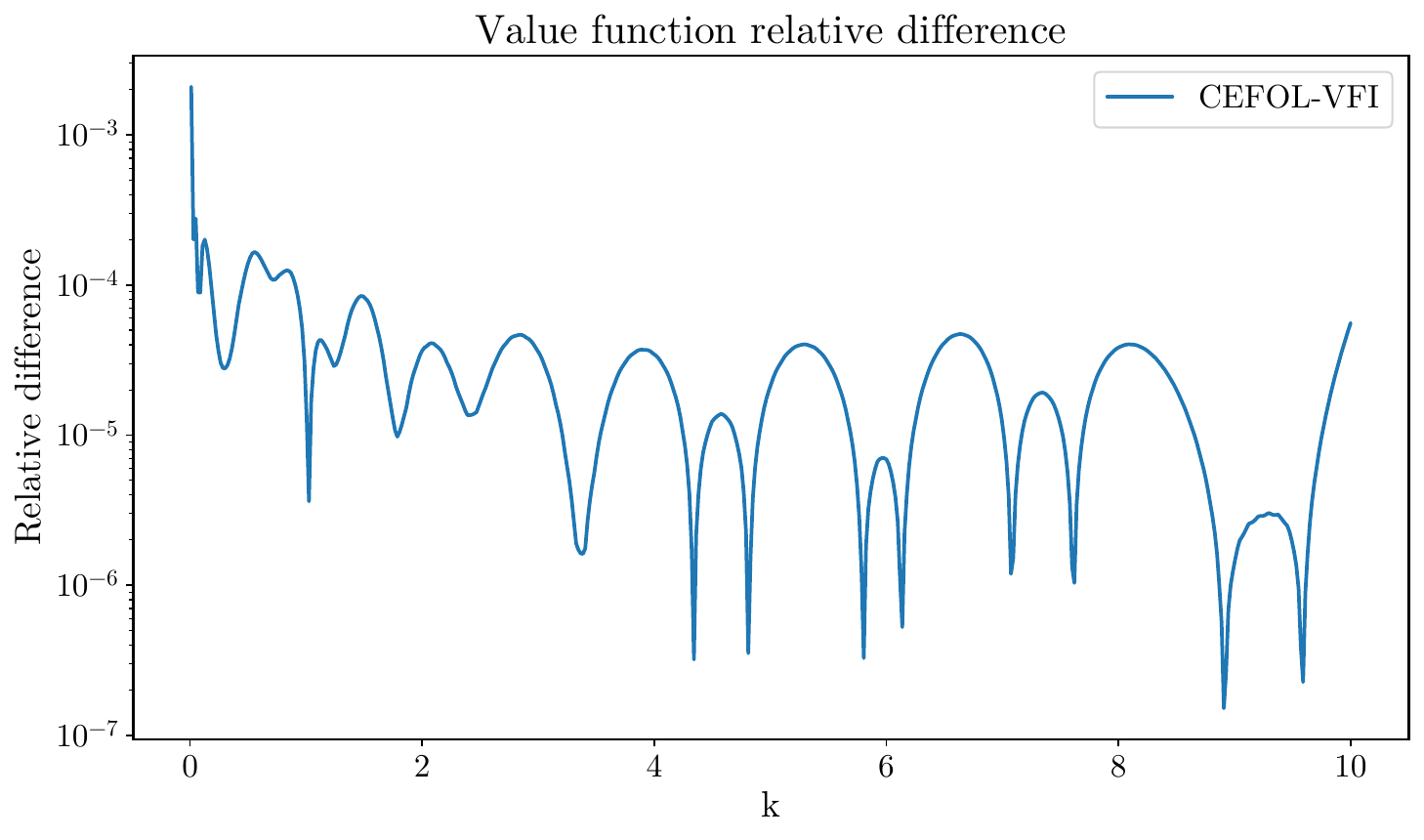}
{Relative difference between the CEFOL value network and the VFI benchmark for the general small-noise model with $\sigma=10$.}
{fig:sn-general-sigma10-value-diff}

Figure~\ref{fig:sn-general-sigma10-policy} reports the consumption policy and the wealth line. Figure~\ref{fig:sn-general-sigma10-policy-diff} reports the relative difference between the learned and VFI consumption policies.

\resultfig{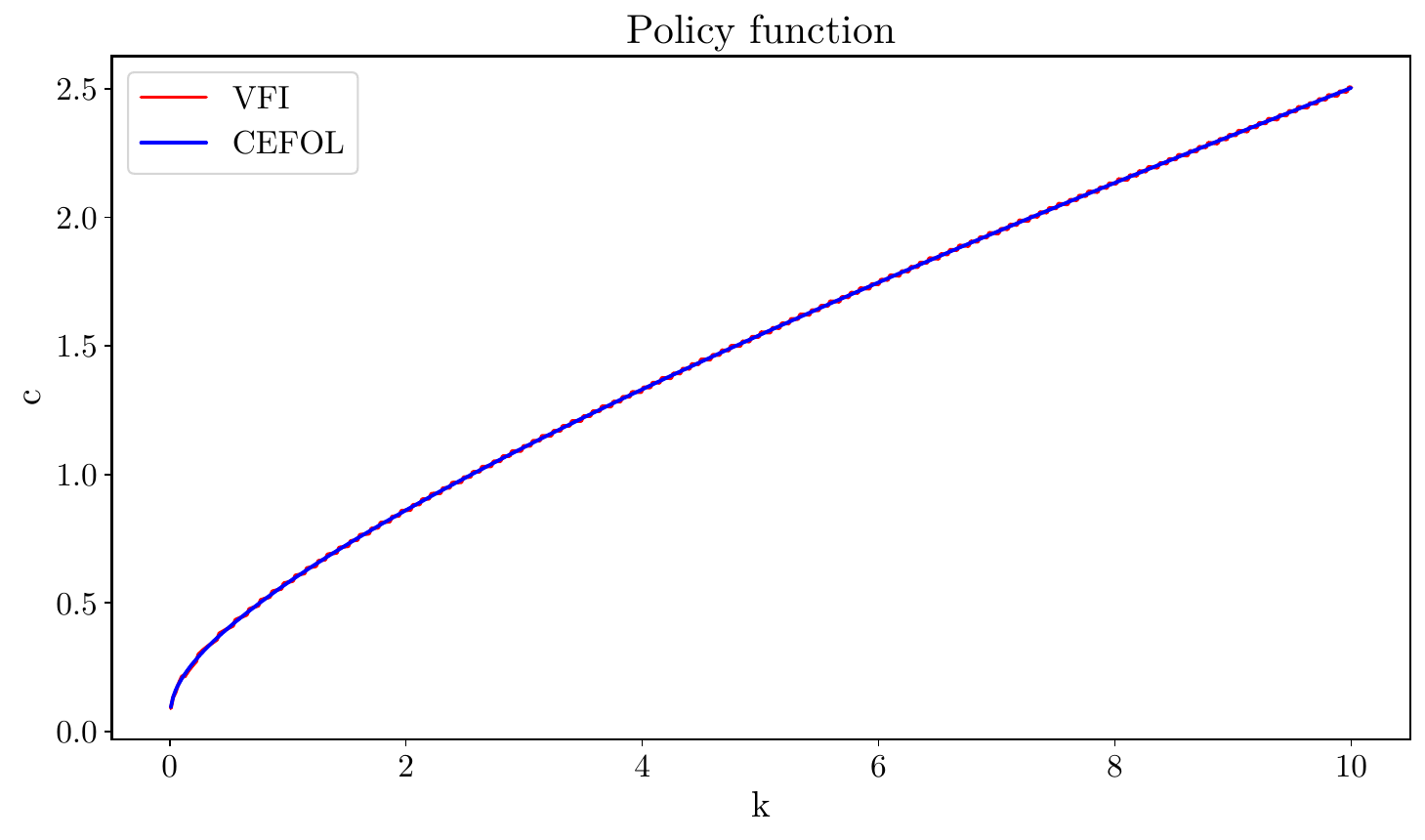}
{Consumption policy for the general small-noise model with $\sigma=10$. The dashed line reports available wealth.}
{fig:sn-general-sigma10-policy}

\resultfig{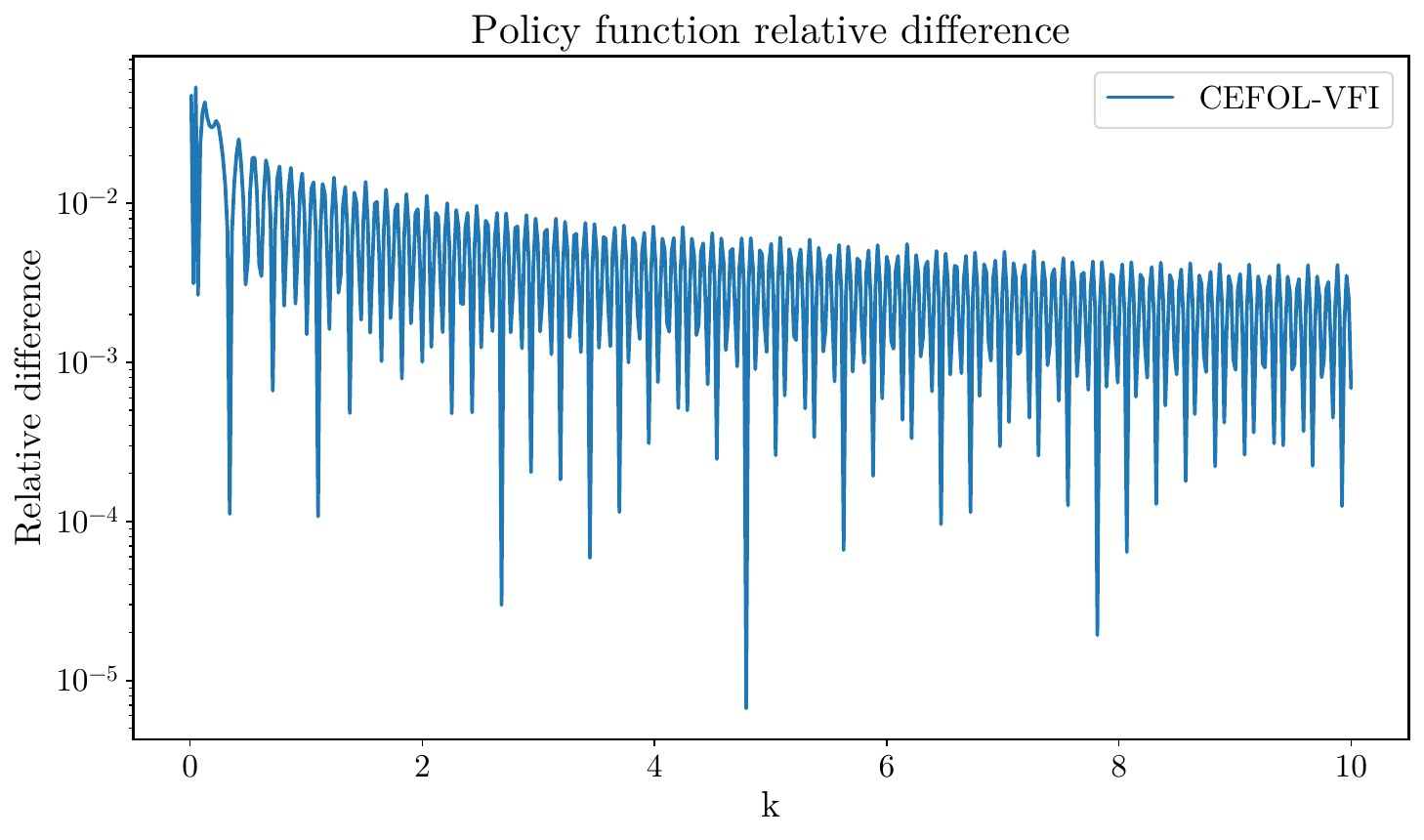}
{Relative difference between the CEFOL consumption policy and the VFI benchmark for the general small-noise model with $\sigma=10$.}
{fig:sn-general-sigma10-policy-diff}

Figures~\ref{fig:sn-general-sigma10-bellman} and \ref{fig:sn-general-sigma10-euler} report fixed-point and Euler-equation diagnostics. Both VFI and CEFOL are shown on the same logarithmic scale.

\resultfig{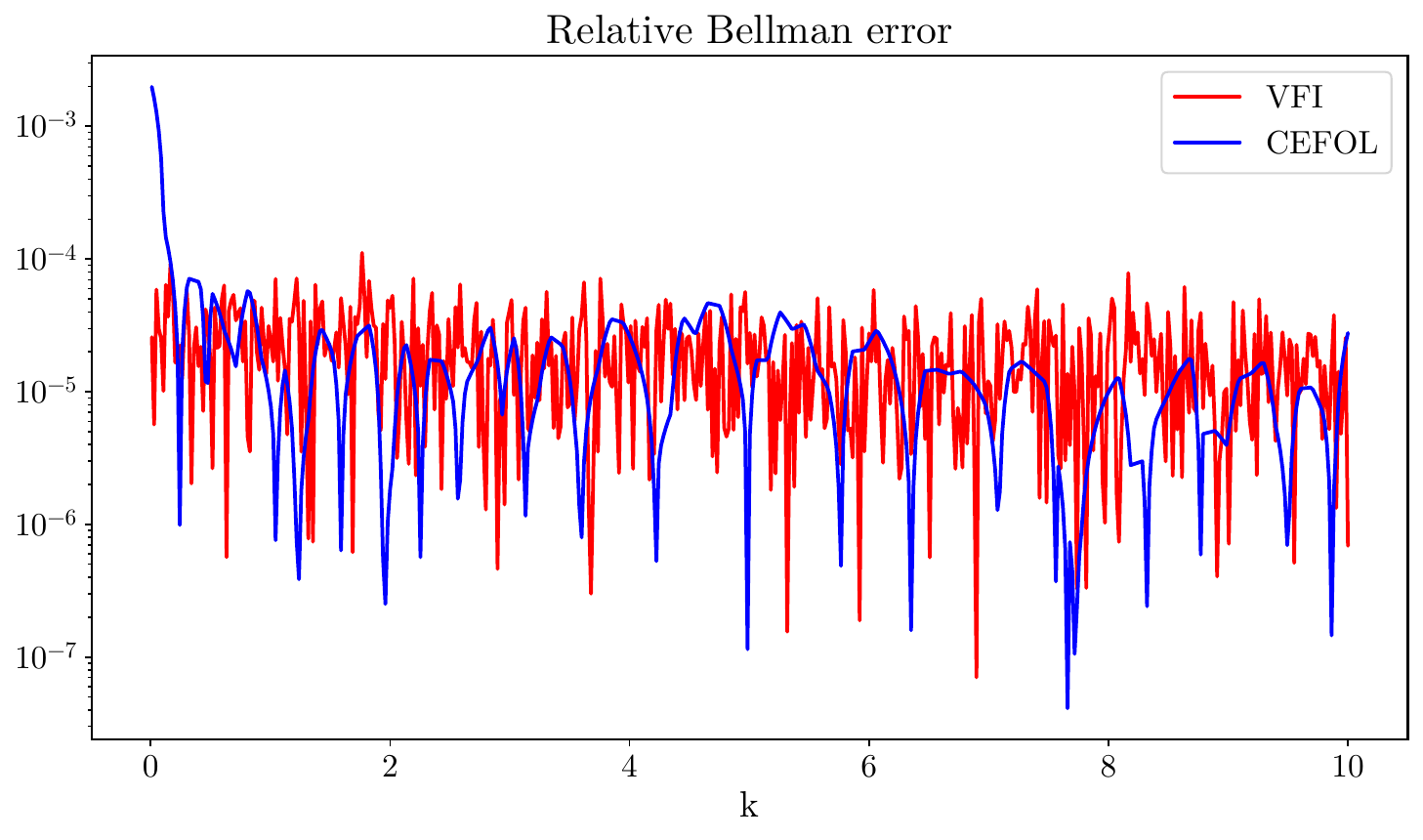}
{Relative Bellman error for the general small-noise model with $\sigma=10$.}
{fig:sn-general-sigma10-bellman}

\resultfig{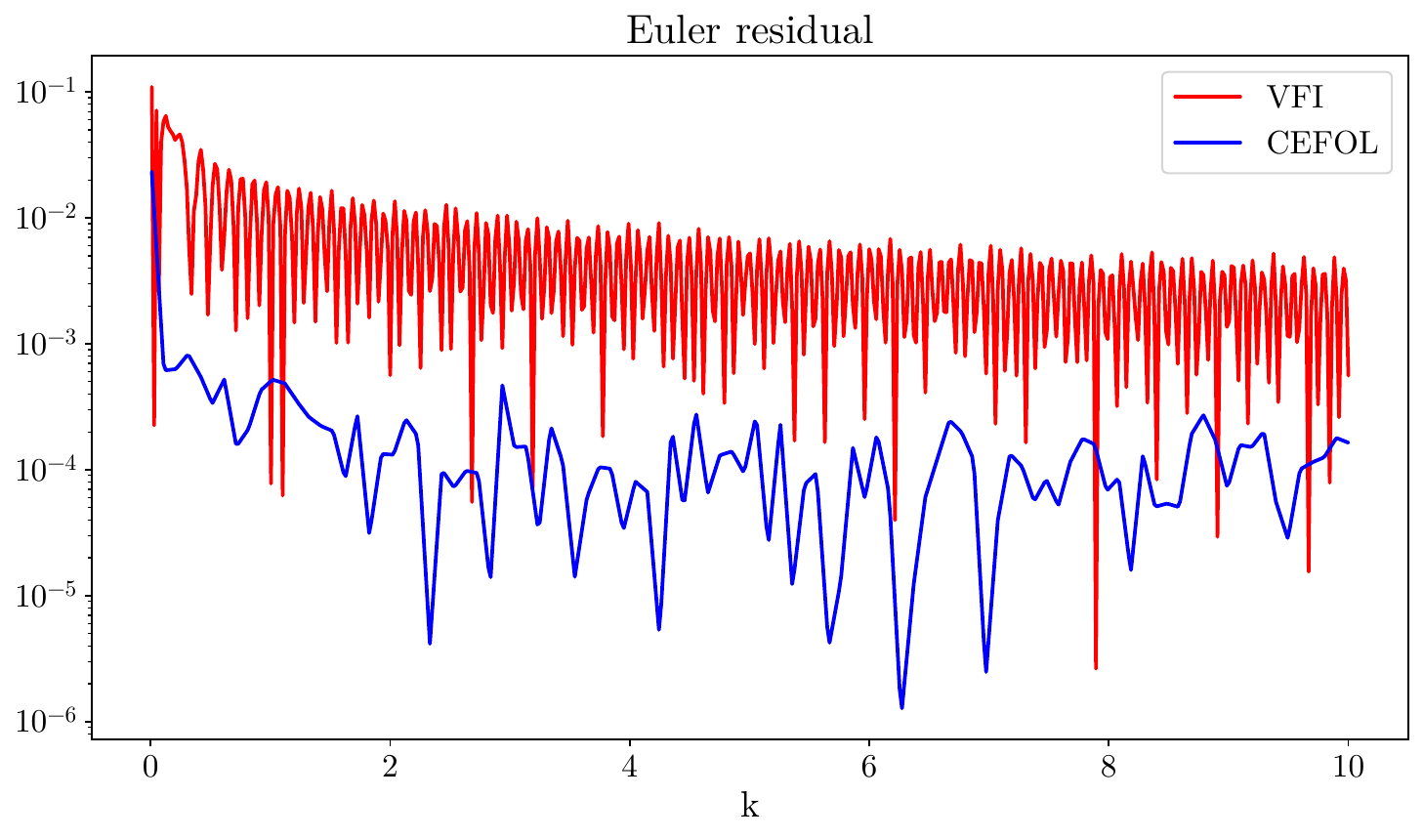}
{Euler residual for the general small-noise model with $\sigma=10$.}
{fig:sn-general-sigma10-euler}

\subsection{DSGE Model with Epstein--Zin Preferences and Stochastic Volatility}
\label{subsec:rs_dsge}

We consider a dynamic stochastic general equilibrium (DSGE) economy featuring recursive preferences and stochastic volatility in the aggregate productivity process, following \citet{caldara2012computing}. Relative to the canonical neoclassical growth framework, the model enriches the preference side through the \citet{epstein1989substitution} recursive utility specification and the shock structure through time-varying second moments. These two features render the model well suited for the joint analysis of macroeconomic quantities and asset prices. The central analytical advantage of this specification is that it disentangles the coefficient of relative risk aversion from the elasticity of intertemporal substitution, while simultaneously allowing the conditional volatility of technology shocks to fluctuate over time.

\subsubsection{Problem Setup}

Let \(k_t\) denote beginning-of-period capital, \(z_t\) current productivity, and \(\sigma_t\) log volatility. At date \(t\), the household chooses current labor \(l_t\) and the normalized consumption ratio \(a_t\). To align the notation with the generic control notation in Section~\ref{subsec:cefol_network_framework}, denote the DSGE control vector by
\begin{equation*}
c_t
=
\begin{pmatrix}
l_t\\
a_t
\end{pmatrix}
\in \mathbb R^2,
\end{equation*}
where \(l_t\) is labor supply and \(a_t\) is the normalized consumption ratio.

Given current state variables and controls, aggregate resources satisfy
\begin{equation*}
\tilde c_t+k_{t+1}
=
e^{z_t}k_t^\zeta l_t^{1-\zeta}
+
(1-\delta)k_t,
\end{equation*}
where \(\zeta\in(0,1)\) denotes the capital share and \(\delta\in[0,1]\) is the physical depreciation rate. To connect the normalized control to actual consumption, define
\begin{equation*}
a_t
=
\frac{
\tilde c_t
}{
e^{z_t}k_t^\zeta l_t^{1-\zeta}
+
(1-\delta)k_t
}.
\end{equation*}
Thus, the household allocates current resources between consumption \(\tilde c_t\) and savings carried into the next period as capital \(k_{t+1}\).

The exogenous technology process \(z_t\) follows a first-order autoregressive specification with heteroskedastic innovations:
\begin{equation*}
z_t
=
\lambda z_{t-1}
+
e^{\sigma_t}\epsilon_t,
\qquad
\epsilon_t\sim\mathcal N(0,1),
\end{equation*}
\begin{equation*}
\sigma_t
=
(1-\rho)\bar\sigma
+
\rho\sigma_{t-1}
+
\eta\omega_t,
\qquad
\omega_t\sim\mathcal N(0,1).
\end{equation*}
The log-volatility process \(\sigma_t\) is itself a stationary AR(1) process with mean \(\bar\sigma\), persistence \(\rho\), and conditional standard deviation \(\eta\). Because \(\sigma_t\) enters multiplicatively in the conditional variance of \(z_t\), the model generates time-varying macroeconomic uncertainty endogenously. This propagation mechanism is essential for producing realistic fluctuations in risk premia and for matching the observed countercyclical behavior of asset-market volatility. Accordingly, the state is
\begin{equation*}
s_t
=
(k_t,z_t,\sigma_t)
\in\mathbb R^3.
\end{equation*}

The decision problem of the representative household is defined by the recursive utility functional. Let $\psi>0$ denote the elasticity of intertemporal substitution and define
\begin{equation*}
\theta
=
\frac{1-\gamma}{1-1/\psi}.
\end{equation*}

The Bellman equation is
\begin{equation*}
V(s_t)
=
\max_{\tilde c_t,l_t}
\left\{
(1-\beta)
\left[
\tilde c_t^\nu(1-l_t)^{1-\nu}
\right]^{\frac{1-\gamma}{\theta}}
+
\beta
\left(
\mathbb E_t
\left[
V(s_{t+1})^{1-\gamma}
\right]
\right)^{\frac{1}{\theta}}
\right\}^{\frac{\theta}{1-\gamma}},
\end{equation*}
where $\nu$ determines the relative weight of consumption versus leisure in the period utility aggregator and $\gamma$ controls relative risk aversion. When $\theta\neq1$, the household's attitude toward risk and willingness to substitute consumption across time are governed by distinct parameters.

The admissible control region is
\begin{equation*}
a_t\in(0,1],
\qquad
l_t\in[0,1),
\end{equation*}
which guarantees strictly positive consumption and leisure. The lower bound \(a_t=0\) is excluded because it implies \(\tilde c_t=0\). For \(\nu\in(0,1)\), the Cobb--Douglas period utility \(U(\tilde c,l)=\tilde c^\nu(1-l)^{1-\nu}\) satisfies \(\partial U/\partial\tilde c\to\infty\) as \(\tilde c\to0\). Likewise, the upper bound \(l_t=1\) is excluded because it implies zero leisure and \(\partial U/\partial l\) becomes unbounded as \(l\to1\).

The solution is represented by the four blocks in Section~\ref{subsec:cefol_network_framework}, implemented here through six DSGE-specific subnetworks:
\begin{align*}
c_t
&=
\begin{pmatrix}
l_t\\
a_t
\end{pmatrix}
=
\begin{pmatrix}
l(s_t;\theta_l)\\
a(s_t;\theta_a)
\end{pmatrix},
\\
m_t
&=
m(s_t;\theta_m)
=
\begin{pmatrix}
m_1(s_t;\theta_{m_1})\\
m_2(s_t;\theta_{m_2})
\end{pmatrix},
\\
\mathcal C_t
&=
\mathcal C(s_t,c_t;\theta_{\mathcal C}),
\\
V_t
&=
V(s_t;\theta_V).
\end{align*}
Both policy components use sigmoid activations and therefore remain in the open interval \((0,1)\). The multiplier components use positive-output transformations to enforce nonnegativity. A target value network, updated through an exponential moving average, provides a stable regression target for the certainty-equivalent network. Apart from this DSGE-specific decomposition of the policy and multiplier blocks, the certainty-equivalent and value blocks are trained as in Section~\ref{subsec:cefol_training_procedure}.

\subsubsection{First-Order Condition and KKT Losses}
\label{subsubsec:dsge_foc_aio_residuals}

To align this DSGE specialization with the general CEFOL formulation in Section~\ref{subsec:cefol_foc_kkt_losses}, we consider two scalar controls, labor \(l_t\) and the consumption ratio \(a_t\), and two inequality constraints:
\begin{equation*}
n_c=2,
\qquad
M_g=2,
\qquad
M_q=0,
\qquad
g_1(s_t,c_t)=l_t,
\qquad
g_2(s_t,c_t)=1-a_t.
\end{equation*}
The controls are ordered as
\begin{equation*}
c_t=(l_t,a_t)',
\end{equation*}
so \(k=1\) indexes labor and \(k=2\) indexes the consumption ratio. The constraint derivatives are
\begin{equation*}
\frac{\partial g_1(s_t,c_t)}{\partial l_t}=1,
\qquad
\frac{\partial g_1(s_t,c_t)}{\partial a_t}=0,
\qquad
\frac{\partial g_2(s_t,c_t)}{\partial l_t}=0,
\qquad
\frac{\partial g_2(s_t,c_t)}{\partial a_t}=-1.
\end{equation*}

The labor and consumption-ratio policies are represented by separate policy networks,
\begin{equation*}
l_t=l(s_t;\theta_l),
\qquad
a_t=a(s_t;\theta_a),
\end{equation*}
while the two inequality constraints have separate multiplier networks,
\begin{equation*}
m_{1,t}=m_1(s_t;\theta_{m_1}),
\qquad
m_{2,t}=m_2(s_t;\theta_{m_2}).
\end{equation*}

For the Epstein--Zin transformation \(f(x)=x^{1-\gamma}\), the general distortion term is
\begin{equation}
\chi_{t+1}^{f}
=
\left(
\frac{
V(s_{t+1})
}{
\mathcal C(s_t,c_t)
}
\right)^{-\gamma}.
\label{eq:dsge_distortion}
\end{equation}
Its neural network representation is
\begin{equation}
\widehat{\chi}_{t+1}^{f}
=
\left(
\frac{
V(s_{t+1};\theta_V)
}{
\mathcal C(s_t,c_t;\theta_{\mathcal C})
}
\right)^{-\gamma}.
\label{eq:dsge_hat_distortion}
\end{equation}

Let
\begin{equation*}
w_t
=
e^{z_t}k_t^\zeta l_t^{1-\zeta}
+
(1-\delta)k_t,
\qquad
\tilde c_t=a_tw_t,
\qquad
k_{t+1}=(1-a_t)w_t,
\end{equation*}
and define
\begin{equation*}
U_t
=
\tilde c_t^\nu(1-l_t)^{1-\nu},
\qquad
U_{\tilde c,t}
=
\nu\tilde c_t^{\nu-1}(1-l_t)^{1-\nu},
\qquad
U_{l,t}
=
-(1-\nu)\tilde c_t^\nu(1-l_t)^{-\nu}.
\end{equation*}

The labor condition is static. Its population stationarity condition is
\begin{equation}
F_{t+1,1}
+
\lambda_{1,t}
=
0,
\label{eq:dsge_labor_stationarity_condition}
\end{equation}
where
\begin{equation}
F_{t+1,1}
=
(1-\beta)
V(s_t)^{1/\psi}
U_t^{-1/\psi}
\left[
U_{l,t}
+
U_{\tilde c,t}
(1-\zeta)
e^{z_t}
k_t^\zeta
l_t^{-\zeta}
\right].
\label{eq:dsge_labor_integrand}
\end{equation}
Although indexed by \(t+1\) to preserve the generic notation, \(F_{t+1,1}\) is \(\mathcal F_t\)-measurable. The corresponding network-based labor stationarity residual is
\begin{equation}
R_{1,t}^{\mathrm{stat}}
=
\widehat F_{t+1,1}
+
m_1(s_t;\theta_{m_1}),
\label{eq:dsge_labor_stationarity_residual}
\end{equation}
where \(\widehat F_{t+1,1}\) replaces \(V(s_t)\) in \eqref{eq:dsge_labor_integrand} by \(V(s_t;\theta_V)\).

The labor-stationarity loss is therefore evaluated directly by MSE:
\begin{equation}
\widehat{\mathcal L}_{S,1}^{\mathrm{MSE}}
(\theta_l,\theta_a,\theta_{m_1})
=
\frac{1}{N}
\sum_{i=1}^{N}
v_{S,1}
\left[
R_{1,t}^{\mathrm{stat},(i)}
\right]^2.
\label{eq:dsge_labor_stationarity_loss}
\end{equation}

For the intertemporal consumption--capital condition, define the gross marginal return on capital by
\begin{equation}
R^k_{t+1}
=
\zeta
e^{z_{t+1}}
k_{t+1}^{\zeta-1}
l_{t+1}^{1-\zeta}
+
1-\delta.
\label{eq:dsge_capital_return}
\end{equation}
The population stationarity condition is
\begin{equation}
\mathbb E_t
\left[
F_{t+1,2}
\right]
-
\lambda_{2,t}
=
0,
\label{eq:dsge_consumption_stationarity_condition}
\end{equation}
where the stochastic stationarity integrand is
\begin{align}
F_{t+1,2}
={}&
w_t
(1-\beta)
V(s_t)^{1/\psi}
U_t^{-1/\psi}
U_{\tilde c,t}
\nonumber\\
&\times
\Bigg[
1
-
\beta
\chi_{t+1}^{f}
\left(
\frac{
V(s_{t+1})
}{
\mathcal C(s_t,c_t)
}
\right)^{1/\psi}
\left(
\frac{U_{t+1}}{U_t}
\right)^{-1/\psi}
\frac{
U_{\tilde c,t+1}
}{
U_{\tilde c,t}
}
R^k_{t+1}
\Bigg].
\label{eq:dsge_consumption_integrand}
\end{align}
Using \eqref{eq:dsge_distortion}, the product of the two value-ratio terms can equivalently be written as
\begin{equation*}
\chi_{t+1}^{f}
\left(
\frac{
V(s_{t+1})
}{
\mathcal C(s_t,c_t)
}
\right)^{1/\psi}
=
\left(
\frac{
V(s_{t+1})
}{
\mathcal C(s_t,c_t)
}
\right)^{\frac{1}{\psi}-\gamma}.
\end{equation*}

In implementation, the sampled integrand is
\begin{align}
\widehat F_{t+1,2}
={}&
w_t
(1-\beta)
V(s_t;\theta_V)^{1/\psi}
U_t^{-1/\psi}
U_{\tilde c,t}
\nonumber\\
&\times
\Bigg[
1
-
\beta
\widehat{\chi}_{t+1}^{f}
\left(
\frac{
V(s_{t+1};\theta_V)
}{
\mathcal C(s_t,c_t;\theta_{\mathcal C})
}
\right)^{1/\psi}
\left(
\frac{U_{t+1}}{U_t}
\right)^{-1/\psi}
\frac{
U_{\tilde c,t+1}
}{
U_{\tilde c,t}
}
R^k_{t+1}
\Bigg].
\label{eq:dsge_hat_consumption_integrand}
\end{align}
Here,
\begin{equation*}
l_{t+1}=l(s_{t+1};\theta_l),
\qquad
a_{t+1}=a(s_{t+1};\theta_a),
\end{equation*}
and
\begin{equation*}
\tilde c_{t+1}
=
a_{t+1}
\left[
e^{z_{t+1}}
k_{t+1}^{\zeta}
l_{t+1}^{1-\zeta}
+
(1-\delta)k_{t+1}
\right].
\end{equation*}

For each future draw \(j\), the network-based consumption--capital stationarity residual is
\begin{equation}
R_{2,t}^{\mathrm{stat},(i,j)}
=
\widehat F_{t+1,2}^{(i,j)}
-
m_2(s_t^{(i)};\theta_{m_2}).
\label{eq:dsge_consumption_stationarity_residual}
\end{equation}

Suppose \(N_z\) is even and let \(J_z=N_z/2\). For each mini-batch state \(s_t^{(i)}\), divide the \(N_z\) future draws into two conditionally independent groups:
\begin{equation}
\overline R_{2,t}^{\mathrm{stat},(i,1)}
=
\frac{1}{J_z}
\sum_{j=1}^{J_z}
R_{2,t}^{\mathrm{stat},(i,j)},
\qquad
\overline R_{2,t}^{\mathrm{stat},(i,2)}
=
\frac{1}{J_z}
\sum_{j=J_z+1}^{N_z}
R_{2,t}^{\mathrm{stat},(i,j)}.
\label{eq:dsge_split_consumption_residuals}
\end{equation}
The all-in-one consumption--capital stationarity loss is
\begin{equation}
\widehat{\mathcal L}_{S,2}^{\mathrm{AiO}}
(\theta_l,\theta_a,\theta_{m_2})
=
\frac{1}{N}
\sum_{i=1}^{N}
v_{S,2}
\overline R_{2,t}^{\mathrm{stat},(i,1)}
\overline R_{2,t}^{\mathrm{stat},(i,2)}.
\label{eq:dsge_consumption_stationarity_loss}
\end{equation}

The overall stationarity loss is
\begin{equation}
\widehat{\mathcal L}_{S}
=
\widehat{\mathcal L}_{S,1}^{\mathrm{MSE}}
+
\widehat{\mathcal L}_{S,2}^{\mathrm{AiO}}.
\label{eq:dsge_stationarity_loss}
\end{equation}

The KKT conditions are
\begin{equation*}
l_t\geq0,
\qquad
\lambda_{1,t}\geq0,
\qquad
l_t\lambda_{1,t}=0,
\end{equation*}
and
\begin{equation*}
1-a_t\geq0,
\qquad
\lambda_{2,t}\geq0,
\qquad
(1-a_t)\lambda_{2,t}=0.
\end{equation*}
The Fischer--Burmeister loss is therefore
\begin{equation}
\begin{aligned}
\widehat{\mathcal L}_{FB}
(\theta_l,\theta_a,\theta_{m_1},\theta_{m_2})
=
\frac{1}{N}
\sum_{i=1}^{N}
\Bigg\{
&v_{FB,1}
\left[
\Phi^{FB}
\left(
l_t^{(i)},
m_1(s_t^{(i)};\theta_{m_1})
\right)
\right]^2
\\
&+
v_{FB,2}
\left[
\Phi^{FB}
\left(
1-a_t^{(i)},
m_2(s_t^{(i)};\theta_{m_2})
\right)
\right]^2
\Bigg\}.
\end{aligned}
\label{eq:dsge_fb_loss}
\end{equation}
Since \(M_q=0\), no equality-constraint term appears in this model.

The total first-order/KKT loss is
\begin{equation}
\widehat{\mathcal L}_{\mathrm{FOC}}
=
\lambda_S
\widehat{\mathcal L}_{S}
+
\lambda_{FB}
\widehat{\mathcal L}_{FB}.
\label{eq:dsge_foc_loss}
\end{equation}
The policy and multiplier networks are updated jointly:
\begin{equation*}
\begin{aligned}
&
(\theta_l,\theta_a,\theta_{m_1},\theta_{m_2})
\\
&\leftarrow
(\theta_l,\theta_a,\theta_{m_1},\theta_{m_2})
-
\alpha_{cm}
\nabla_{\theta_l,\theta_a,\theta_{m_1},\theta_{m_2}}
\widehat{\mathcal L}_{\mathrm{FOC}}.
\end{aligned}
\end{equation*}
During this policy--multiplier step, the value and certainty-equivalent networks are held fixed. They are subsequently updated using their respective residual losses, as described in Section~\ref{subsec:cefol_training_procedure}.

\subsubsection{Numerical Results}

We report DSGE results for the calibration $(\beta,\gamma,\psi)=(0.9,5.0,0.5)$. The direct value network, the one-step temporal-difference expansion, and the certainty-equivalent-implied value are closely aligned. The consumption-ratio and labor policies vary smoothly over the evaluation range. The relative Bellman error remains below $10^{-4}$ throughout the evaluation range, and the FOC residual remains below $10^{-3}$ across the capital range. In all figures, capital $k_t$ is varied over $[0.5k_{ss},1.5k_{ss}]$, while productivity and volatility are held at their steady-state values.

\paragraph{Results: $\beta=0.9$, $\gamma=5.0$, $\psi=0.5$}

This calibration combines high risk aversion with low intertemporal substitutability.

\resultfig{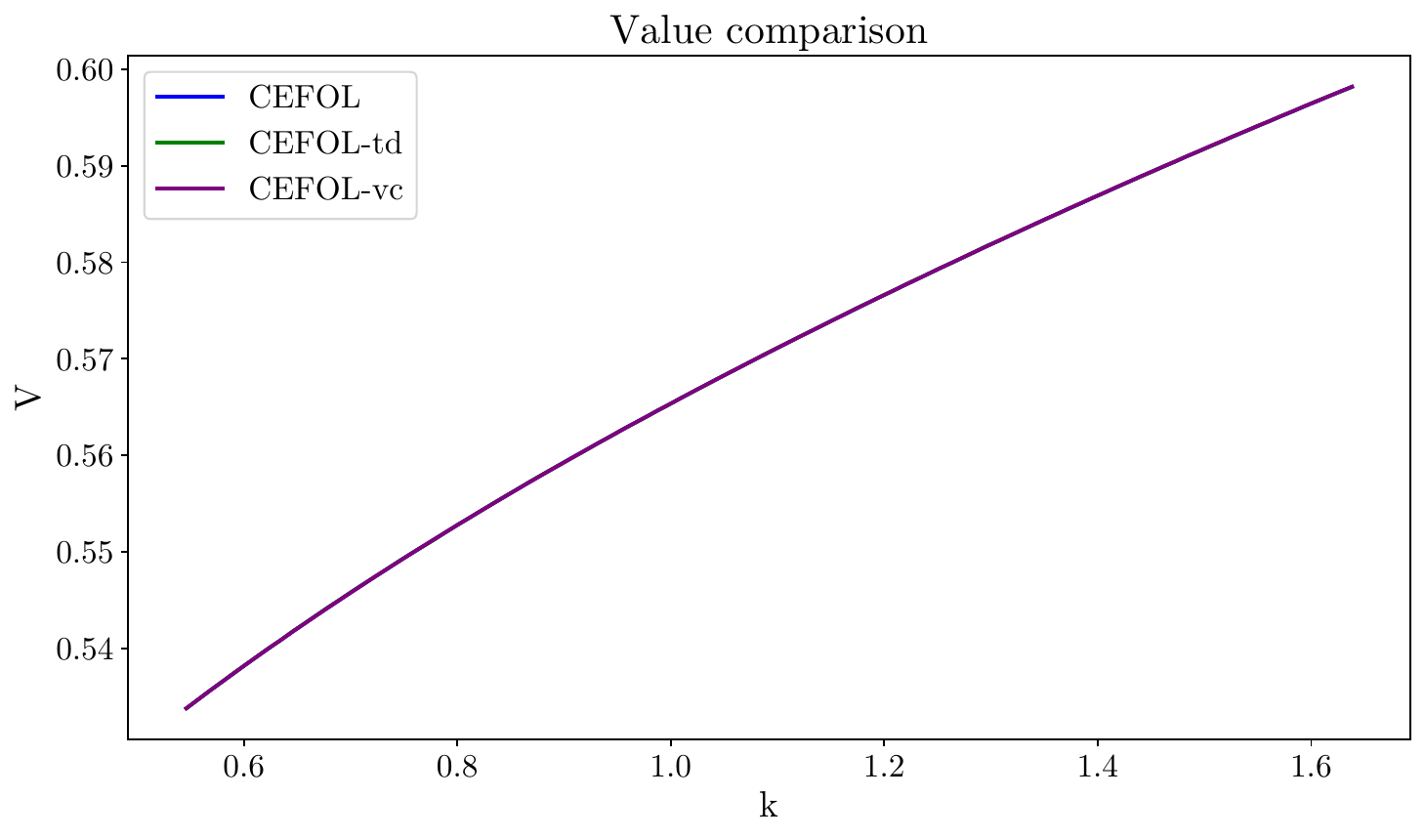}
{Value function comparison for the DSGE model. Capital $k_{t}$ varies from $0.5 k_{s s}$ to $1.5 k_{s s}$, while productivity $z_{t}$ and volatility $\sigma_{t}$ are fixed at their steady-state values.}
{fig:dsge-beta09-gamma5-psi05-value}

\resultfig{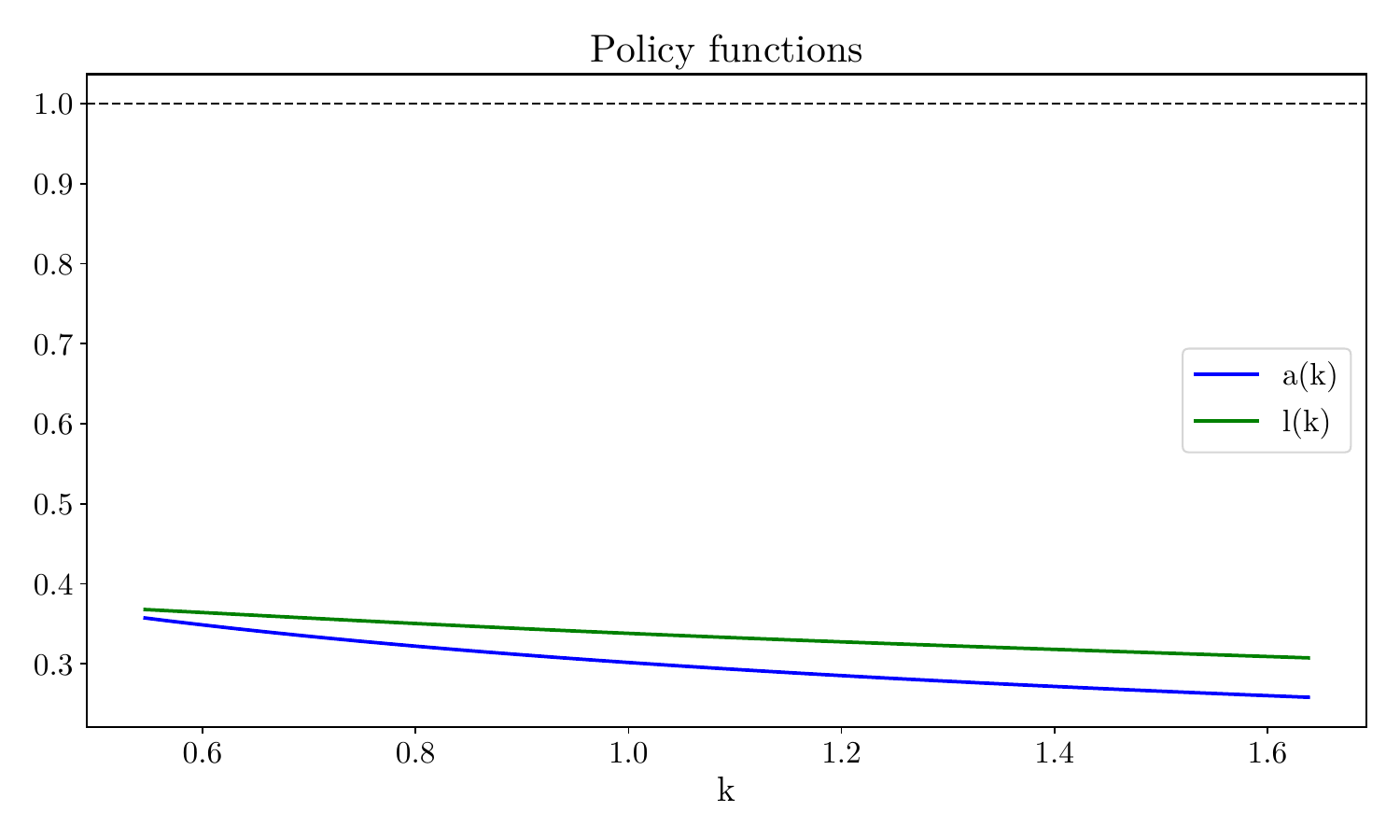}
{Policy functions for the DSGE model. The figure reports the consumption ratio $a_{t}$ and labor supply $l_{t}$ as functions of capital $k_{t}$, with $z_{t}$ and $\sigma_{t}$ fixed at their steady-state values.}
{fig:dsge-beta09-gamma5-psi05-policy}

\resultfig{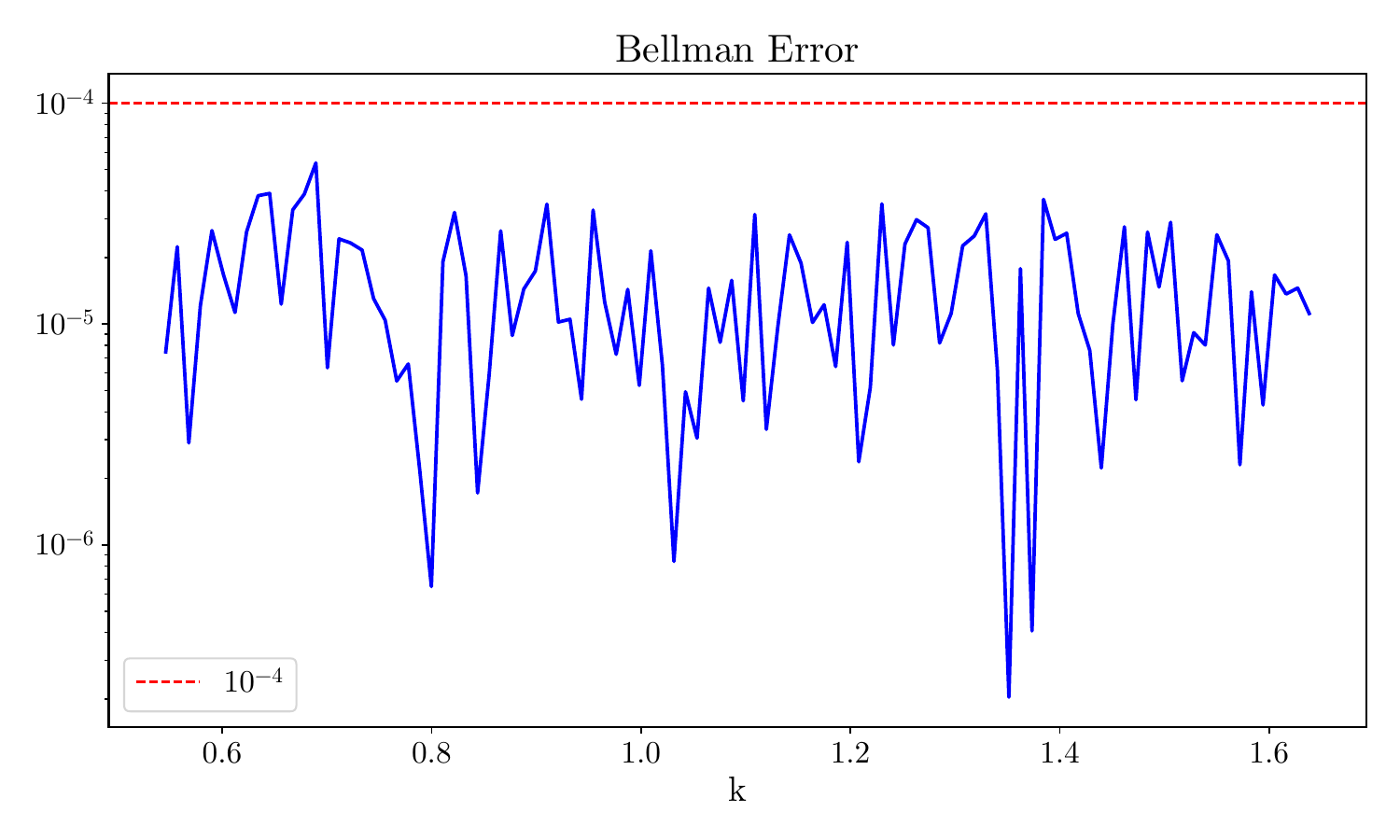}
{Bellman error for the DSGE model. Capital $k_{t}$ varies from $0.5 k_{s s}$ to $1.5 k_{s s}$, while the other state variables are fixed at their steady-state values.}
{fig:dsge-beta09-gamma5-psi05-bellman}

\resultfig{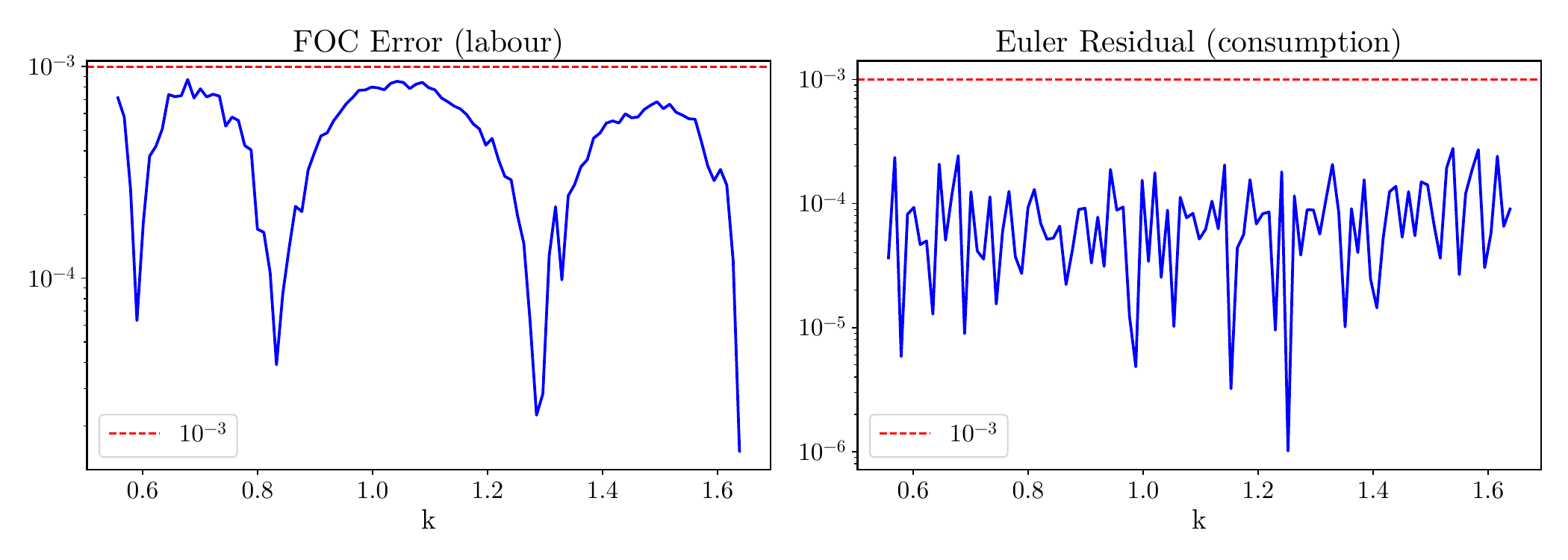}
{FOC residuals of labor and consumption for the DSGE model. Capital $k_{t}$ varies from $0.5 k_{s s}$ to $1.5 k_{s s}$, with productivity and volatility fixed at their steady-state values.}
{fig:dsge-beta09-gamma5-psi05-euler}

\section{Conclusion}
\label{sec:conclusion}

This paper develops the certainty-equivalent first-order Learning algorithm (CEFOL), a simulation-based deep learning method for solving discrete-time dynamic programming problems with recursive utility. The method is motivated by a key difficulty of recursive utility: nonlinear certainty-equivalent values enter both the Bellman equation and the model-specific optimality conditions. In such problems, policy learning cannot be reduced to standard expected-utility Euler residuals or Bellman residuals without explicitly accounting for the nonlinear certainty-equivalent term.

CEFOL’s core idea is to approximate the state-control certainty-equivalent value using a neural network and embed this approximation into first-order, stationarity, Euler, and KKT residual conditions. The algorithm therefore provides a policy-focused learning framework: the policy and multiplier networks are disciplined by model-specific first-order and KKT residuals, while the value and certainty-equivalent networks maintain recursive Bellman consistency and certainty-equivalent value accuracy. Classical Euler residuals are included as a special case of the broader first-order residual system.

CEFOL accommodates general equality and inequality constraints on the controls through a unified KKT framework, in which the multiplier network and the Fischer--Burmeister loss jointly enforce feasibility, nonnegativity, and complementarity. This formulation handles occasionally binding constraints without penalty functions or problem-specific reformulations, making the method directly applicable to borrowing limits, portfolio-share restrictions, market-clearing conditions, and other economically motivated restrictions.

The paper also presents several neural-network architectures for implementing CEFOL. The baseline four-network architecture combines a value network, a policy network, a multiplier network, and a certainty-equivalent network. A five-network variant decomposes the certainty-equivalent block into a conditional-expectation network and a nonlinear difference network, which can be useful when the conditional expectation of next period value and the certainty-equivalent value have different numerical scales. A compact three-network variant removes the separate value update and evaluates the next period values through an implicit recursive mapping. Across these architectures, the first-order/KKT residual remains the central policy-learning objective.

To make first-order residual loss computationally tractable, CEFOL adapts the all-in-one expectation operator to recursive utility. By using conditionally independent future shock draws, the algorithm estimates squared conditional first-order moments without explicitly computing an inner conditional expectation at every state-control pair. This makes the method mesh-free and simulation-based, while still allowing out-of-sample diagnostic evaluation through more accurate nested simulation when desired.

The numerical experiments support the effectiveness of CEFOL across all model classes considered. In the risk-sensitive and Epstein--Zin consumption-saving models, the learned value, certainty-equivalent, and policy components are mutually consistent with the recursive Bellman and first-order conditions across alternative robustness and preference calibrations. In the small-noise robust-control model, the learned solutions closely reproduce the relevant VFI benchmarks in the general nonlinear case. In the DSGE model, CEFOL remains effective in a setting with a multidimensional state vector, multiple continuous controls, stochastic volatility, and inequality constraints. Across the experiments, value-function comparisons, Bellman diagnostics, policy functions, and first-order-condition residuals consistently indicate that the learned solutions are accurate over the economically relevant state region.

\bibliographystyle{dcu} 
	\bibliography{reference_final}
	

\end{document}